 \documentclass{aa}  

\usepackage[]{txfonts}
\usepackage{natbib}  
\usepackage[figuresright]{rotating}
\usepackage{graphicx}
\usepackage{color}

\begin{document}

   \title{Simultaneous \textit{Kepler}/K2 and XMM-Netwon observations of superflares in the Pleiades}


   \author{M. G. Guarcello\inst{1} \and G. Micela\inst{1} \and S. Sciortino\inst{1} \and J. L\'{o}pez-Santiago\inst{2} \and C. Argiroffi\inst{1,3} \and F. Reale\inst{1,3} \and E. Flaccomio\inst{1} \and J. D. Alvarado-G\'{o}mez\inst{4} \and V. Antoniou\inst{4} \and J. J. Drake\inst{4} \and I. Pillitteri\inst{1} \and L. M. Rebull\inst{5,6} \and J. Stauffer\inst{6}}

   \institute{INAF - Osservatorio Astronomico di Palermo, Piazza del Parlamento 1, I-90134, Palermo, Italy\\
              \email{mario.guarcello@inaf.it}
                \and
                Dpto. de Astrof\'{i}sica y Cencias de la Atm\'{o}sfera, Universidad Complutense de Madrid, 28040 Madrid, Spain  USA
                \and
               Dip. di Fisica e Chimica, Universit\'{a} di Palermo, Piazza del Parlamento 1, 90134, Palermo, Italy
                \and
                Harvard-Smithsonian Center for Astrophysics, 60 Garden Street, Cambridge, MA 02138,
                \and
                Infrared Science Archive (IRSA), Infrared Processing and Analysis Center (IPAC), California Institute of Technology, 1200 E. California Blvd., Pasadena, CA 91125, USA
                \and
                Spitzer Science Center (SSC), Infrared Processing and Analysis Center (IPAC), California Institute of Technology, 1200 E. California Blvd., Pasadena, CA 9112, USA
              }

   \date{}

 
  \abstract
   {Flares are powerful events ignited by a sudden release of magnetic energy which triggers a cascade of interconnected phenomena, each resulting in emission in different electromagnetic bands. In fact, in the Sun flares are observed across the whole electromagnetic spectrum. Multi-band observations of stellar flares are instead rare. This limits our ability to extend what we learn from solar flares to the case of flares occurring in stars with different properties.}
   {With the aim of studying flares in the 125-Myr-old stars in the Pleiades observed simultaneously in optical and X-ray light, we obtained new XMM-Newton observations of this cluster during the observations of \textit{Kepler} K2 Campaign 4. The objective of this paper is to characterize the most powerful flares observed in both bands and to constrain the energy released in the optical and X-ray, the geometry of the loops, and their time evolution. We also aim to compare our results to existing studies of flares occurring in the Sun and stars at different ages.}
   {We selected bright X-ray/optical flares that occurred in 12 known members of the Pleiades from their K2 and XMM-Newton light curves. The sample includes ten K-M stars, one F9 star, and one G8 star. Flare average properties were obtained from integrated analysis of the light curves during the flares. The time evolution of the plasma in the magnetic loops is constrained with  time-resolved X-ray spectral analysis. }
   {Most of the flares studied in this work emitted more energy in optical than in X-rays, as in most solar flares, even if the Pleiades flares output a larger fraction of their total energy in X-rays than typical solar flares do. Additionally, the energy budget in the two bands is weakly correlated. We also found comparable flare duration in optical and X-rays and observed that rapidly rotating stars (e.g., with rotation period shorter than 0.5 days) preferentially host short flares. We estimated the slope of the cooling path of the flares in the log(EM)-versus-log(T) plane. The values we obtained are affected by large uncertainties, but their nominal values suggest that the flares analyzed in this paper are mainly due to single loops with no sustained heating occurring during the cooling phase. We also observed and analyzed oscillations with a period of 500$\,$s during one of the flares.}


   \maketitle
%

\section{Introduction}
\label{sec_intro}


Stellar flares are very powerful magnetic phenomena occurring after a rearrangement of the magnetic field topology and the resulting reconnection of magnetic field lines in stellar coronae \citep[e.g.,][]{FletcherDHJK2011SSRv}. During flares, a large amount of energy previously stored in the magnetic field is released, accelerating a flow of high-speed electrons up to mega-electronvolt energies. This energy can be deposited both locally at the reconnection site and onto the underlying dense chromosphere at the loop footpoints by the electron beams moving downward along the field lines, heating the surrounding plasma to $\geq$10 million degrees. Very quickly, the pressure in the heated plasma exceeds the ambient chromospheric pressure, making the heated plasma expand and fill the overlying magnetic tubes. The plasma then gradually cools down as soon as heating decreases and stops. \par

      This general scenario is supported by extensive observations of solar flares across all accessible wavelengths. Flares are in fact intrinsic multi-wavelength phenomena, releasing energy in several bands of the electromagnetic spectrum \citep{Benz2008LRSP.5.1B}: accelerated electrons emit gyrosynchrotron radiation and nonthermal hard X-rays upon Coulomb collisions with particles in the chromospheric plasma \citep[e.g.,][]{HoyngDMR1981ApJ.246L.155H}; the heated region of the chromosphere and upper photosphere surrounding the loop footpoints radiates in optical and UV bands \citep[e.g.,][]{Neidig1989SoPh.121.261N}; and the evaporated plasma confined in the magnetic loop emits soft X-rays. 
      
Observed correlations between the energy released in different bands can probe the complex interplay between the phenomena occurring during flares. For instance, the correlation observed between the energy released in optical and hard X-rays in solar flares was taken as proof of the connection between the energy deposited in the chromosphere and the heating of the hot spots in the deeper layers \citep{MatthewsVHN2003AA.409.1107M,Metcalf2003ApJ.595.483M,HudsonWM2006SoPh.234.79H}. However, an important unanswered question remains as to whether this and other results obtained from solar flares can be directly extended to stellar flares. This cannot be taken for granted given that the energy released by flares occurring in the Sun and in stars across the Hertzsprung-Russel diagram can differ by orders of magnitude.  Flares in young and active stars can be particularly energetic. For instance, in pre-main sequence (PMS) stars, flares typically release $\sim$10$^{34}\,$ergs of energy in soft X-rays \citep[e.g.,][]{FavataFRM2005ApJs}, with the brightest events reaching $\sim$10$^{36}\,$ergs, while the total irradiance measurements for some of the largest solar flares indicate radiated energies of between 10$^{31}$ and 10$^{32}\,$ergs. Although technically challenging, multi-wavelength observations of flares occurring in stars with different properties are therefore required in order to fully understand the physics of flares and to determine whether or not the ensemble of phenomena in play change during stellar evolution and the gradual transition to lower magnetic activity levels as a result of spin-down. \par
      
In this paper we analyze one of the rare simultaneous optical and X-ray observations of stellar flares thanks to the XMM/Newton observations of the Pleiades taken during the \textit{Kepler}/K2 Campaign 4. The aim is to calculate and compare flare properties in both bands. In Sect. \ref{sec_pleiades} we describe previous X-ray and optical observations in the time domain of the Pleiades, and in Sects \ref{k2_data_sec} and \ref{xray_data_sec} we present the \textit{Kepler} and X-ray data, respectively. A sample of 12 stars hosting flares analyzed in detail in this paper is described in Sect. \ref{sample_sec}. We then first derive and analyze flare properties averaged over flare duration (Sect. \ref{glob_flare}) and subsequently analyze the time evolution of the emitting plasma (Sect. \ref{time_res}). Finally, in Sect. \ref{thats_all_folks} we compare our results to those of existing studies of flares. \par

\section{Existing studies on optical and X-ray variability of the Pleiades}
\label{sec_pleiades}

\begin{figure}[!h]
        \centering      
        \includegraphics[width=8cm]{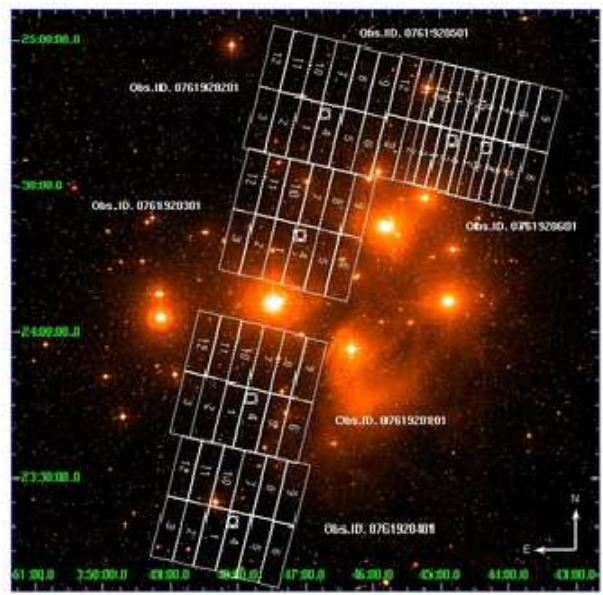}
        \caption{DSS2-red $2^\circ \times2^\circ$ image centered on the Pleiades. The detector footprints of the XMM pointings analyzed here and their associated Obs.IDs. are shown.}
        \label{xmm_fields}
        \end{figure}

The Pleiades open cluster (also named M45, Melotte~22, and NGC~1432; the name "the Seven Sisters" is used to indicate its brightest members), located in the constellation of Taurus, is the brightest stellar cluster in the sky. It is therefore not surprising that it is one of the most studied astronomical targets, with dedicated literature including more than one thousand published papers, and continues to be considered a cornerstone for understanding stellar physics. With an age of 125$\,$Myr \citep{StaufferSK1998ApJ} and a mean distance of 134$\pm$6$\,$pc (from the Gaia Data Release 2, indicating a mean parallax of 7.5$\pm$0.3$\,$mas, \citealp{Gaia2016AA595A2G}), the Pleiades is a rich cluster, with 2107 stars identified by \citet{BouyBSB2015AA577A148B} as having a high probability of being members from an analysis of accurate multi-epoch proper motion measurements and multi wavelength photometry. Other recent compilations of candidate members were obtained by \citet{StaufferHFA2007ApJS172663S}, \citet{LodieuDH2012MNRAS.422.1495L}, \citet{BouyBMC2013AA554A.101B}, and \citet{SarroBBB2014AA563A45S}. A $2^\circ \times2^\circ$ DSS2-red image of the Pleiades is shown in Fig. \ref{xmm_fields}, with  the XMM fields analyzed in this work indicated. \par

A number of authors have analyzed the optical variability in the Pleiades, classifying several types of variable stars. In recent years, the revolution in optical time-domain astronomy due to the unprecedented photometric precision of the \textit{Kepler} telescope has  led to several discoveries in this open cluster. Using K2 \citep{Howell2014PASP.126.398H} data, \citet{RebullSBC2016AJa} determined the rotation periods of cluster members; \citet{RebullSBC2016AJb} analyzed the light curves of those members with multiple periodic signals; \citet{StaufferRBH2016AJ} focused on the angular momentum evolution and magnetic dynamo of young low-mass stars using Pleiades members as a template; and \citet{White2017MNRAS.471.2882W} analyzed the variability of the Seven Sisters, classifying six of them as slowly pulsating B-stars, while the variability observed in the remaining star (Maia) was attributed to rotational modulation of a large spot. Spot coverage and its connection with stellar properties and rotational periods was studied by \citet{Savanov2017ARep.61.996S} on a sample of 759 confirmed members. \par

    The Pleiades cluster is also a prime target to study stellar X-ray variability across a wide range of stellar spectral types. Evidence for solar-like cyclic activity in the low-mass Pleiades stars was found by \citet{Schmitt1993AA.277.114S}, comparing ROSAT All Sky Survey data with Einstein/IPC data in order to cover a time baseline of about 10 years. \citet{GagneCS1995ApJ} observed X-ray variability with an amplitude larger than a factor two over $\sim$1$\,$yr in approximately 25\% of Pleiades members. Large amplitude variability over a large timescale and flares in X-rays were also observed by \citet{Micela1996ApJS.102.75M} using ROSAT observations. ROSAT observations have also been  analyzed by \citet{Marino2003AA.406.629M} to verify that X-ray variability is a common feature in late-type Pleiades members over both short (hours) and medium (months) timescales. \par

X-ray flares occurring in 20 Pleiades stars observed with ROSAT were studied by \citet{Stelzer2000AA.356.949S}, finding decay times ranging between 0.3 and 12.6$\,$ksec, peak luminosity in the range 29.8$\leq$log(L$_{\rm X}$[erg/s])$\leq$31.2, and total emitted energies in the range 32.6$\leq$log(E$_{\rm X}$[erg])$\leq$34.4. X-ray flares in the Pleiades stars HII~1032, HII~1100, HII~1516 and HII~1280 have been observed with XMM by \citet{Briggs2003MNRAS.345.714B}. In these four flares, the observed peak luminosity ranged between 29.6$\leq$log(L$_{\rm X}$[erg/s])$\leq$30.8 in the 0.3-4.5$\,$keV energy band and they were characterized by a decay time of 2--2.9 ks. The two strongest flares were observed in HII~1032 and HII~1100, with peak plasma temperatures and loop emission measures (EM) of T$_{\rm peak}$=39$^{46}_{32}\,$MK and EM=3.4$\times10^{53}\,$cm$^{-3}$ in HII~1032, T$_{\rm peak}$=24$^{27}_{22}\,$MK and EM=1.4$\times10^{53}\,$cm$^{-3}$ in HII~1100. The light curve of HII~1100 also shows a peculiar dip of 600$\,$s occurring just before the onset of the X-ray flare. \par

\section{Data and light curves}
 
\subsection{Kepler/K2 observations}
\label{k2_data_sec}

The original \textit{Kepler} mission \citep{Borucki2010Sci.327.977B} was designed with the primary purpose of detecting planets using the method of transits. Thanks to its superb photometric precision in the 420--900$\,$nm band pass, the mission has not only allowed us to identify hundreds of confirmed and thousands of candidate exoplanets, but has also provided an inestimable collection of light curves of stars across most of the Hertzsprung-Russell diagram. It is therefore not surprising that \textit{Kepler} data has a huge impact not only on exoplanetary science, but also on several other fields of astronomy, such as asteroseismology \citep{Chaplin2010ApJ.713L.169C}, variability of low mass stars \citep[e.g.,][]{McQuillan2014ApJS.211.24M}, eclipsing binaries \citep{Prsa2011AJ.141.83P}, flares and superflares \citep{Maehara2012Natur.485.478M,Davenport2016ApJ}, and even variability in active galactic nuclei \citep{Pei2014ApJ.795.38P}. \par

        The \textit{Kepler} primary mission ended with the loss of two reaction wheels, which prevented the telescope from maintaining  stable pointing. The mission was extended by aligning the telescope along its orbital plane, minimizing the torque from solar radiation pressure. The extended \textit{Kepler}/K2 mission has been designed to cover a series of fields along the ecliptic, each for $\sim$80 days  \citep{Howell2014PASP.126.398H}. The K2 mission has provided new opportunities to study stellar variability. For instance, the original \textit{Kepler} field did not contain stellar clusters younger than 1$\,$Gyr, while K2 fields include PMS stars and young clusters (e.g., Upper Sco, Taurus Molecular Cloud, the Pleiades, M32, the Hyades, Praesepe, and others). \par 

The Pleiades have been included in the K2 Campaign 4 and observed from 2015 February 8 to 2015 April 20, providing light curves for a total of 1020 high-probability Pleiades members, {spread over most of the cluster field}. In this paper we analyze the pre-search data conditioning light curves generated by the K2 project in the long-cadence mode (time resolution of $\sim$30$\,$min), and obtained from the Mikulski Archive for Space Telescopes. We included only data points not flagged as affected by thruster firings and any other bad data flags.

\subsection{XMM-Newton observations}
\label{xray_data_sec}

    The six X-ray observations of the Pleiades analyzed in this paper (P.I. Jeremy Drake) were obtained on 10, 11, 13, 17, 18, and 26 February 2015  with XMM-Newton using the European Photon Imaging Camera (EPIC) during the fourth \textit{Kepler}/K2 campaign. EPIC observations were performed 
simultaneously with three detectors (PN, MOS1, and MOS2), which are approximately co-pointed but have different shapes, FoVs, and sensitivities (with the PN detector being the most sensitive). The observations shown in Fig. \ref{xmm_fields} were designed in order to avoid the brightest stars in the field and to optimize the number of K2 targets potentially observable. All these observations were performed with a roll angle of 256$^\circ$, using the medium filter, but with different exposure time as shown in Table \ref{obs_table}. \par
    
    \begin{table}
    \caption{XMM observations of the Pleiades}
    \label{obs_table}
    \centering                       
    \begin{tabular}{|c|c|c|c|}
    \hline
    \multicolumn{1}{|c|}{Obs.ID$^1$} &
    \multicolumn{1}{|c|}{RA} &
    \multicolumn{1}{|c|}{DEC} &
    \multicolumn{1}{|c|}{Texp} \\
    \hline
    \multicolumn{1}{|c|}{} &
    \multicolumn{1}{|c|}{J2000} &
    \multicolumn{1}{|c|}{J2000} &
    \multicolumn{1}{|c|}{ksec} \\
    \hline
    101  & 03:47:43.87 & +23:47:55.9 & 53  \\
    201  & 03:46:37.62 & +24:46:38.6 & 58  \\
    301  & 03:47:00.08 & +24:21:30.7 & 60  \\
    401  & 03:47:59.81 & +23:22:26.3 & 74.9\\
    501  & 03:44:43.13 & +24:40:57.4 & 89.9\\
    601  & 03:44:13.00 & +24:39:21.0 & 83.6\\
    \hline
    \multicolumn{4}{l}{$^1$ Only the last three digits after 0761920- are shown.} \\
    \end{tabular}
    \end{table}

\begin{figure*}[!h]
        \centering      
        \includegraphics[width=7cm]{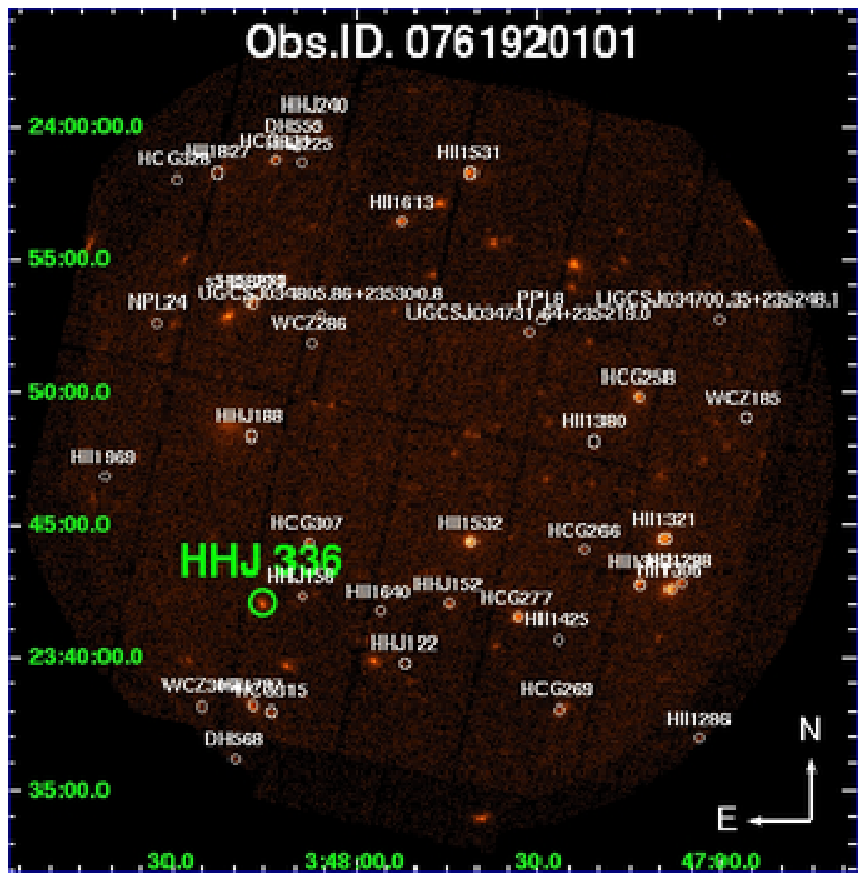}
        \includegraphics[width=7cm]{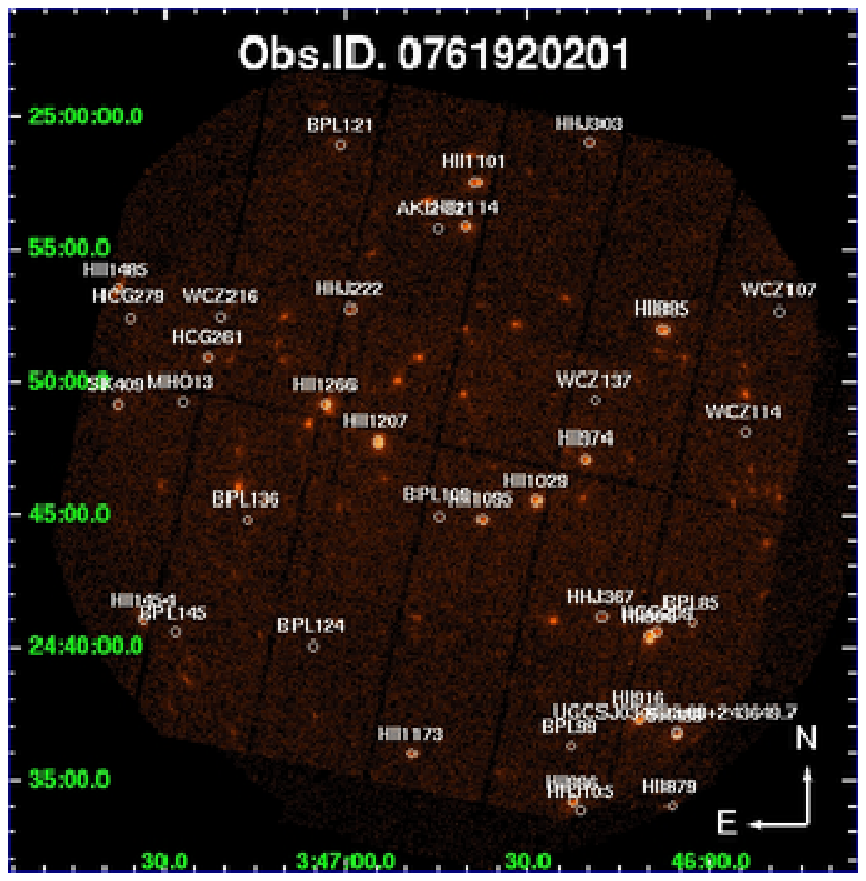}
        \includegraphics[width=7cm]{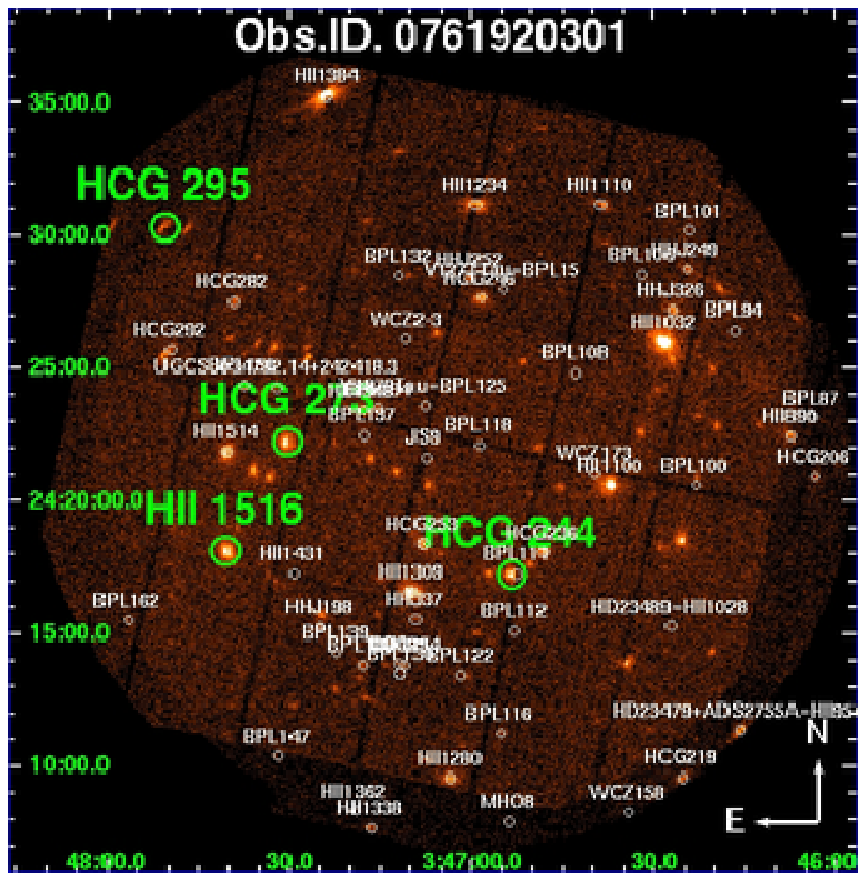}
        \includegraphics[width=7cm]{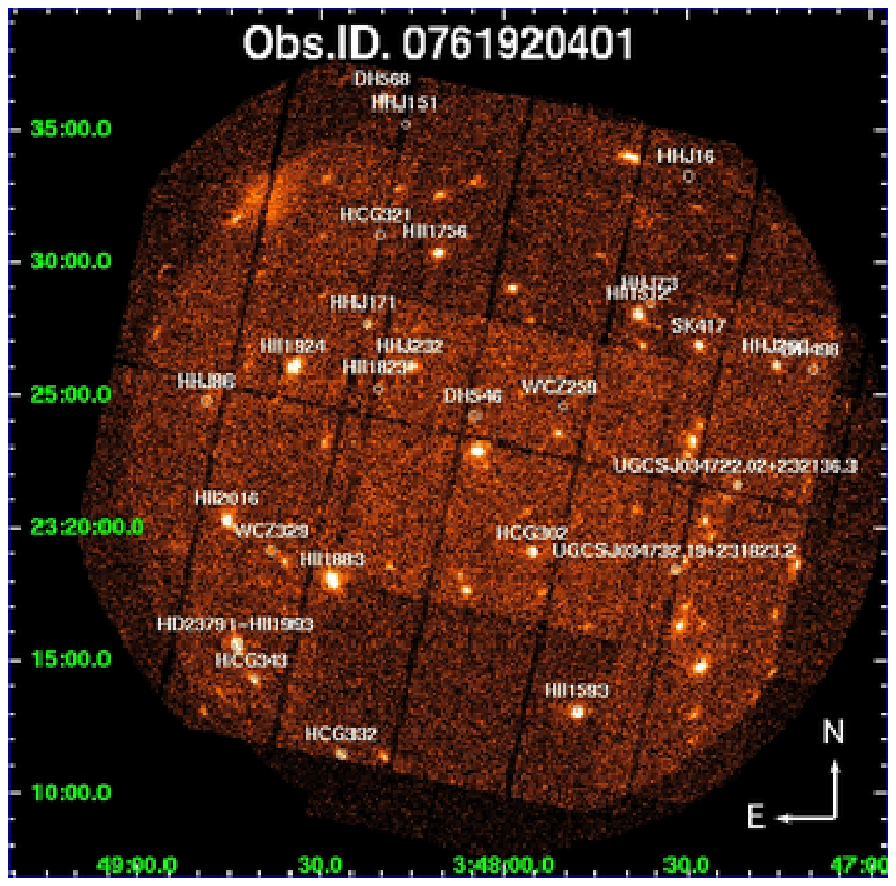}
        \includegraphics[width=7cm]{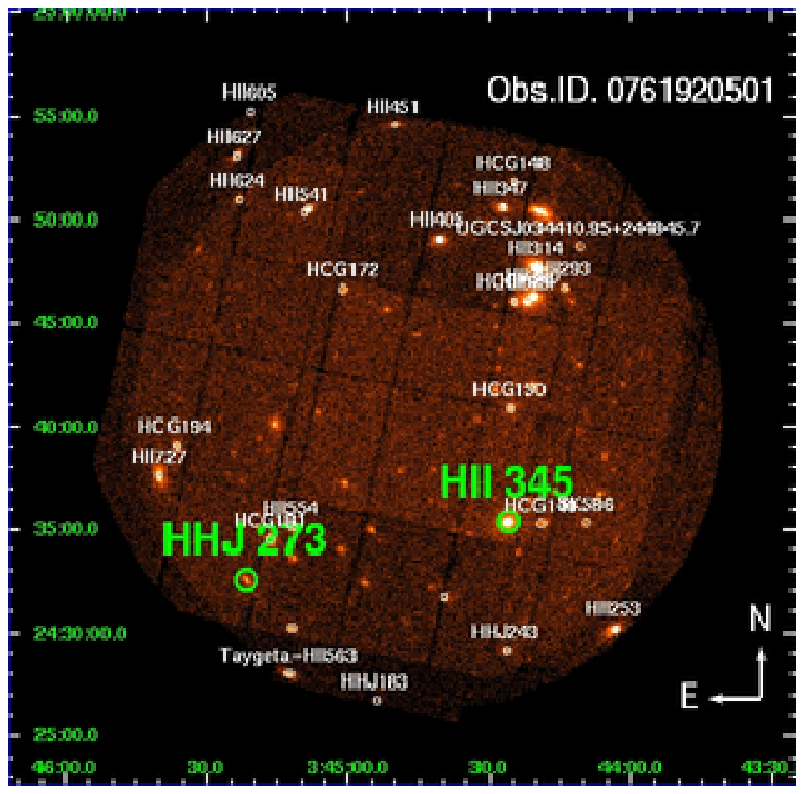}
        \includegraphics[width=7cm]{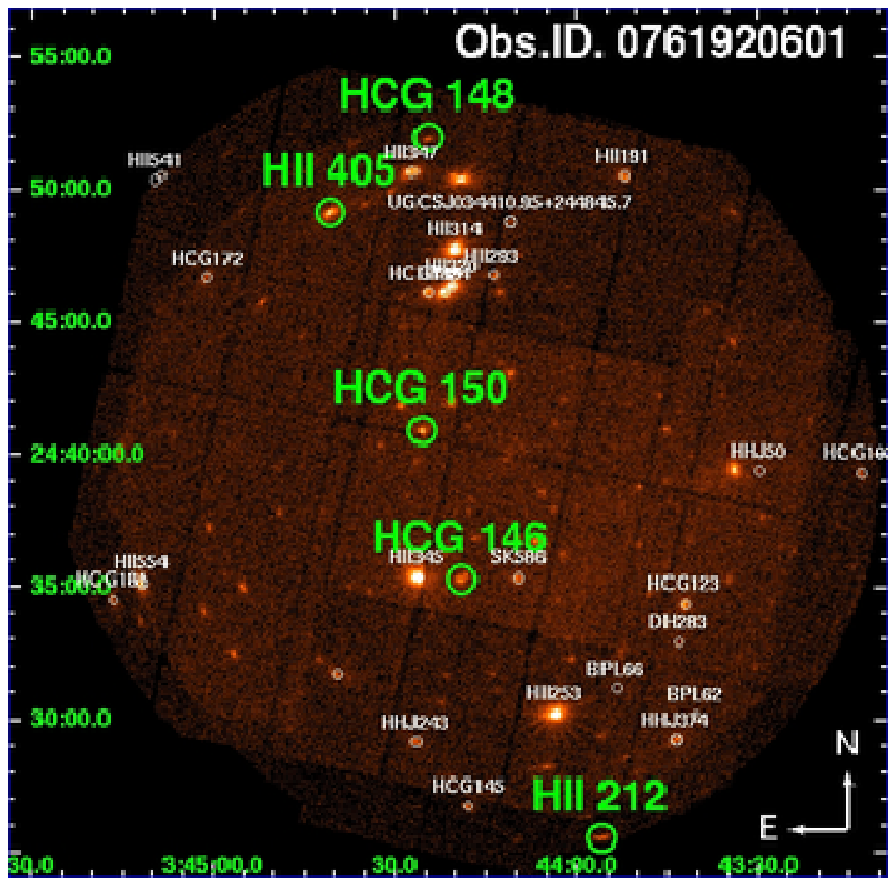}
        \caption{Combined MOS1, MOS2, and PN background-filtered images of the six XMM fields. Large green circles mark the sources with bright flares discussed in this paper, while the small circles denote all the K2 targets falling in these fields. Images were smoothed using a Gaussian smoothing function with a kernel radius of 2 pixels.}
        \label{combined_fields}
        \end{figure*}

    The MOS1, MOS2, and PN images of the six fields are shown in Fig.~\ref{combined_fields}, where we marked the positions of the sources hosting the bright flares studied in this paper and those of all the K2 targets. \par
    
    Images were processed using SAS v.15 adopting the standard pipeline. Events were filtered in three steps. First, we removed events that triggered more than four adjacent pixels ({\it pattern}$\geq$12) and with flags different from 0\footnote{Flag values encode various event conditions, such as near hot pixels or events outside the field of view. The condition FLAG=0 is the most conservative event quality screening condition, and particularly necessary for PN images.}. In the second step, we selected the ``bad time intervals'' as those with intense background photon noise by inspecting the background light curve.  We then filtered out events detected during these time intervals. The fraction of the total integration time with high background noise level varies in these observations from a minimum of 10-20\%  (MOS and PN, respectively) in observation 0761920301, to a maximum of 30-46\% in observation 0761920601. In the last step, images were produced by filtering the events in three energy bands: soft (0.3-0.8$\,$keV), broad (0.3-7.9$\,$ keV), and hard (2-7.9$\,$keV). \par 
    
Source detection was performed with the wavelet convolution code PWXDetect, initially developed for ROSAT and $Chandra$ images and aimed at finding local maxima in the wavelet-convolved image at different wavelet scales \citep{Damiani1997ApJ.483.350D,Damiani1997ApJ.483.370D}. For the scope of this paper, an accurate process of source detection aimed at improving completeness of faint sources was unnecessary. We therefore adopted a detection threshold of 4.6$\sigma$ for all images, at which one spurious source due to background fluctuations is expected to contaminate the total detected source sample. An improved source detection and validation of detected sources will be performed in a forthcoming paper. \par

\subsection{Sample selection and light curves}
\label{sample_sec}

    \begin{table}
    \caption{Properties of selected stars.}
    \label{sample_table}
    \centering                       
    \begin{tabular}{|c|c|c|c|c|c|}
    \hline
    \multicolumn{1}{|c|}{Star} &
    \multicolumn{1}{|c|}{SpT} &
    \multicolumn{1}{|c|}{Teff} &
    \multicolumn{1}{|c|}{R$^{(1)}_{*}$} &    
    \multicolumn{1}{|c|}{P$^{(2)}_{\rm rot}$} &
    \multicolumn{1}{|c|}{Distance} \\
    \hline
    \multicolumn{1}{|c|}{} &
    \multicolumn{1}{|c|}{} &
    \multicolumn{1}{|c|}{K} &
    \multicolumn{1}{|c|}{R$_{\odot}$} &
    \multicolumn{1}{|c|}{days}  &
    \multicolumn{1}{|c|}{pc} \\
    \hline
     HII~212 &  M0       & 3909$\pm$125& $0.50_{0.43}^{0.55}$ & 4.49 & $133\pm1$   \\
     HHJ~273 & (M5)      & 3178$\pm$50 & $0.23_{0.21}^{0.25}$ & 1.15 & $142\pm2$   \\
     HHJ~336$^b$ & (M4)      & 3180$\pm$50 & $0.23_{0.21}^{0.25}$ & 0.37 & $157\pm7$   \\
     HCG~244$^b$& (M4)      & 3280$\pm$50 & $0.27_{0.25}^{0.29}$ & 0.66 & $137\pm2$   \\
     HII~345 &  G8       & 5150$\pm$125& $0.77_{0.74}^{0.79}$ & 0.84 & $135\pm1$   \\
     HII~1516& (K8)      & 3921$\pm$125& $0.50_{0.47}^{0.55}$ & 0.31 & $135\pm1$   \\
     HCG~146 & (M4)      & 3167$\pm$50 & $0.23_{0.21}^{0.25}$ & 0.68 & $135\pm2$   \\
     HCG~273 & (M2)      & 3508$\pm$50 & $0.35_{0.33}^{0.37}$ & 2.76 & $136\pm1$   \\
     HCG~150 & (M4)      & 3287$\pm$50 & $0.28_{0.26}^{0.29}$ & 1.08 & $127\pm1$   \\
     HCG~295 & (M2)      & 3412$\pm$50 & $0.31_{0.30}^{0.33}$ & 1.42 & $137\pm2$   \\
     HII~405 &  F9       & 6219$\pm$125& $1.07_{1.02}^{1.13}$ & 1.91 & $135\pm1$   \\
     HCG~148$^b$ & (M4)      & 3135$\pm$50 & $0.22_{0.21}^{0.24}$ & 0.22 & $143\pm3$   \\%
    \hline
    \multicolumn{6}{l}{(1): Stellar radii; (2): Rotation periods \citep{RebullSBC2016AJa}.} \\
    \multicolumn{6}{l}{$^b$: binary star} \\
    \end{tabular}
    \end{table}
X-ray light curves of all detected sources were extracted adopting circular extraction regions with a radius of 400 pixels (20$^{\prime\prime}$) centered on each source position. When we needed to account for source crowding, extraction regions were shrunk to the largest radius at which there was no overlap with adjacent source regions. We also visually inspected the defined extraction regions in order to avoid contamination from PSF wings of nearby bright sources. Additionally, following the prescription of the SAS guide, background regions were chosen with the same size as the extraction regions, close to source positions, avoiding PSF wings of nearby sources, and, in the case of PN images, and sharing the the same distance from the read out node of the associated source, in order to minimize the impact of varying response along the chip. \par

        We selected X-ray flares in the combined MOS+PN light curves using the Maximum Likelihood (ML) method \citep{Scargle1982ApJ,CaramazzaFMR2007AA,AlbaceteColomboFMS2007}. The method consists of dividing the whole light curve into time blocks during which the count rate can be considered constant within a given confidence level. We then visually inspected the K2 light curves of the stars where X-ray flares occurred, selecting 12 objects with bright flares observed both in optical and X-rays. \par
        
Table \ref{sample_table} lists the main properties of the stars that hosted the selected flares. All these stars are bona fide cluster members, with membership probability approaching unity as obtained by \citet{BouyBMC2013AA554A.101B}. Effective temperatures are taken from \citet{StaufferRBH2016AJ}, adopting the conversion tables for PMS stars of \citet{PecautMamajek2013ApJS}. Stellar radii are calculated from the 125-Myr MIST isochrone \citep{Choi2016ApJ} using the properties of these stars. Rotation periods are calculated from K2 light curves by \citet{RebullSBC2016AJa}. Two of these stars, namely HHJ~336 and HCG~244, are multi periodic with a detected second period of 0.35 and 0.24 days, respectively \citep{RebullSBC2016AJb}. Spectral types derived from spectroscopy are available for three stars (values without brackets in Table \ref{sample_table}) and are obtained from \citet{Trumpler1921LicOB,ProsserSK1991AJ,MartinRZ1996ApJ} and \citet{StaufferSK1998ApJ}. Spectral types within brackets in Table 2 are derived from available photometry and effective temperature ($\rm T_{eff}$) values using the conversions into spectral types derived by \citet{PecautMamajek2013ApJS}. Individual distances and their uncertainties are obtained from the parallaxes and their errors in the Gaia DR2 catalog \citep{GaiaCollaboration2018AA.616A.1G}. All the stars in this sample are single stars, with only HCG~244 and HCG~148 being listed as photometric binary stars by \citet{PinfieldDJS2003MNRAS} and HHJ~336 whose multi-periodic K2 light curve is explained in terms of being a binary system.  \par

    \begin{figure*}[]
        \centering      
        \includegraphics[width=11.5cm]{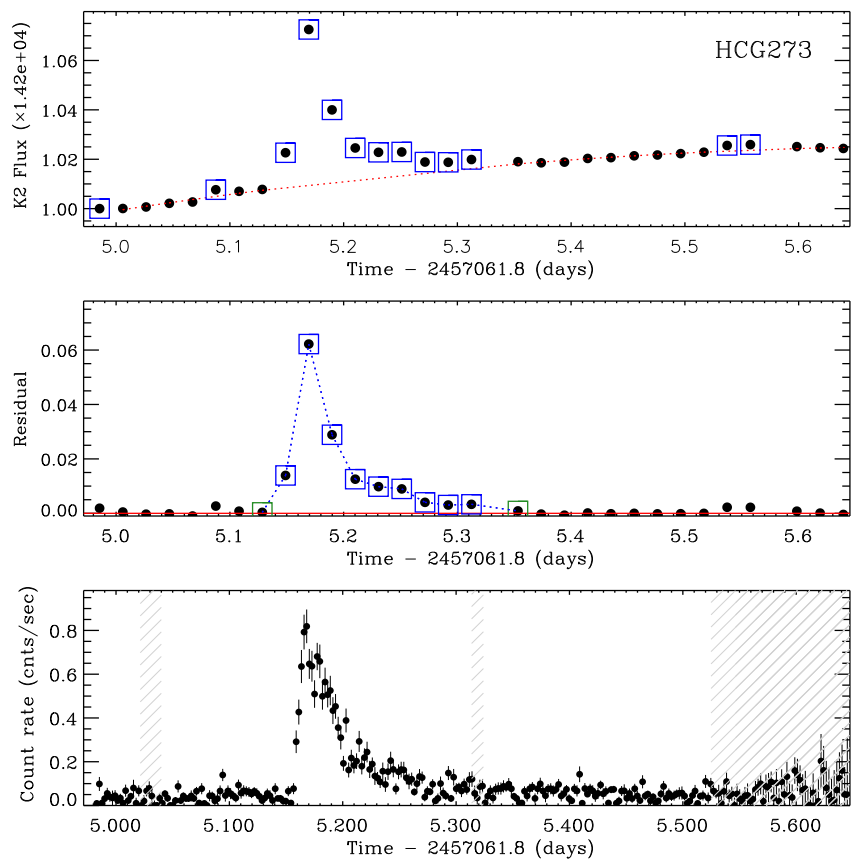}
        \caption{Light curves of HCG~273. Top Panel: The K2 light curve. The error bars of K2 data points are smaller than the plot symbols. The red dotted line marks the polynomial used to fit the quiescent light curve and is limited to the portion of the K2 light curve for which the fit is performed; the blue squares mark the points corresponding to the flare. Central Panel: Residuals of the K2 light curve from the best-fit polynomial function limited in the portion of the light curve where the fit is performed (limited by the vertical dotted lines). The green squares mark the K2 points just before and after the flare. Bottom panel: The combined background-subtracted MOS+PN X-ray light curve. Shaded regions are ``bad time intervals'' in at least one EPIC detector.}
        \label{HHJ336_lc}
        \end{figure*}

\begin{figure}[]
        \centering      
        \includegraphics[width=9cm]{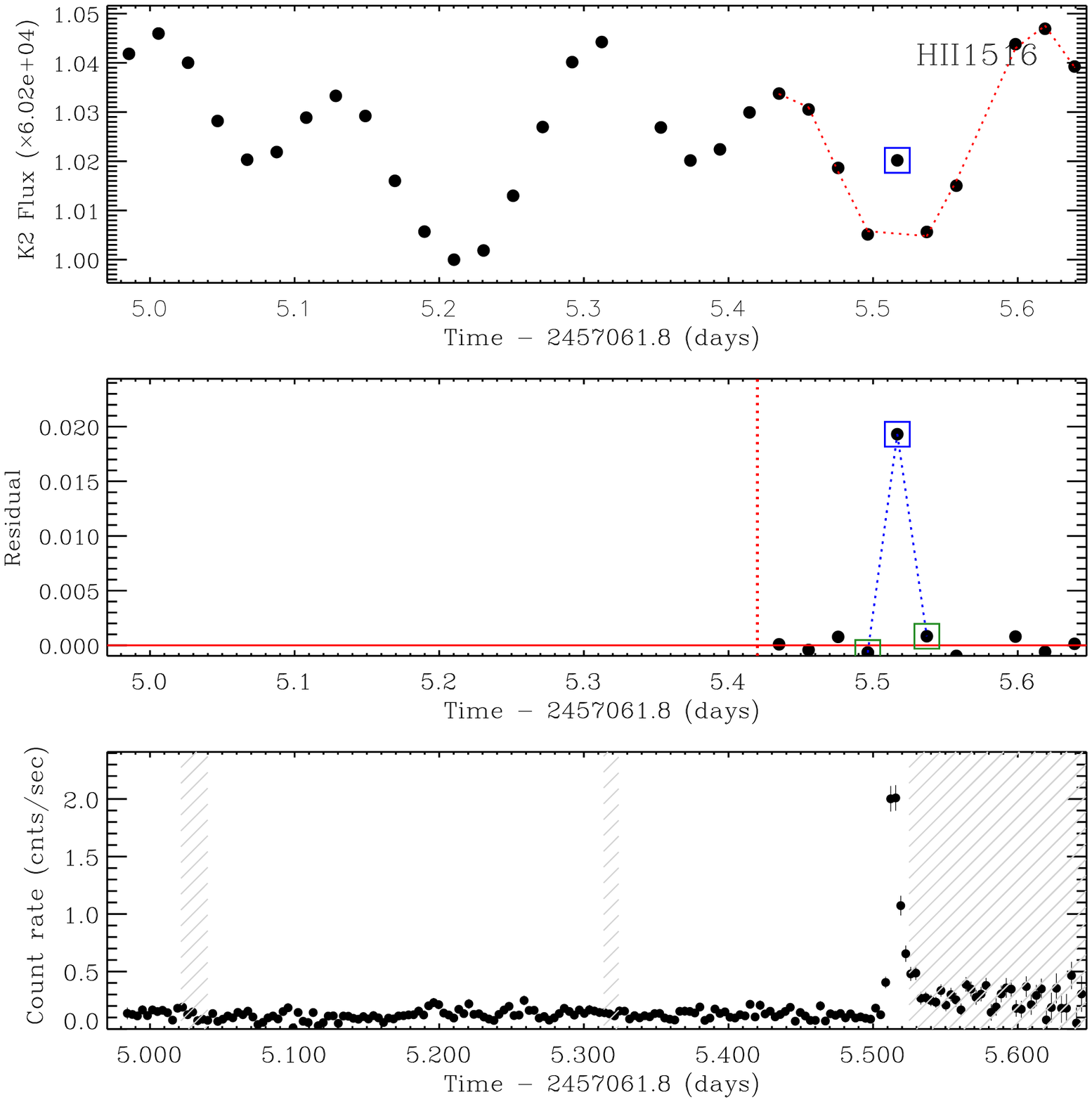}
        \caption{Light curves of HII~1516. Panel layout and content as in Fig.~\ref{HHJ336_lc}}
        \label{HII1516}
        \end{figure}

\begin{figure}[]
        \centering      
        \includegraphics[width=9cm]{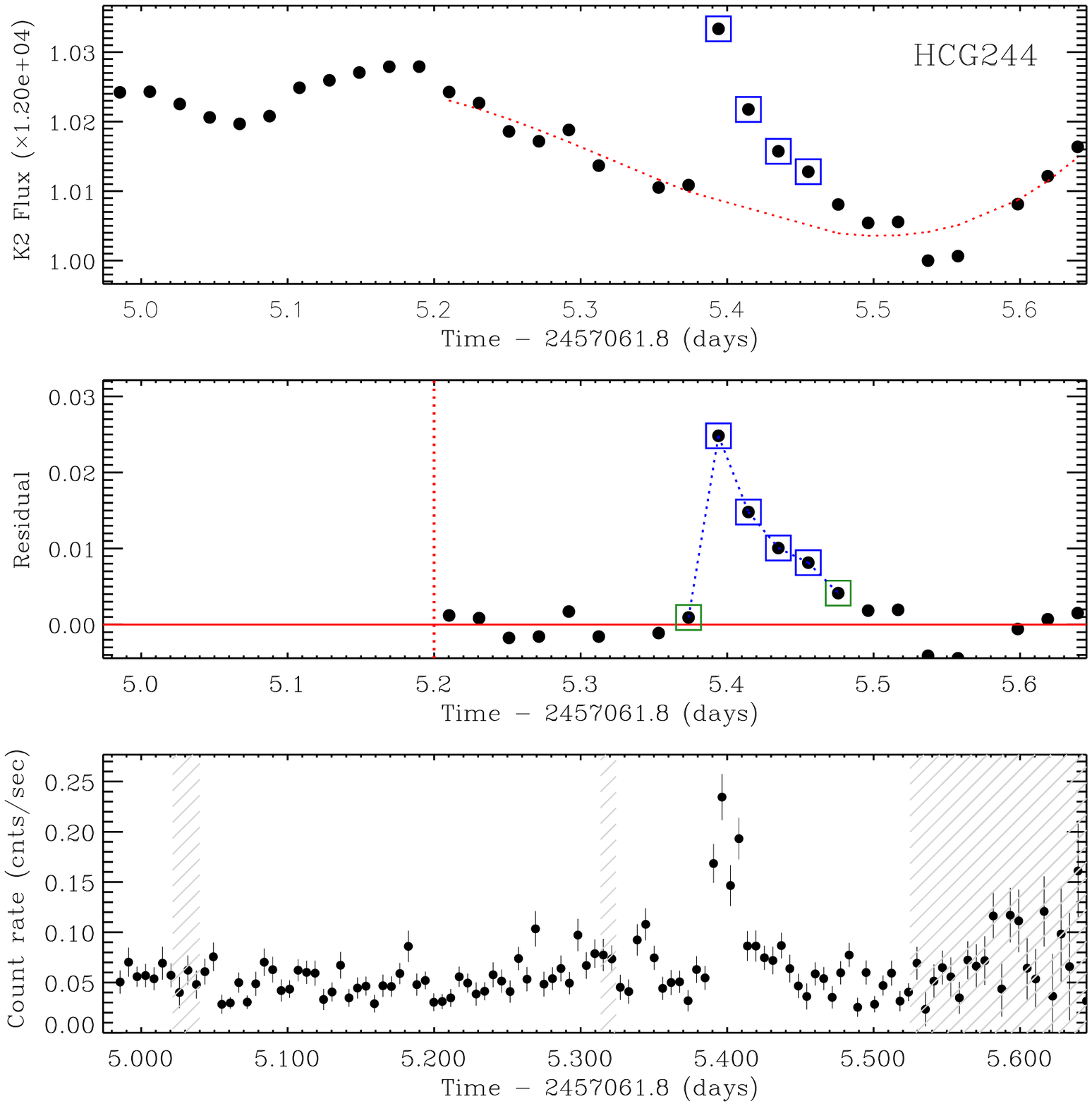}
        \caption{Light curves of HCG~244. Panel layout and content as in Fig.~\ref{HHJ336_lc}}
        \label{HCG244}
        \end{figure}

\begin{figure}[]
        \centering      
        \includegraphics[width=9cm]{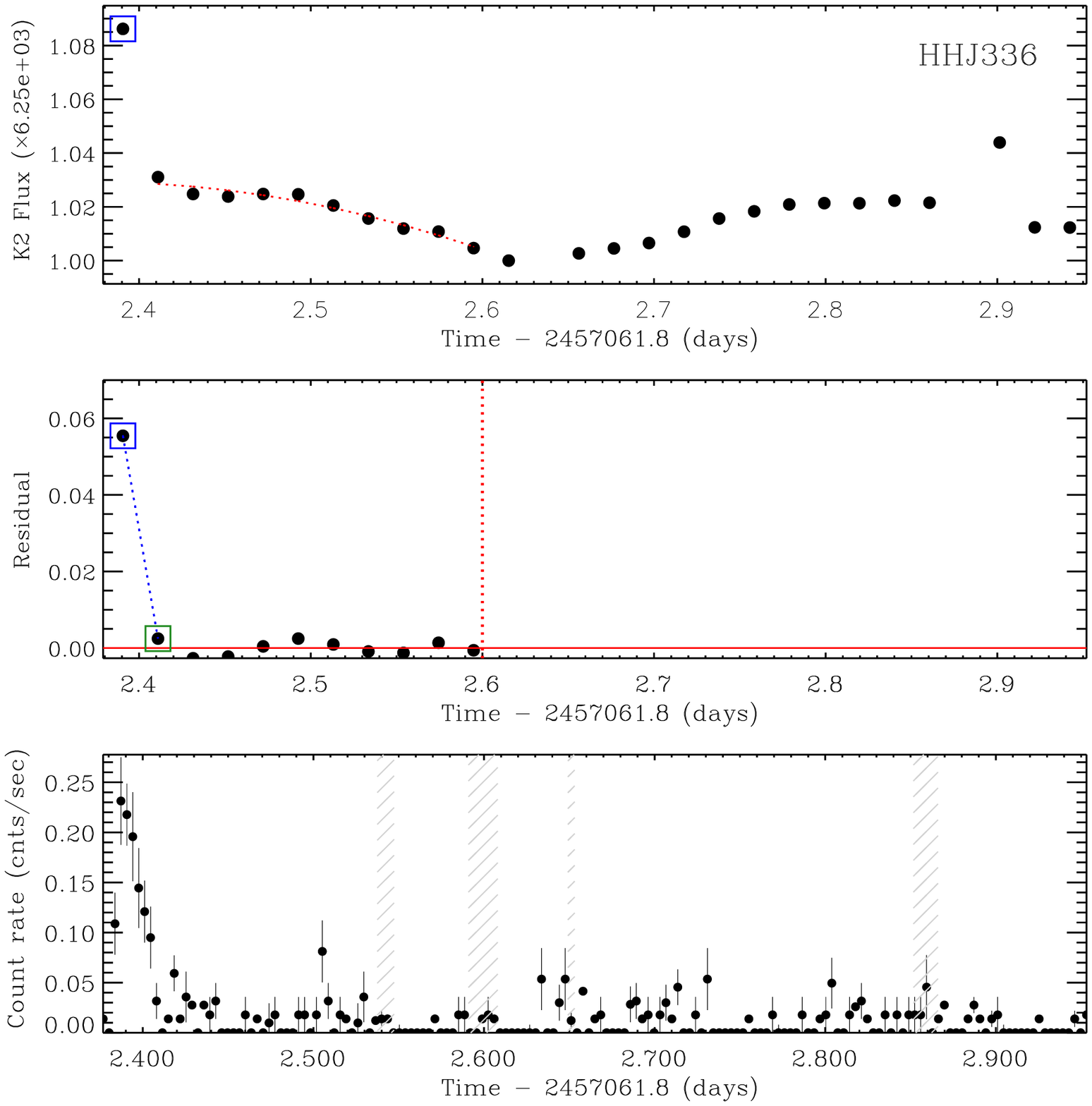}
        \caption{Light curves of HHJ~336. Panel layout and content as in Fig.~\ref{HHJ336_lc}}
        \label{HCG273}
        \end{figure}

\begin{figure}[]
        \centering      
        \includegraphics[width=9cm]{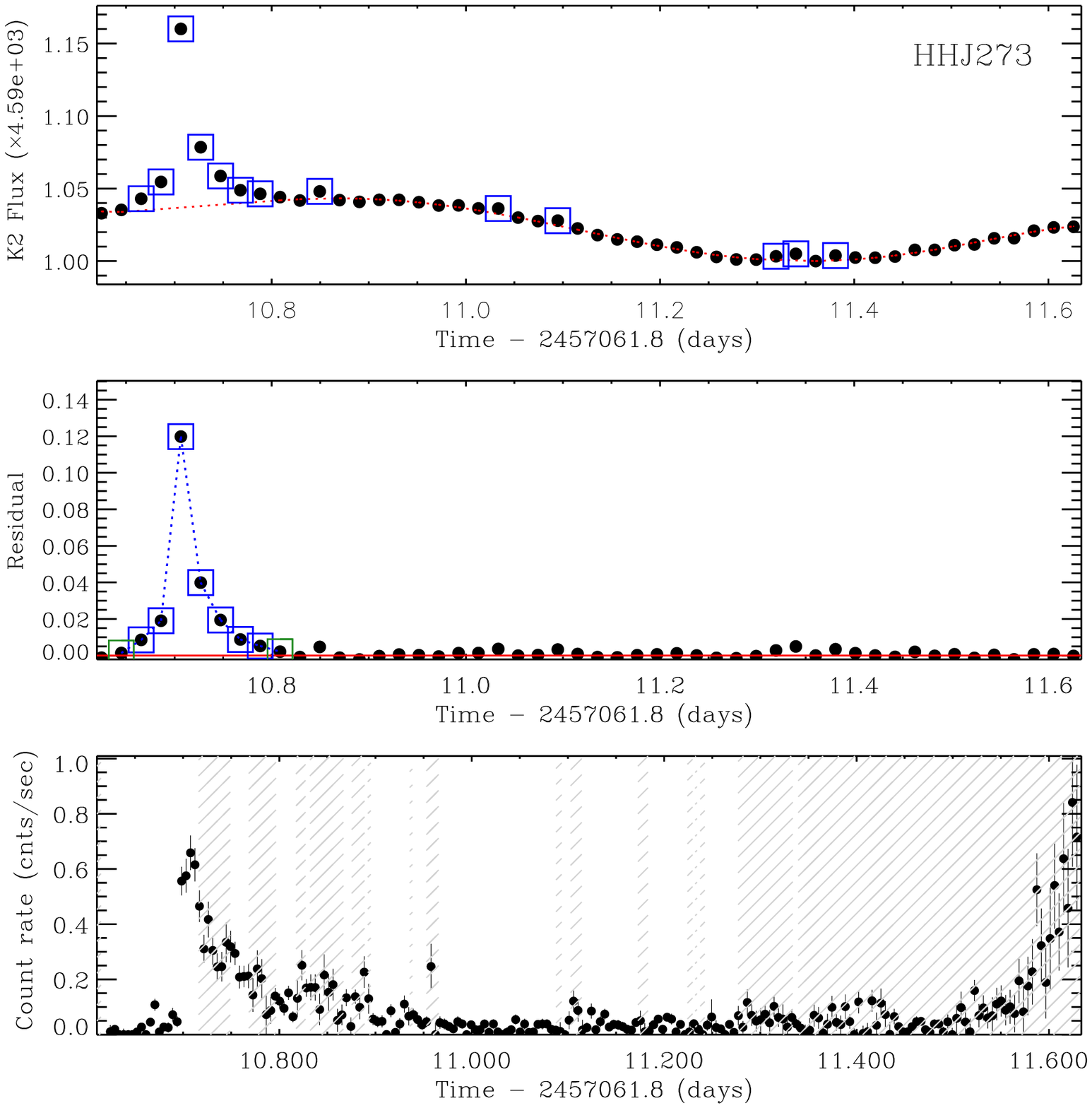}
        \caption{Light curves of HHJ~273. Panel layout and content as in Fig.~\ref{HHJ336_lc}}
        \label{HHJ273}
        \end{figure}

\begin{figure}[]
        \centering      
        \includegraphics[width=9cm]{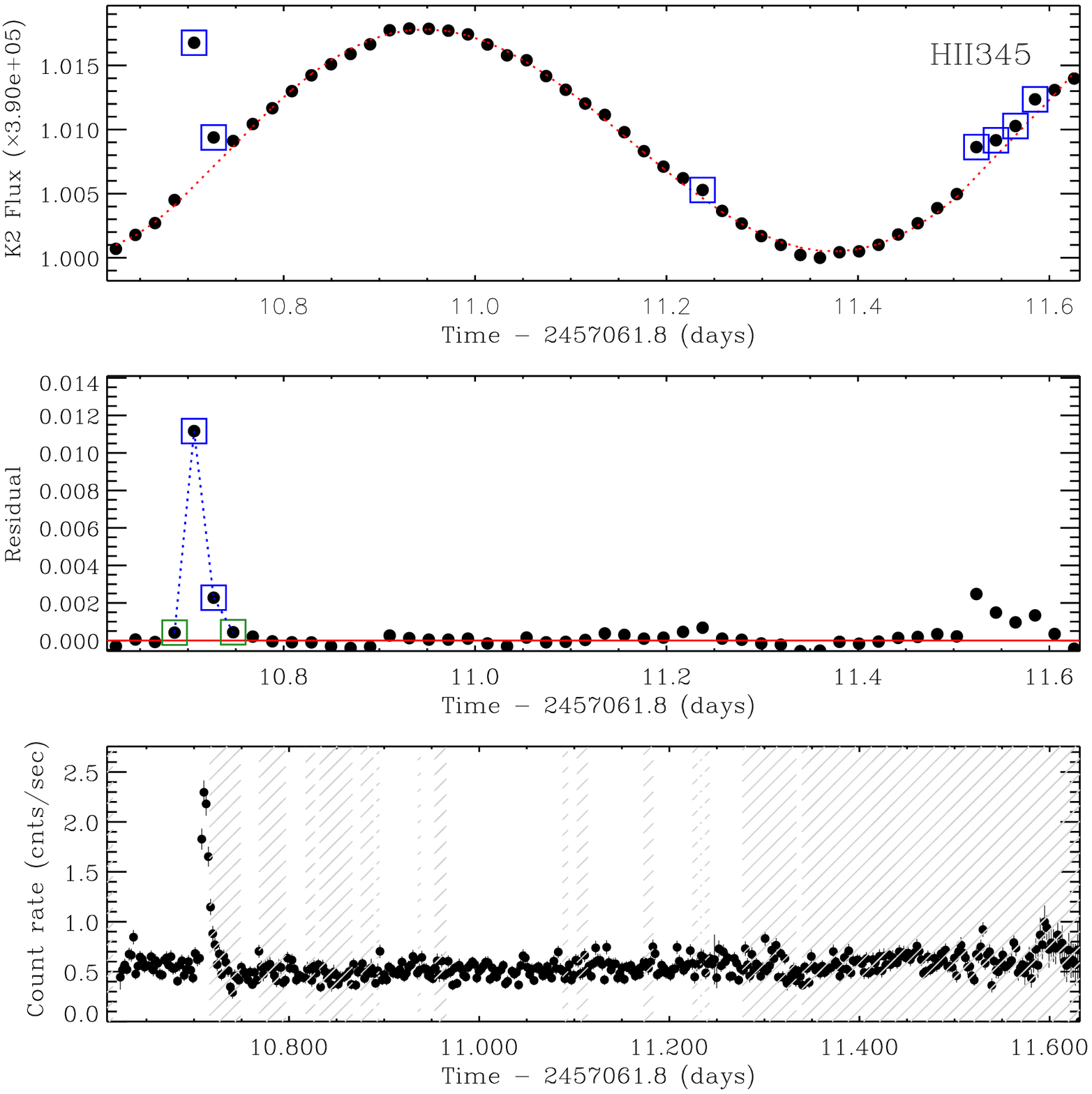}
        \caption{Light curves of HII~345. Panel layout and content as in Fig.~\ref{HHJ336_lc}}
        \label{HII345}
        \end{figure}

\begin{figure}[]
        \centering      
        \includegraphics[width=9cm]{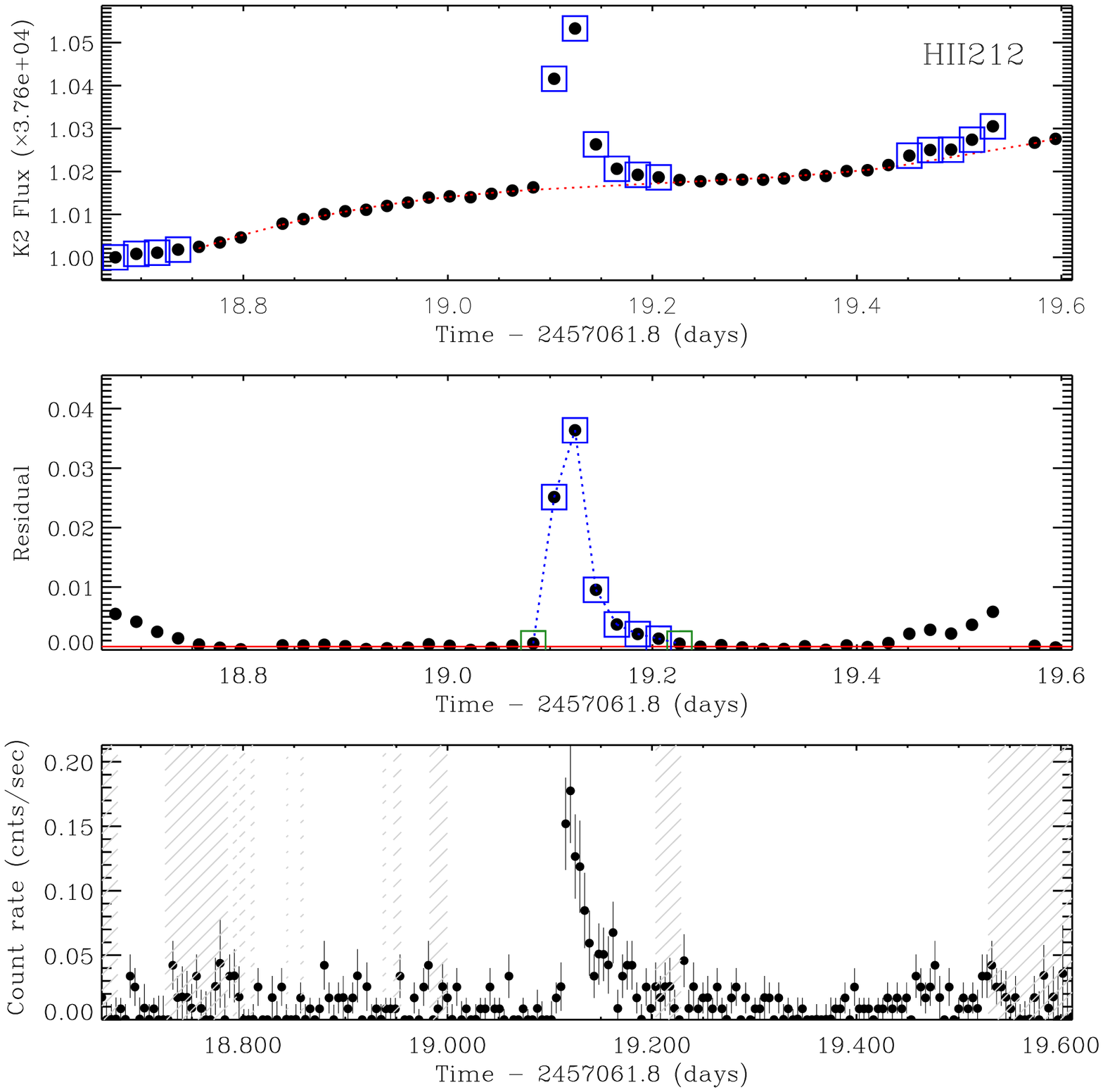}
        \caption{Light curves of HII~212. Panel layout and content as in Fig.~\ref{HHJ336_lc}}
        \label{HII212}
        \end{figure}

\begin{figure}[]
        \centering      
        \includegraphics[width=9cm]{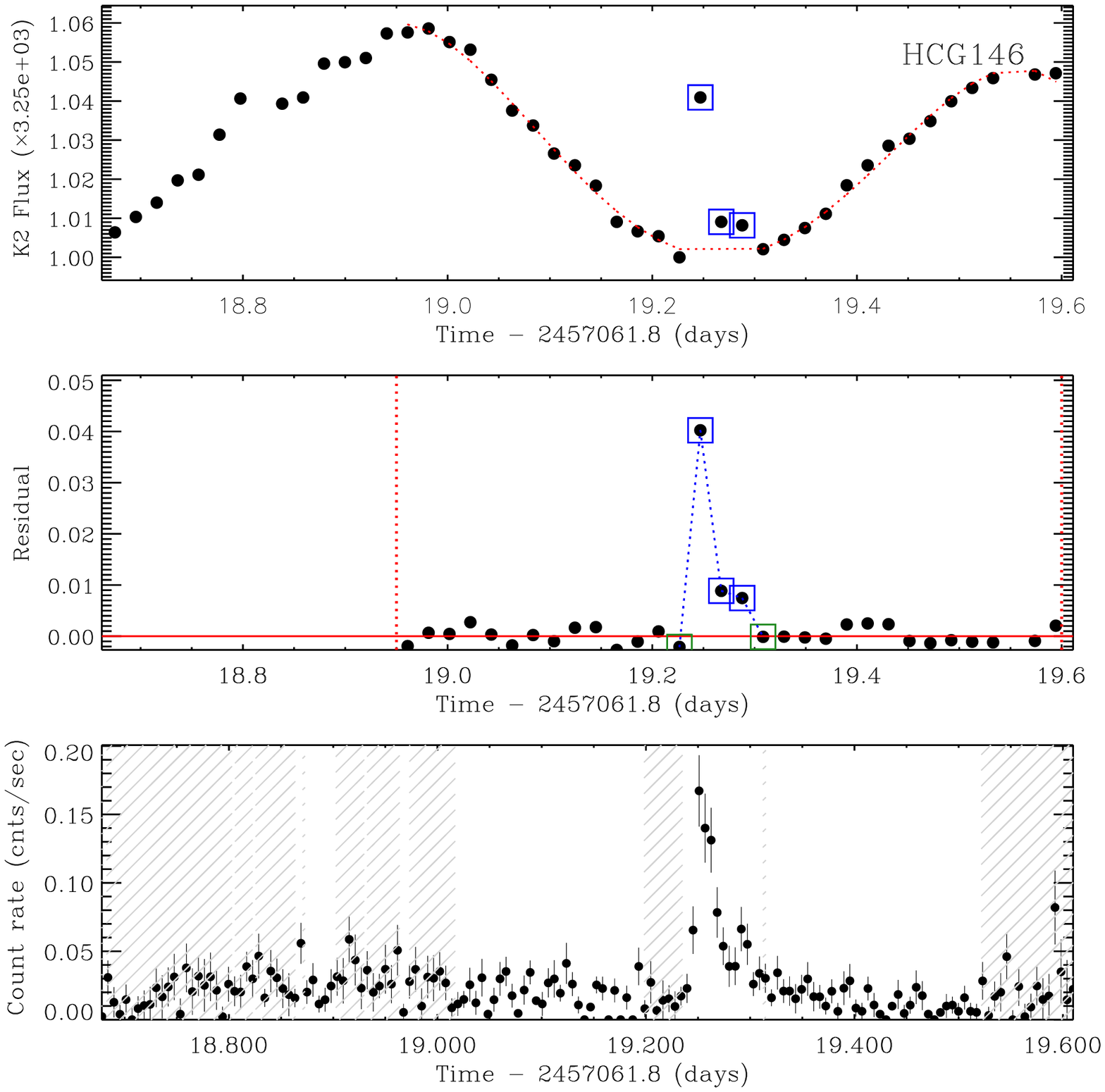}
        \caption{Light curves of HCG~146. Panel layout and content as in Fig.~\ref{HHJ336_lc}}
        \label{HCG146}
        \end{figure}

\begin{figure}[]
        \centering      
        \includegraphics[width=9cm]{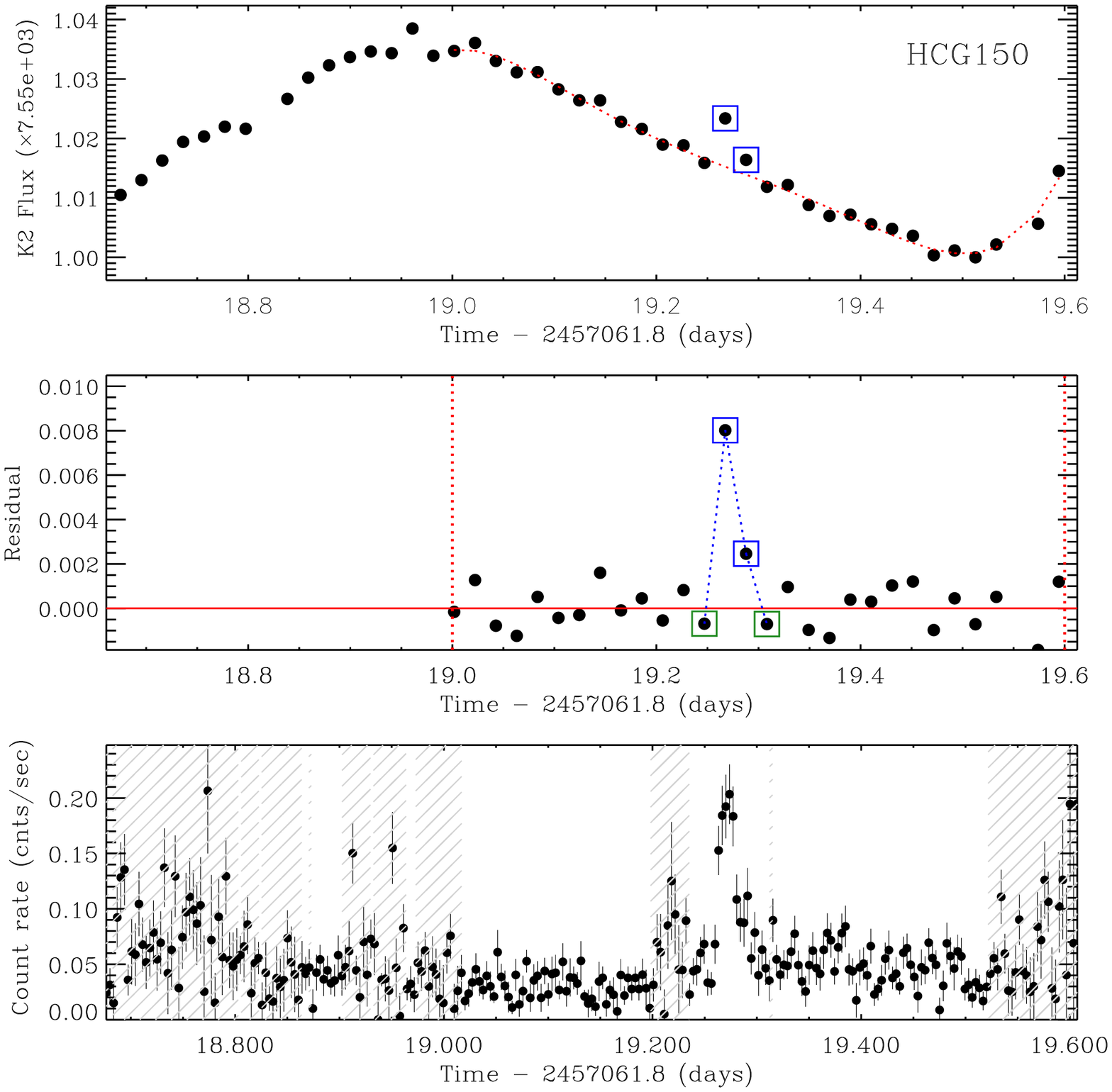}
        \caption{Light curves of HCG~150. Panel layout and content as in Fig.~\ref{HHJ336_lc}}
        \label{HCG150}
        \end{figure}

\begin{figure}[]
        \centering      
        \includegraphics[width=9cm]{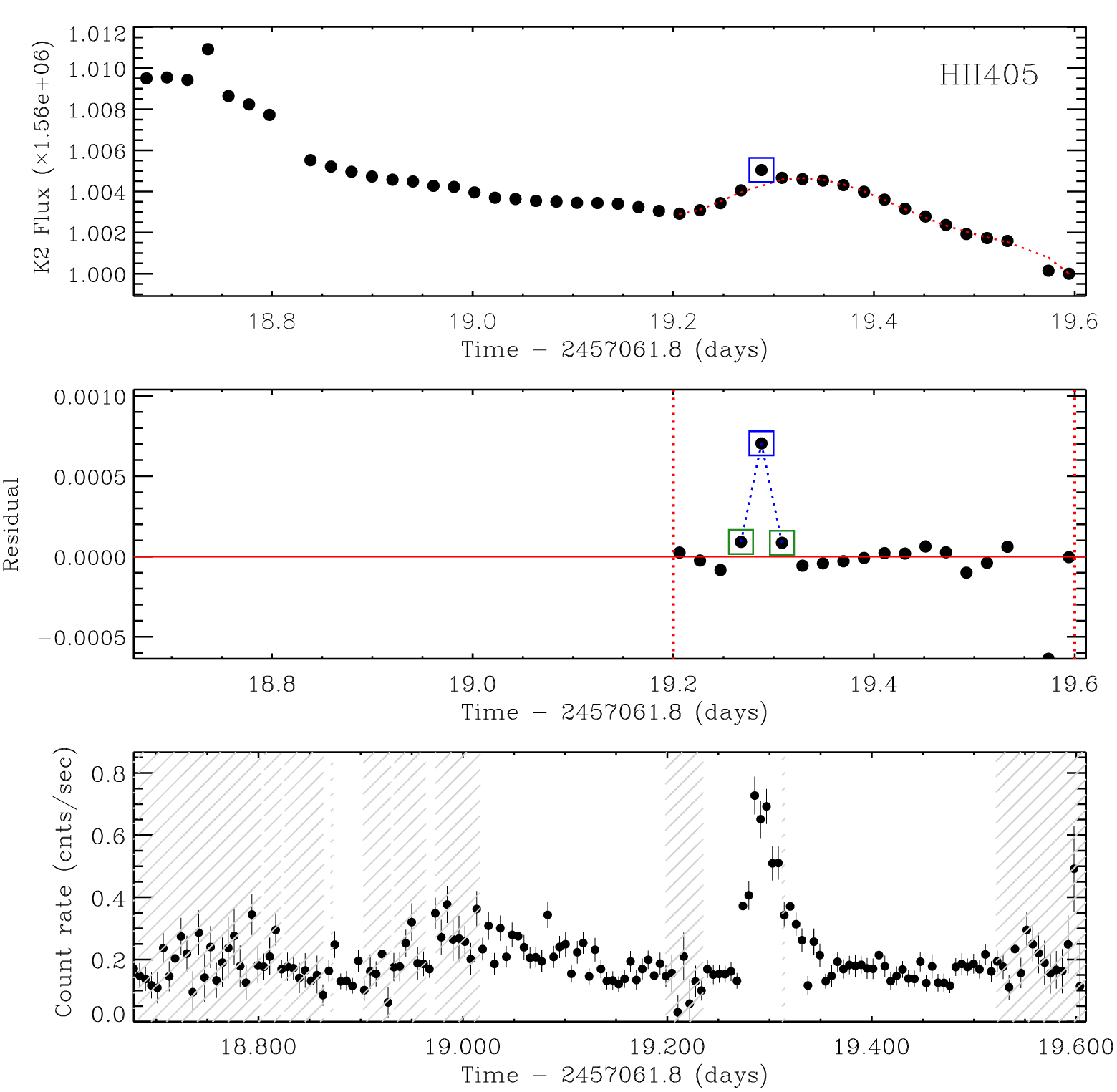}
        \caption{Light curves of HII~405. Panel layout and content as in Fig.~\ref{HHJ336_lc}}
        \label{HII405}
        \end{figure}

\begin{figure}[]
        \centering      
        \includegraphics[width=9cm]{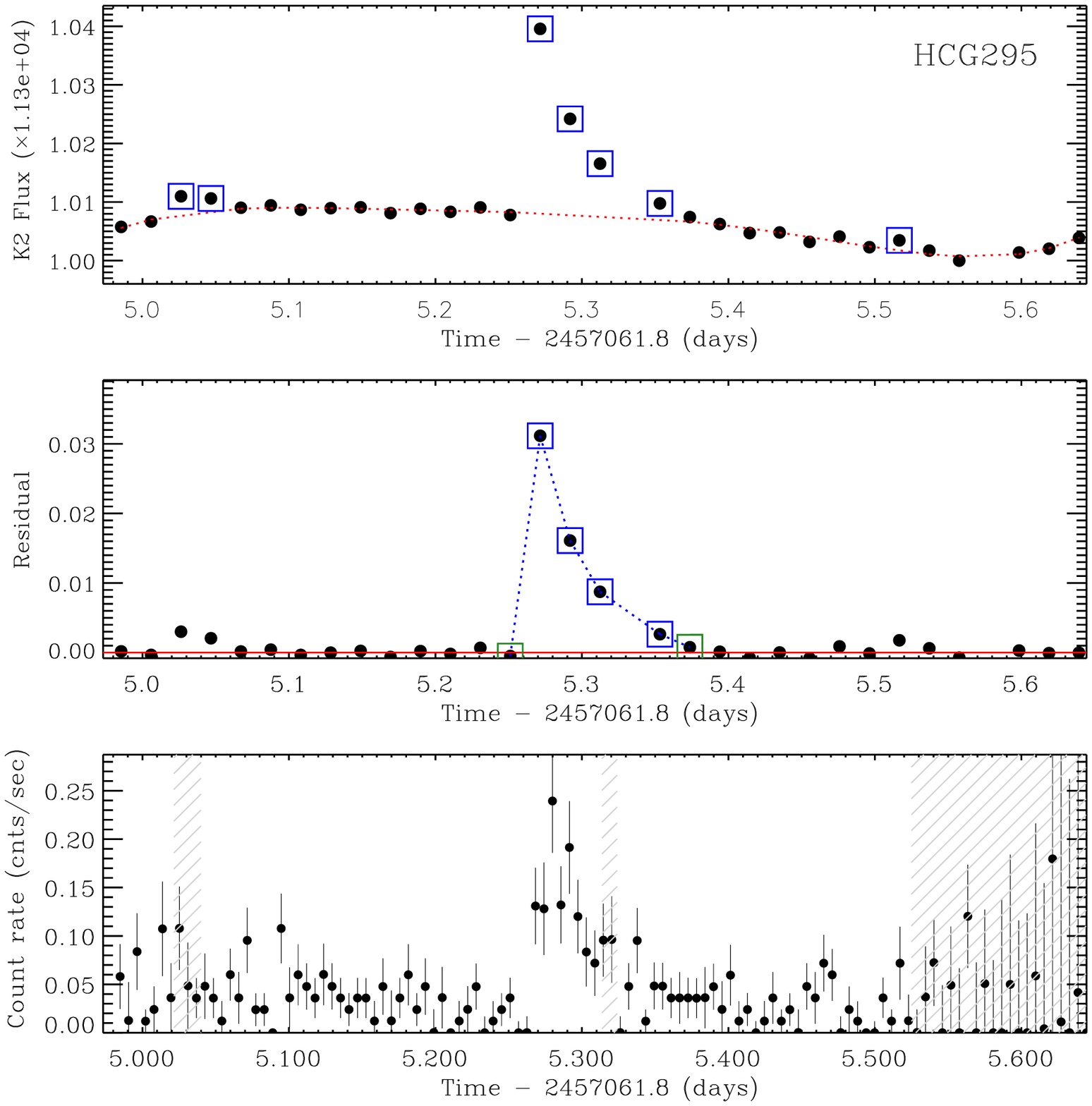}
        \caption{Light curves of HCG~295. Panel layout and content as in Fig.~\ref{HHJ336_lc}}
        \label{HCG295}
        \end{figure}

\begin{figure}[]
        \centering      
        \includegraphics[width=9cm]{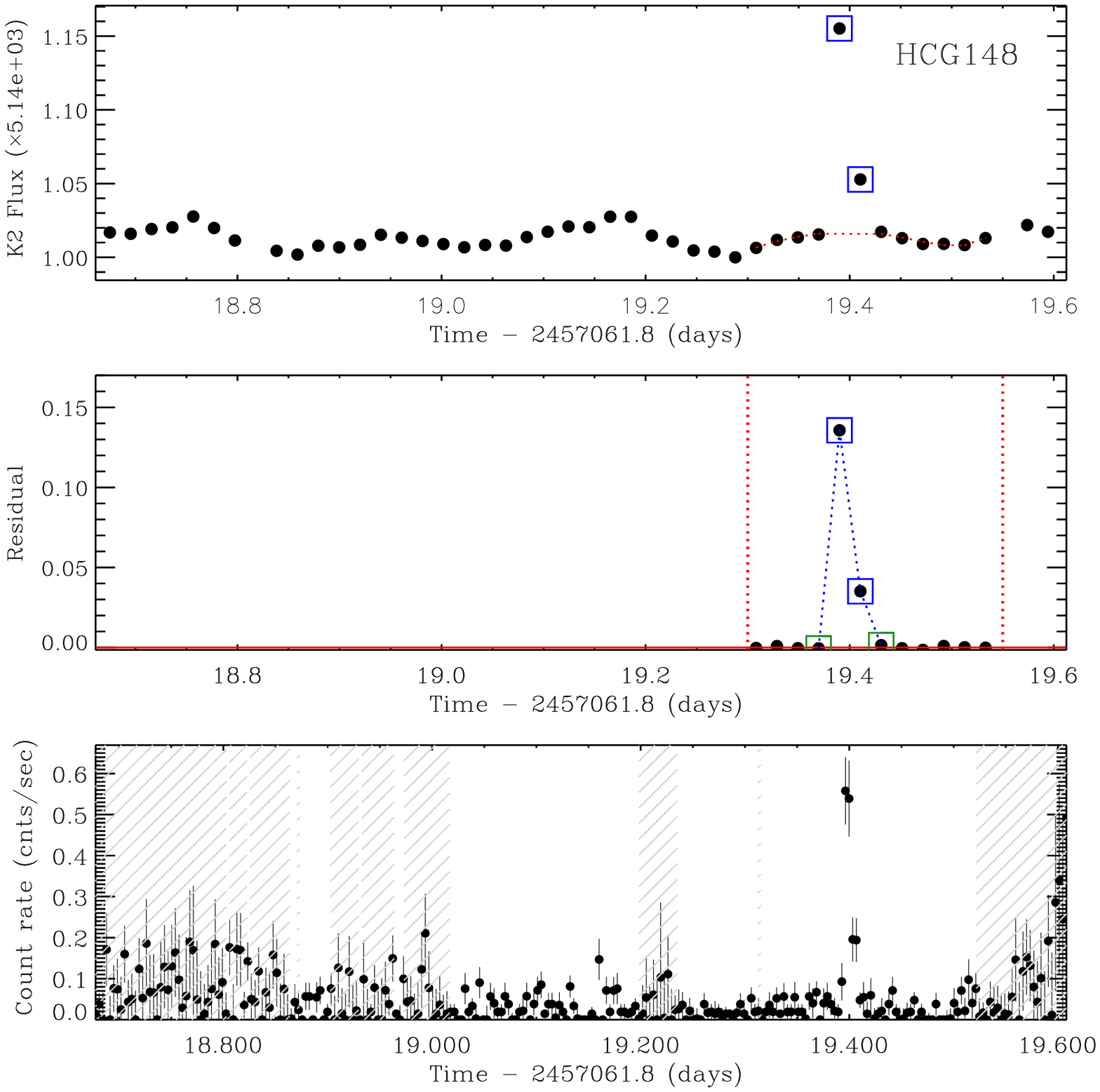}
        \caption{Light curves of HCG~148. Panel layout and content as in Fig.~\ref{HHJ336_lc}}
        \label{HCG148}
        \end{figure}

Figures \ref{HHJ336_lc}--\ref{HCG148} show the light curves of the selected stars. The top panel of each figure shows the K2 light curve limited to the time window of the XMM observations. To improve visualization, each K2 light curve is normalized to the minimum value observed in this interval. In order to define the quiescent level, we performed a fit of the portion of the K2 light curve in the time window of the XMM observations with a polynomial of the fifth order. In each figure we also show the portion of the light curve used to perform the fit, which was restricted ad-hoc in those cases where it was not possible to obtain a good fit over the entire light curve. The fit was performed recursively: points discrepant by more than $2\sigma$ were removed from the fit before the fit was then calculated again and the process repeated until no more discrepant points were found. The final sets of discrepant points typically define the optical flares and are marked with blue boxes in these figures. The residuals of the fit are shown in the central panels. The bottom panel of each figure shows the merged background-subtracted MOS+PN light curve, with the bin size chosen according to source X-ray brightness\footnote{600 s for sources with a total number of counts $\leq$100; 400 s for 100<counts$\leq$2000; 300 s for 2000<counts$\leq$4000; 200 s for 4000<counts$\leq$5000; 100 s for sources with more than 5000 total detected counts.}. The bad time intervals on both MOS and PN detectors are also marked. We removed events detected during these intervals for source detection but not for the selection and analysis of these bright flares. \par

The K2 flares in our sample did not occur at particular rotation phases. Setting the rotational phase $\phi=0.5$ at the minimum of the K2 light curves (i.e., when the main starspots are close to the center of the stellar disc), five flares occurred at 0.25$\leq \phi \leq$0.5, while four flares occurred at 0.85$\leq \phi \leq$0.95. Furthermore, none of the observed physical properties of the flares depend on the rotational phase at which they occurred. This may be due to the small number of flares in our sample, even if our results agree with those obtained on larger samples of stellar flares observed in K2 light curves \citep[e.g.,][]{Doyle2018MNRAS.480.2153D}. \par

Among the remaining 57 K2 targets falling in the XMM fields, 13 stars show that probable optical flares (``probable'' since in all cases but one the flare consists of one or two K2 data points) occurred during the XMM observations. Two of them are not detected in X-rays (BPL~108 and BPL~118); six stars (NPL~24, HII~1114, HII~1234, HHJ~203, HCG~302, HCG~123) are detected in X-rays but there is no X-ray flare observed; in 4 stars (BPL~85, HCG~194, HHJ~232, HII~314) an X-ray flare is observed but is too weak to be analyzed; in one case (HII~314) both the optical and X-ray variability are very complicated, with no evident single-loop flares. The full list of K2 targets falling in the XMM fields and not included in Table \ref{sample_table} is shown in Appendix \ref{allk2_sec}.

\subsection{Flare duration}
\label{duration_sect}   
        
        Flare decay time can provide information about the flaring structures \citep[e.g.,][]{RealeBPS1997AA}. The decay time can be derived directly by folding the flare light curves with an exponential function. In our case this could only be done successfully in three cases: the X-ray flares in HCG~273 (Fig. \ref{HHJ336_lc}), HHJ~336 (Fig. \ref{HCG273}), and HHJ~273 (Fig. \ref{HHJ273}). The resulting decay times of these flares are 2.6$\pm$0.6$\,$ks, 2.9$\pm$0.6$\,$ks, and 2.3$\pm$1.8$\,$ks, respectively. \par
        
An alternative estimate of flare decay times can be obtained under the assumption of a constant release of energy during flare decay, as the ratio between the total emitted energy and the peak luminosity \citep{Flaccomio2018AA.620A.55F}. However, the peak luminosity is averaged over the peak time bin, underestimating the real peak value. Thus, this ratio is typically larger than the real flare decay time. \par   
    
\begin{figure*}[]
        \centering      
    \includegraphics[width=6cm]{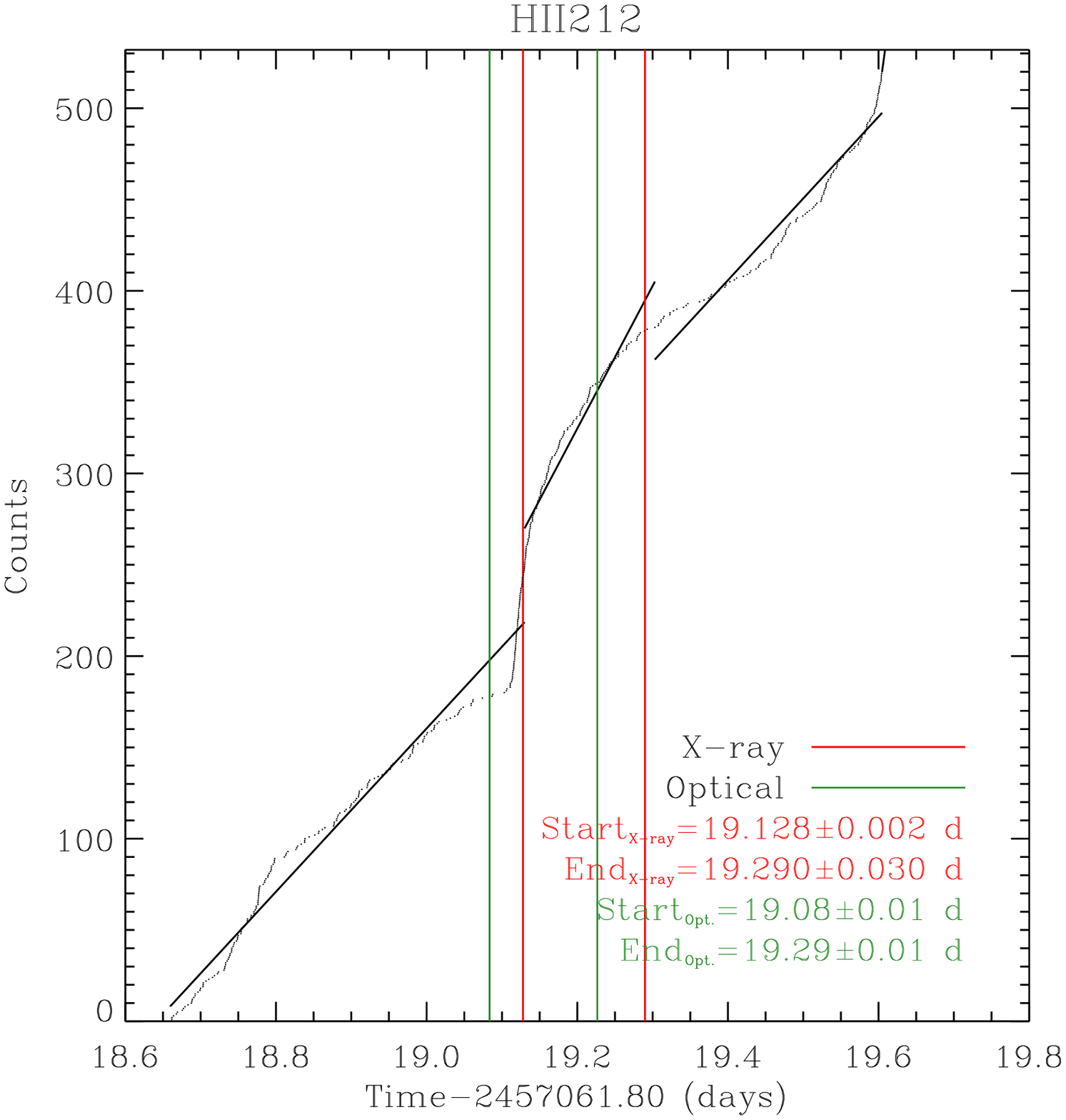}
    \includegraphics[width=6cm]{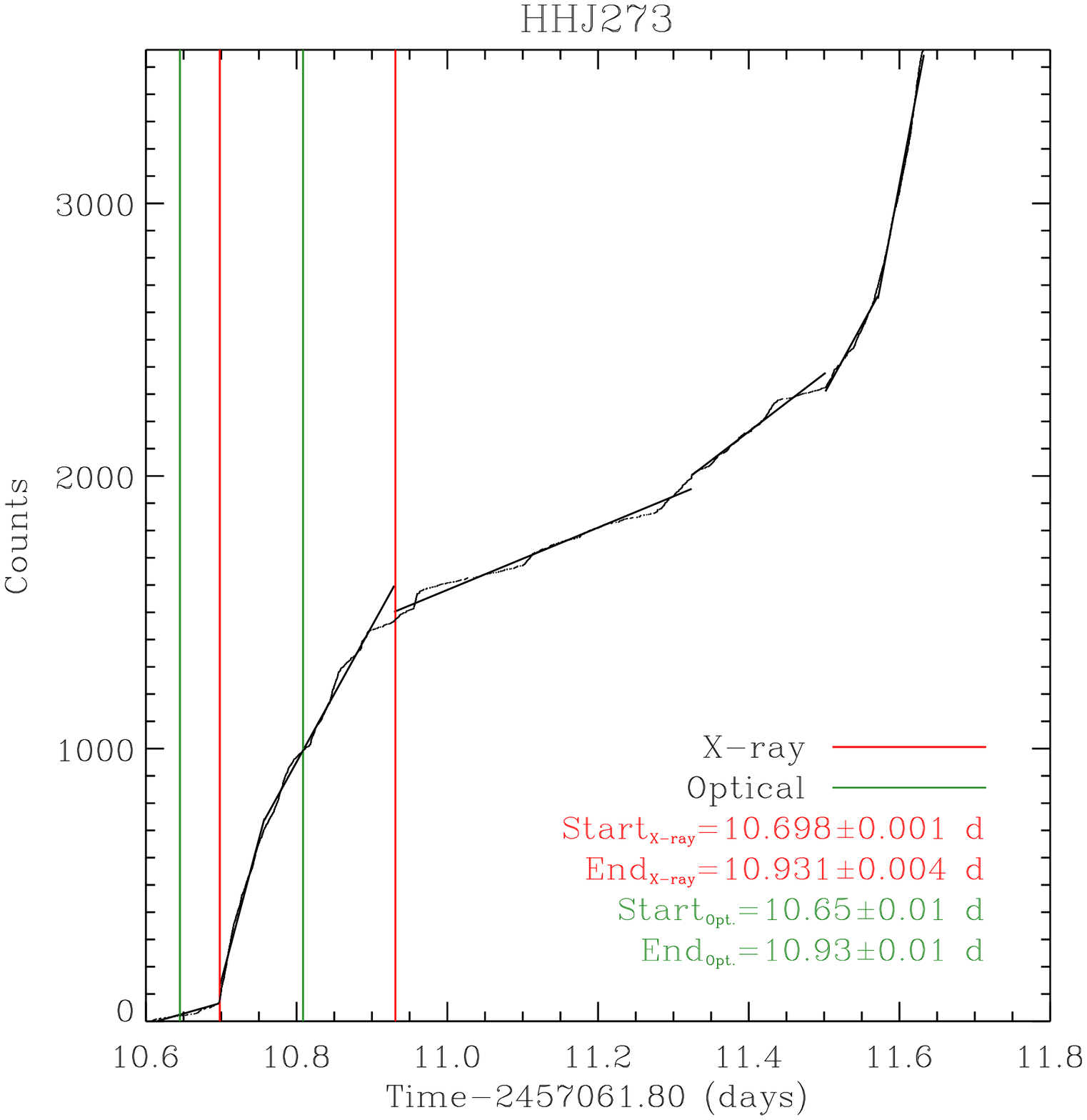}
    \includegraphics[width=6cm]{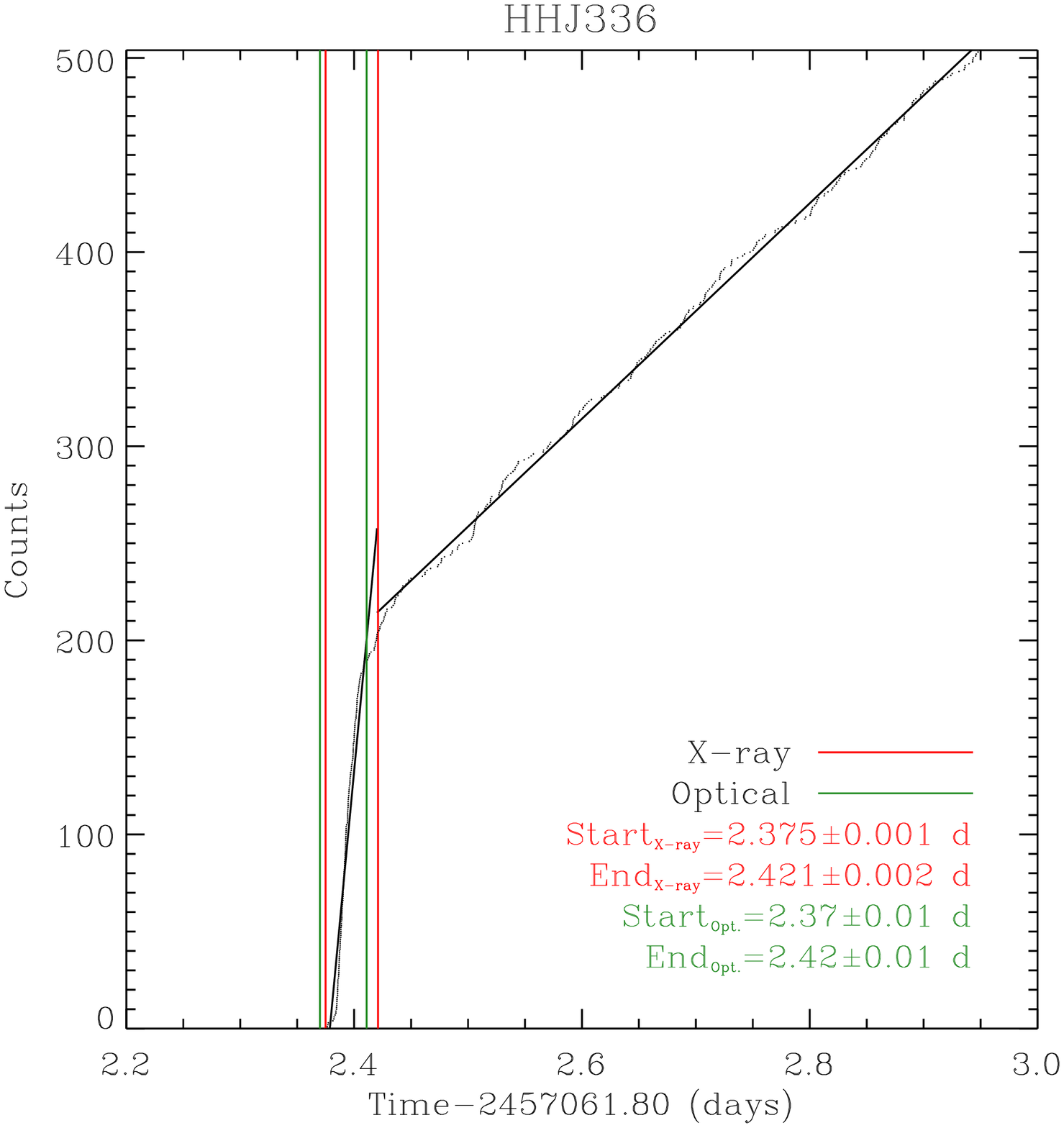}
    \includegraphics[width=6cm]{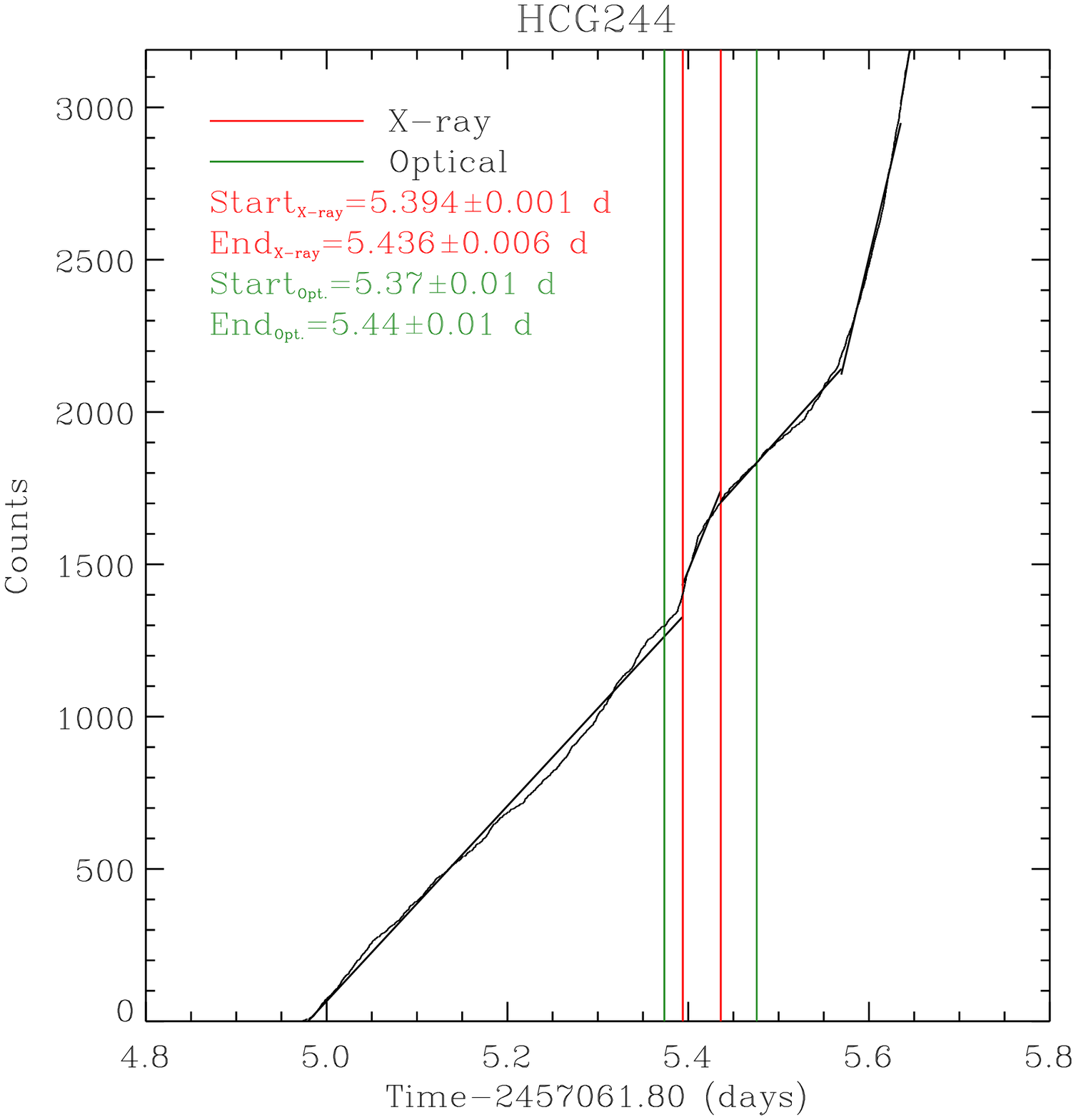}
    \includegraphics[width=6cm]{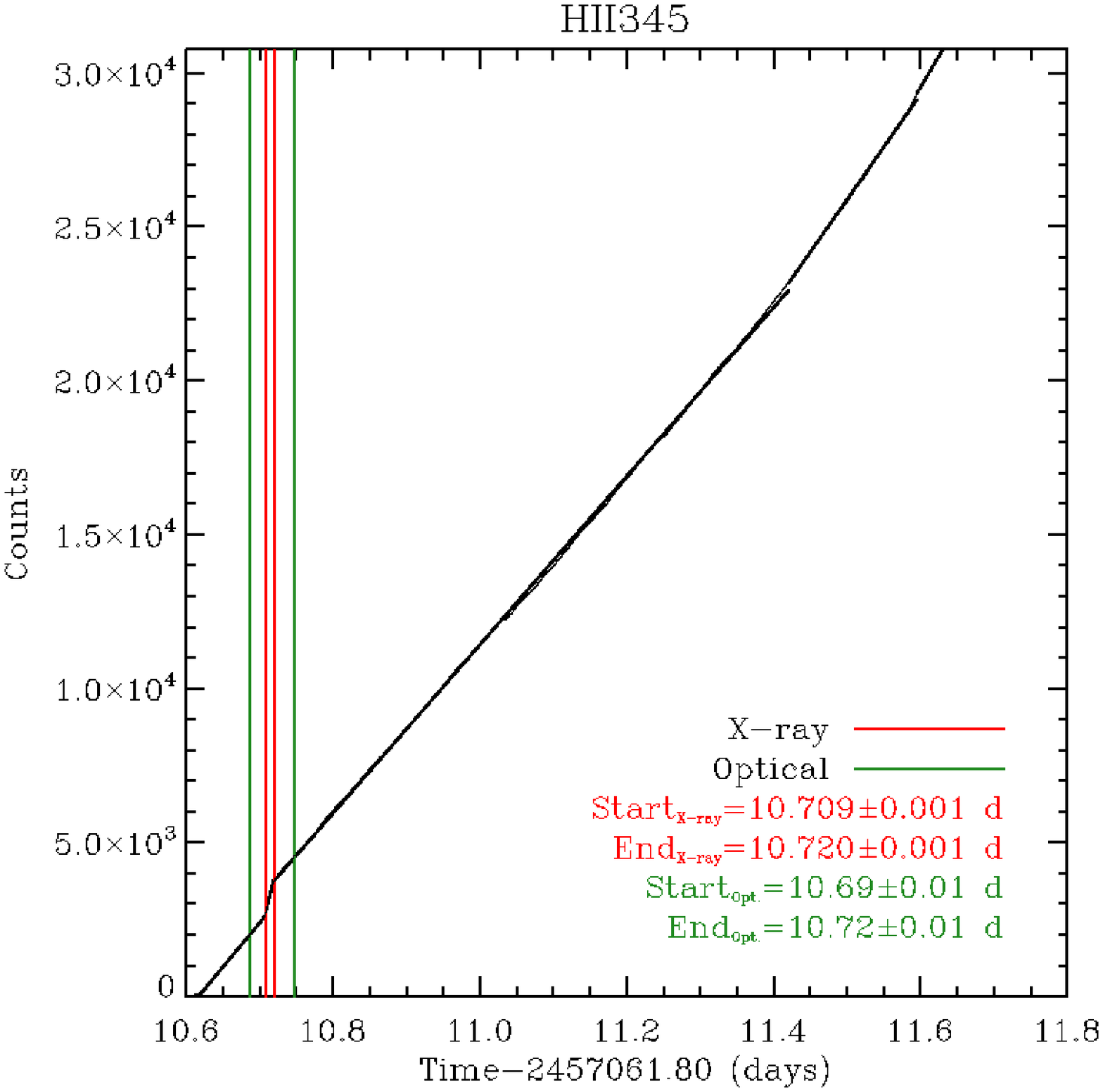}
    \includegraphics[width=6cm]{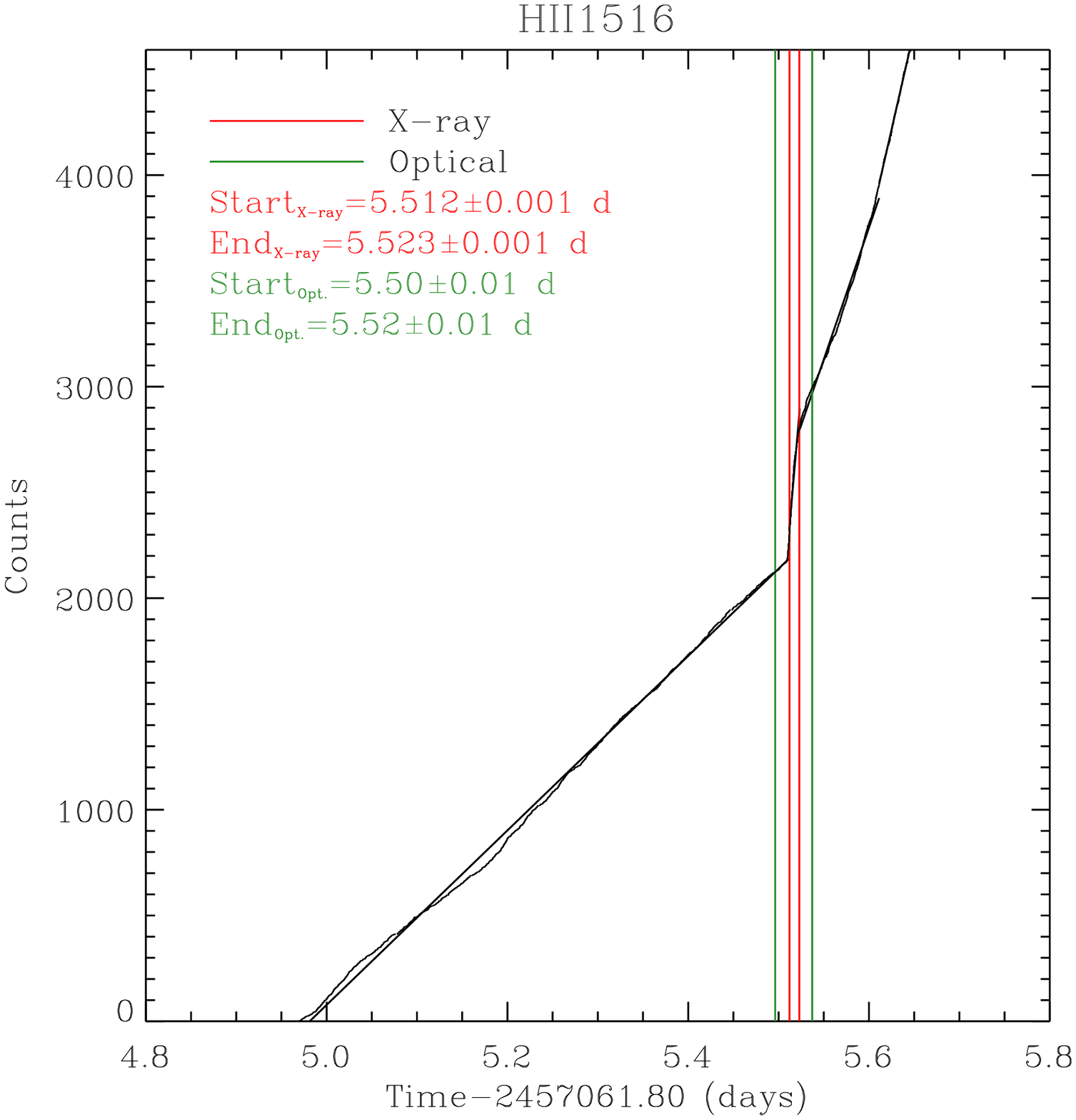}
    \includegraphics[width=6cm]{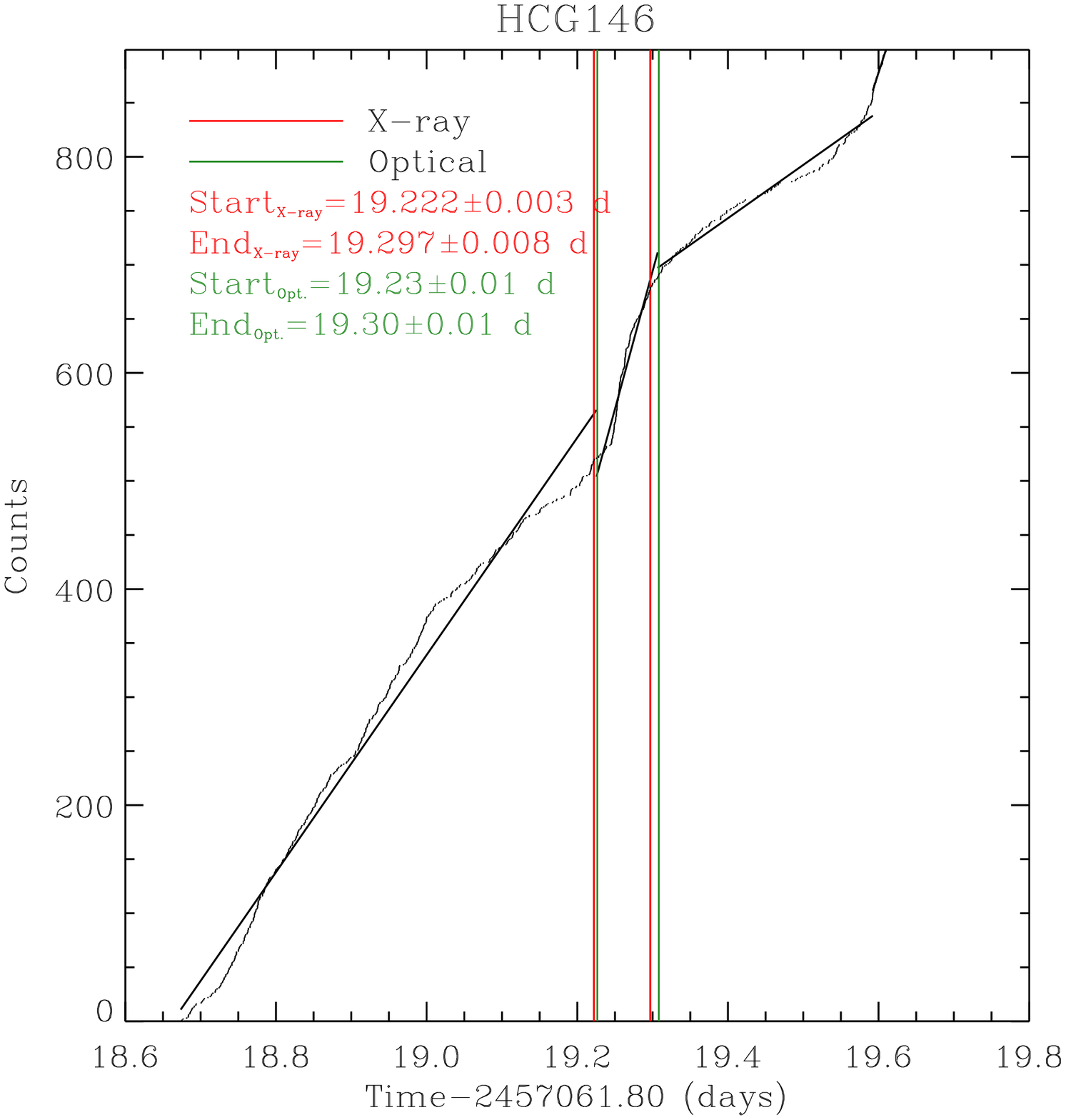}
    \includegraphics[width=6cm]{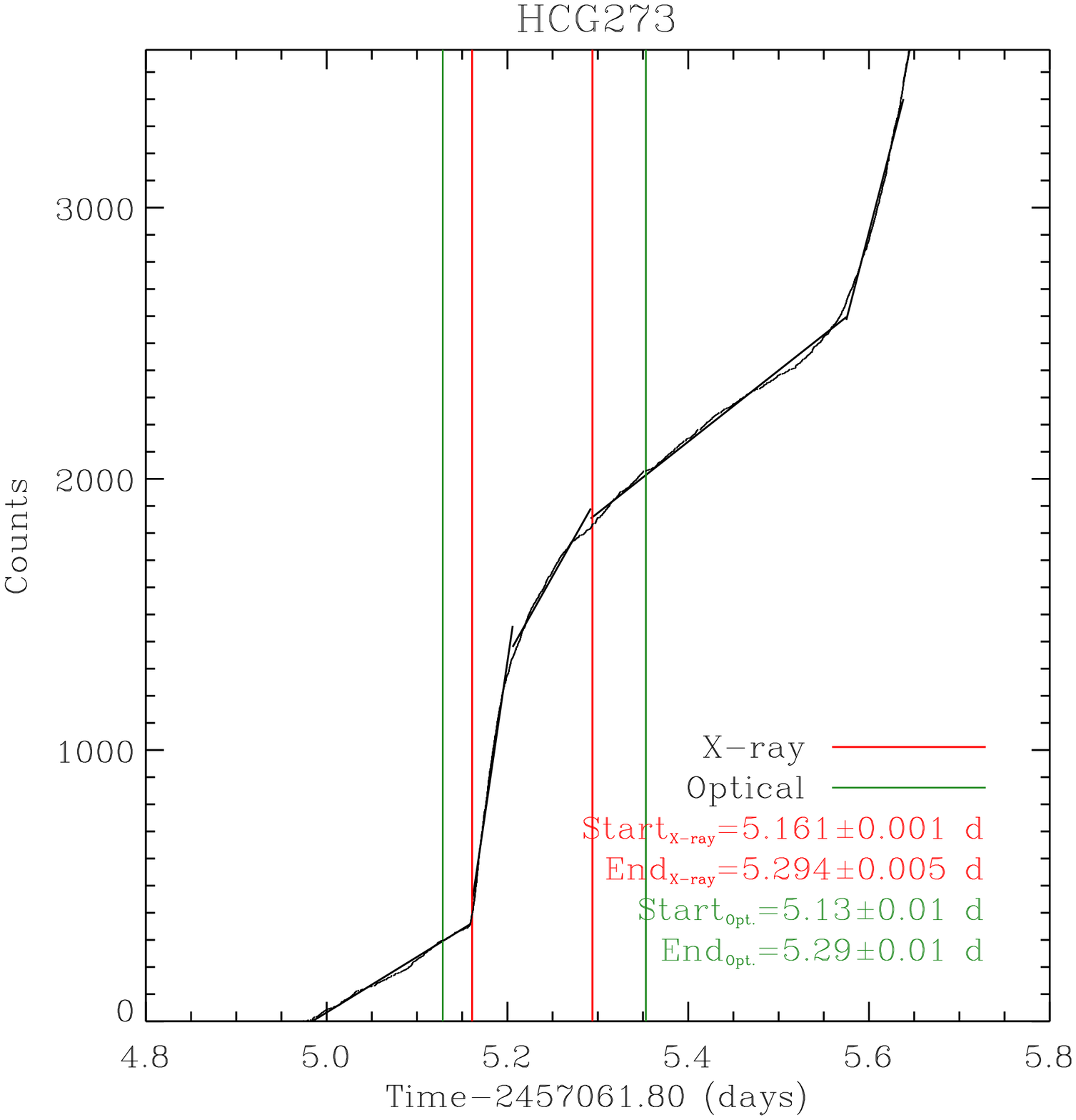}
    \includegraphics[width=6cm]{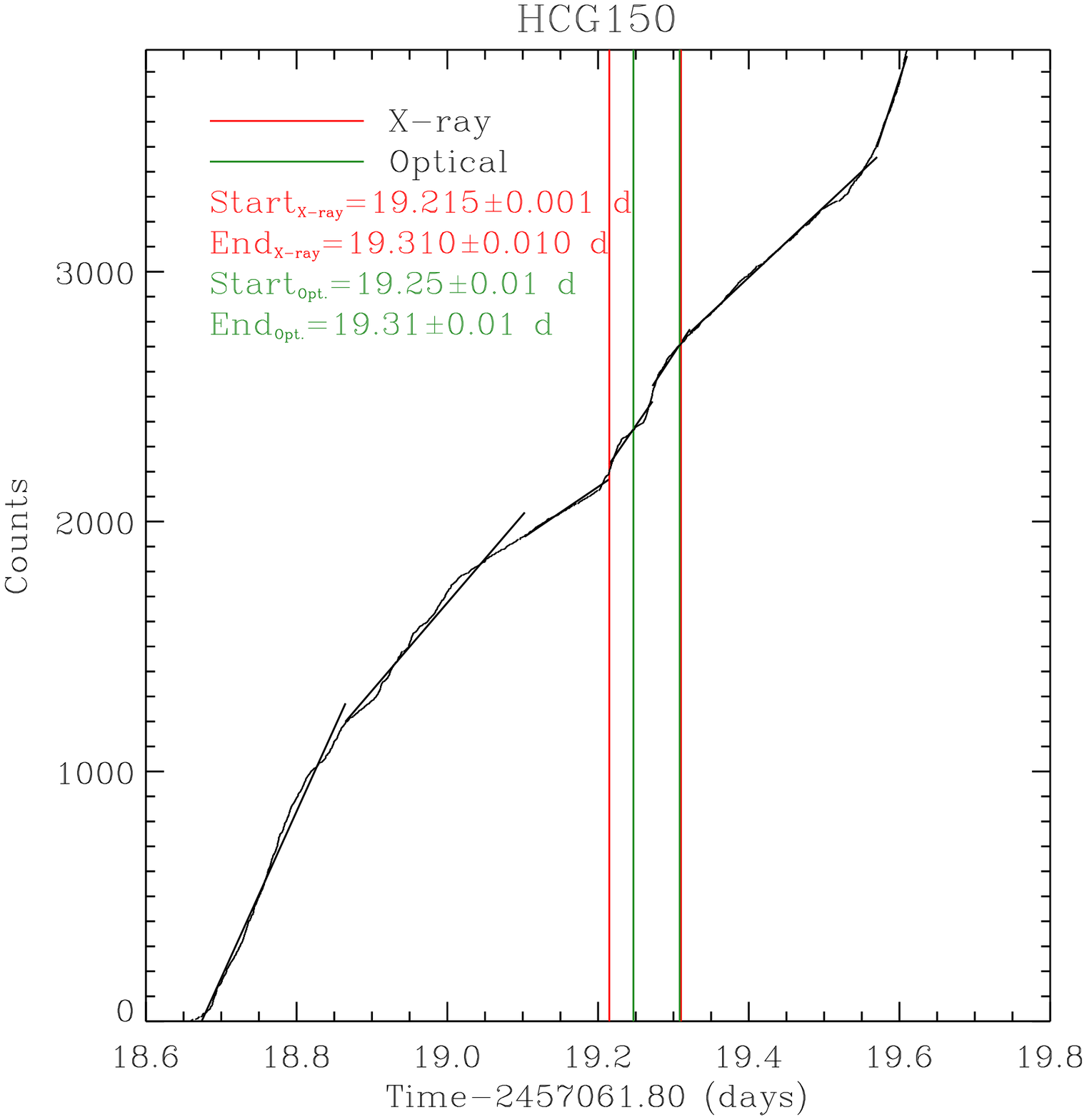}
    \includegraphics[width=6cm]{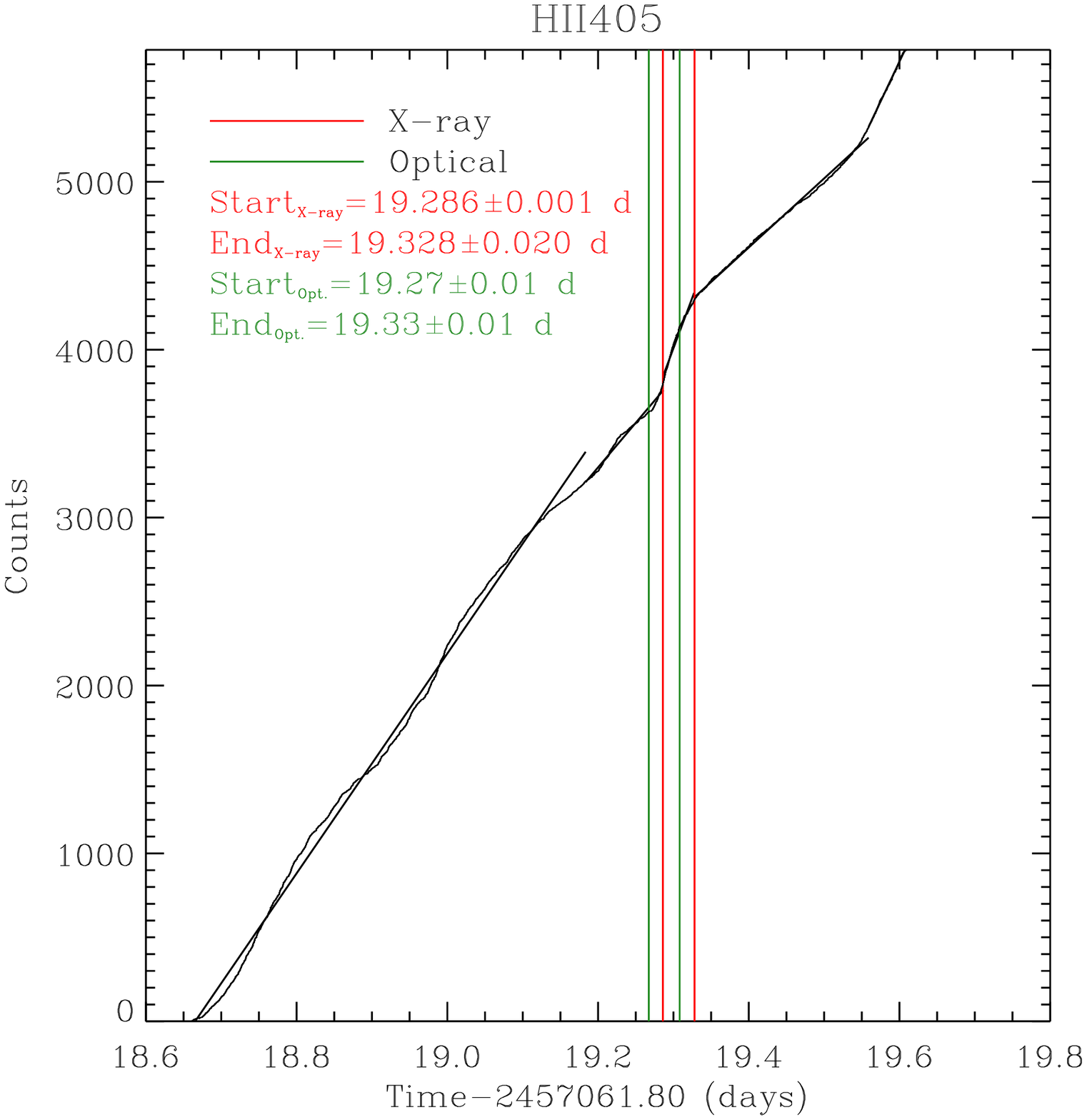}
    \includegraphics[width=6cm]{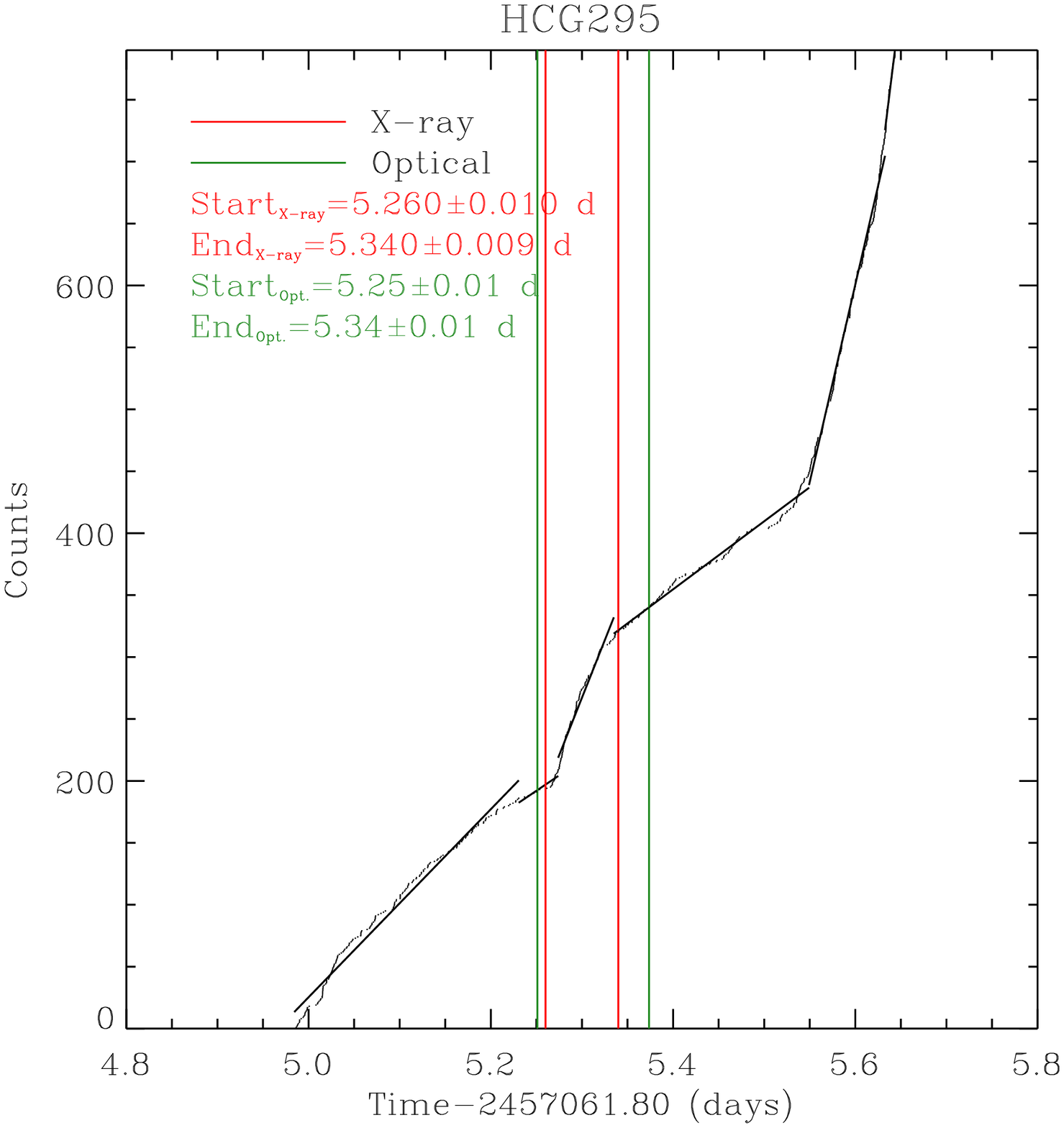}
    \includegraphics[width=6cm]{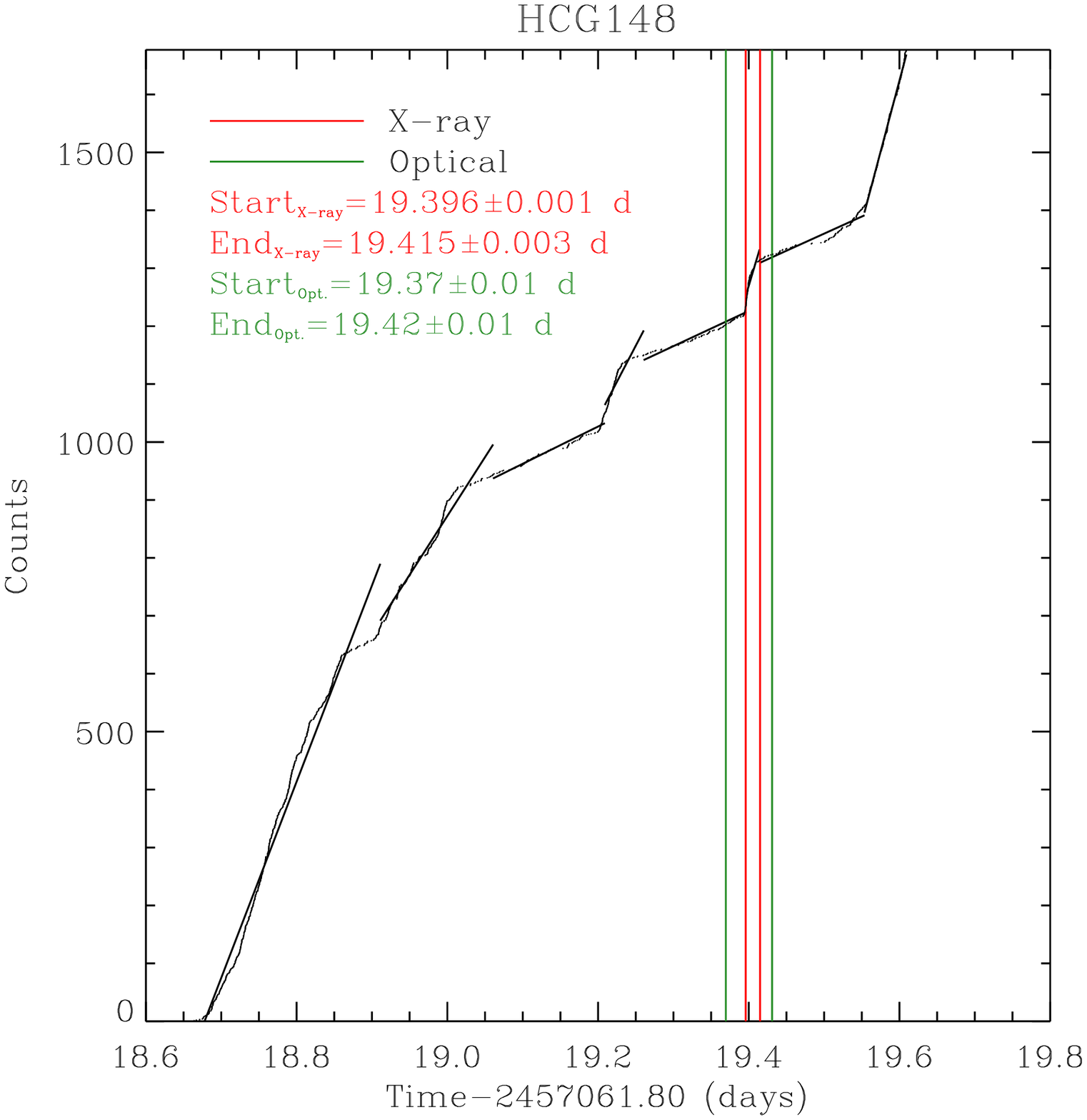}
    \caption{Sequence of the arrival time of the detected X-ray photons in the broad band in the 12 flares analyzed in this paper. The green lines mark the start and end times of the optical flare, while red lines delimit the X-ray flare.}
        \label{phot_arrival_img}
        \end{figure*}

    We therefore decided to characterize these flares by their total duration. In the K2 light curves, this is calculated as the difference between the time of the K2 data points just before and after the flares, that is, those marked with green squares in Figs. \ref{HHJ336_lc}--\ref{HCG148}, with an adopted uncertainty of 21.2 (15$\times$$\sqrt{2}$) minutes. In the XMM light curves, the brightness and variability of the background prevented us from adopting the same approach. In fact, despite the brightness of the flares studied in this paper, it was not always possible to distinguish the tail of flare decay from background fluctuations, even by dividing the light curve into time blocks. We therefore estimated the duration of the X-ray flares directly from the arrival sequence of detected photons, as shown in Fig. \ref{phot_arrival_img}. In this figure, each panel shows the detection sequence of all the X-ray photons in the broad band for each star in our sample. In order to isolate the flares from the quiescence and thus derive the start and end times of each flare, we defined in these sequences the time intervals characterized by an almost constant count rate adopting the procedure described below. \par

We first considered the entire sequence of photon detection times and performed a linear fit of the counts versus time curve. In order to test whether the detection sequence was compatible with that describing a constant count rate, we calculated the two-sided Kolmogorov-Smirnov statistic between the observed sequence and that derived from the linear fit. If the resulting significance of the K-S statistic is larger than an adopted threshold (meaning that the observed sequence of detections is not compatible with a constant count rate), we removed the last data point and repeated both the fit and the K-S test. The procedure was repeated until the K-S statistic indicated that the sequence of photon detections was compatible with a constant count rate. When this happened, all photons were collected into a time block. The test was then repeated from the first photon detected after the time block. In this way, the sequence of detections of X-ray photons was split into a sequence of broken lines, as shown in Fig. \ref{phot_arrival_img}. The intervals corresponding to the X-ray flares are easily identified as they are steeper than the intervals collecting quiescence photons. In order to quantify the uncertainties on this estimate of flare start and end times, we repeated the procedure adopting different values of the threshold for the K-S statistic, varying it by about 10\%. Table \ref{flaretau_table} lists the resulting mean flare duration, the associated uncertainties, and the delay of the X-ray flare with respect the optical flare (i.e., t$\rm_{delay}$=t$\rm_{xray\_start}$-t$\rm_{kep\_start}$). In most (eight) cases the optical flare onset precedes that of the X-ray flare, in three cases t$\rm_{delay}$ is compatible with zero within errors, while only in one case (HCG~150) does the X-ray flare precede the optical flare. However, we note that in this star, residual background fluctuations may have affected the estimate of the initial rise time (see Fig. \ref{HCG150}). This result is compatible with the general model of stellar flares \citep{Benz2008LRSP.5.1B}, with the optical event occurring in the chromosphere/upper photosphere typically preceding the X-ray flare in the magnetic loop, and it has been observed in solar flares \citep[e.g.,][]{MartinezOliveros2012ApJ.753L.26M} and in stellar flares \citep[e.g.,][]{Gudel2002ApJ.580L.73G,Flaccomio2018AA.620A.55F}. \par

    \begin{table}
    \caption{Flare duration}
    \label{flaretau_table}
    \centering                       
    \begin{tabular}{|c|c|c|c|}
    \hline
    \multicolumn{1}{|c|}{Name} &
    \multicolumn{1}{|c|}{t$_{\rm kep}$} &
    \multicolumn{1}{|c|}{t$_{\rm xray}$}  &
    \multicolumn{1}{|c|}{t$_{delay}$} \\    
    \hline
    \multicolumn{1}{|c|}{} &
    \multicolumn{1}{|c|}{ksec} &
    \multicolumn{1}{|c|}{ksec} &
    \multicolumn{1}{|c|}{minutes}\\
    \hline
    HII~212    &   12.4$\pm$1.3  &       14$\pm$3   & 64.0$\pm$15.3\\
    HHJ~273    &   14.1$\pm$1.3  &     20.1$\pm$0.4 & 76.1$\pm$15.1\\
    HHJ~336    &    3.6$\pm$1.3  &      4.0$\pm$0.2 & 7.2$\pm$15.1\\
    HCG~244    &    8.8$\pm$1.3  &      3.6$\pm$0.5 & 29.3$\pm$15.1\\
    HII~345    &    5.3$\pm$1.3  &      1.0$\pm$0.1 & 33.1$\pm$15.1\\
    HII~1516   &    3.5$\pm$1.3  &      1.0$\pm$0.1 & 22.6$\pm$15.1\\
    HCG~146    &    7.1$\pm$1.3  &      6.5$\pm$0.7 & -6.6$\pm$15.6\\
    HCG~273    &   19.4$\pm$1.3  &     11.5$\pm$0.4 & 46.8$\pm$15.1\\
    HCG~150    &    5.3$\pm$1.3  &      8.2$\pm$0.9 & -46.1$\pm$15.1\\
    HCG~295    &   10.6$\pm$1.3  &      7.0$\pm$1   & 12.8$\pm$20.8\\
    HII~405    &    3.5$\pm$1.3  &        4$\pm$2   & 26.7$\pm$15.1\\
    HCG~148    &    5.3$\pm$1.3  &      1.6$\pm$0.3 & 38.0$\pm$15.1\\
    \hline                      
    \multicolumn{4}{l}{t$\rm_{delay}$=t$\rm_{xray\_start}$-t$\rm_{kep\_start}$} \\
    \end{tabular}
    \end{table}

    \begin{figure}[]
        \centering      
        \includegraphics[width=9cm]{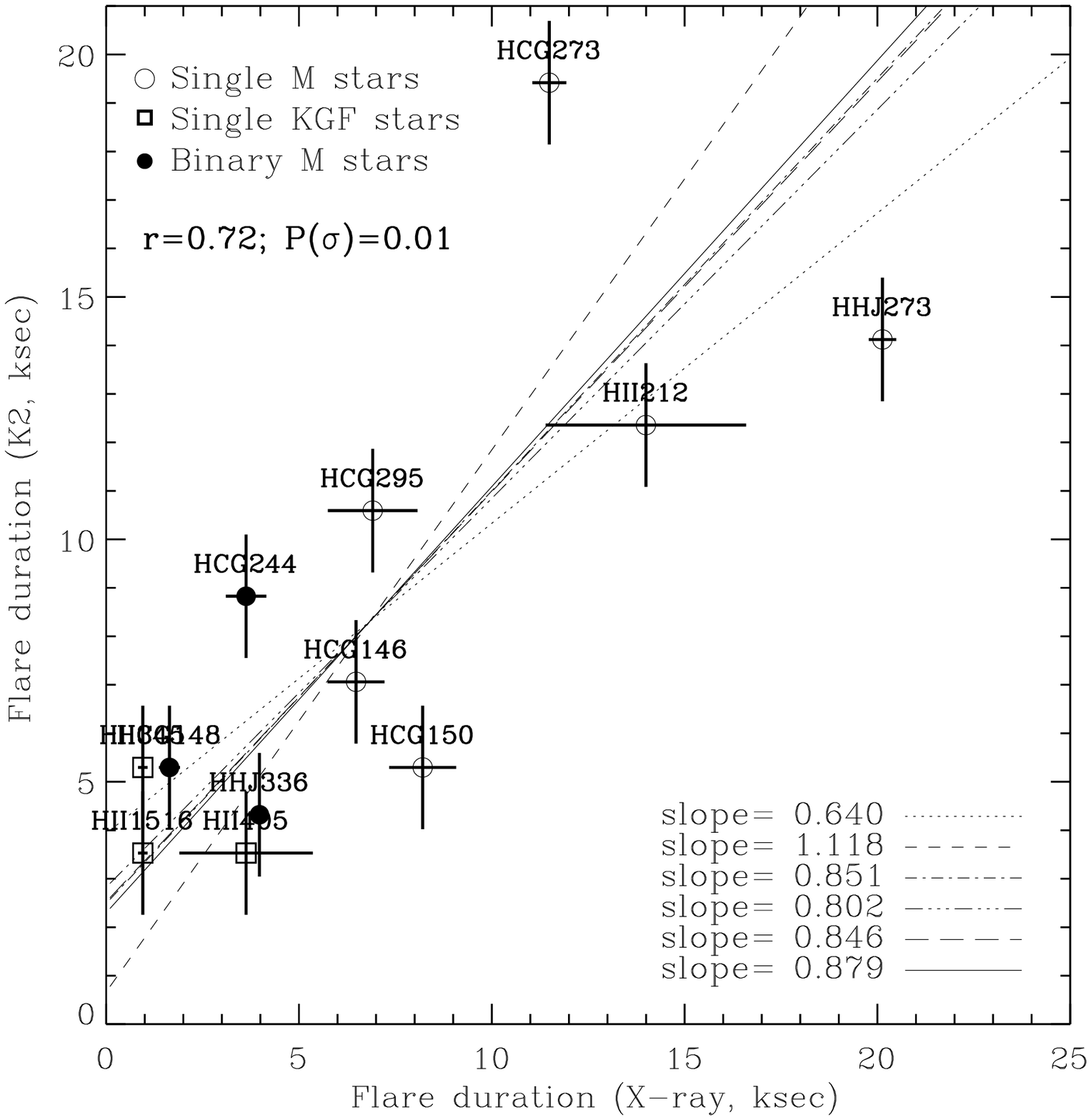}
        \caption{Flare duration in optical vs. X-ray light. For each data point, we show the name of the star. The lines mark six linear fits performed with the \emph{IDL} routine \emph{SIXLIN}. The resulting slopes are indicated in the bottom-right corner (from top to bottom, the fitting methods are: Ordinary Least Squares Y vs. X, Ordinary Least Squares X vs. Y, Ordinary Least Squares Bisector, Orthogonal Reduced Major Axis, Reduced Major-Axis,  Mean ordinary Least Squares). In the left corner we show the results from a Spearman correlation. We observe significant correlation between the flare duration in the two bands, and can see that the flares occurring in the KGF stars were shorter than those in M stars.}
        \label{duration_plot}
        \end{figure}

Figure \ref{duration_plot} compares the duration of the flares in optical and X-ray light. Given the small number of data points, we used the \emph{IDL} routine \emph{SIXLIN} to perform a linear fit of the points in Fig. \ref{duration_plot} adopting six different statistical methods for the linear regression. The resulting slopes, shown in the figure, are typically slightly smaller than one. Most of the flares in our sample therefore have similar duration in the two bands, even if typically slightly longer in X-rays, as observed in most solar flares. In Fig. \ref{duration_plot} we used different symbols to mark single M stars, single KGF stars, and binary stars (all three binary stars in our sample are M stars). Flare duration seems to be different for stars in these three groups, with single M stars showing the longest flares, while binary stars and single KGF stars hosting the shortest flares. This difference is more important in X-rays than in optical, with all X-ray flares shorter than 5 days occurred in binary stars or single KGF stars. Given the small number of stars populating the ``KGF'' sample, our data do not allow us to state whether this is a real flare property or due to a selection effect, since the determination of flare end time may be hindered by high quiescence level and the three KGF stars are the only examples where the X-ray quiescent level is brighter than log(L$\rm_X\,[erg/s]$)>29.4). \par 

\begin{figure*}[]
        \centering      
        \includegraphics[width=9cm]{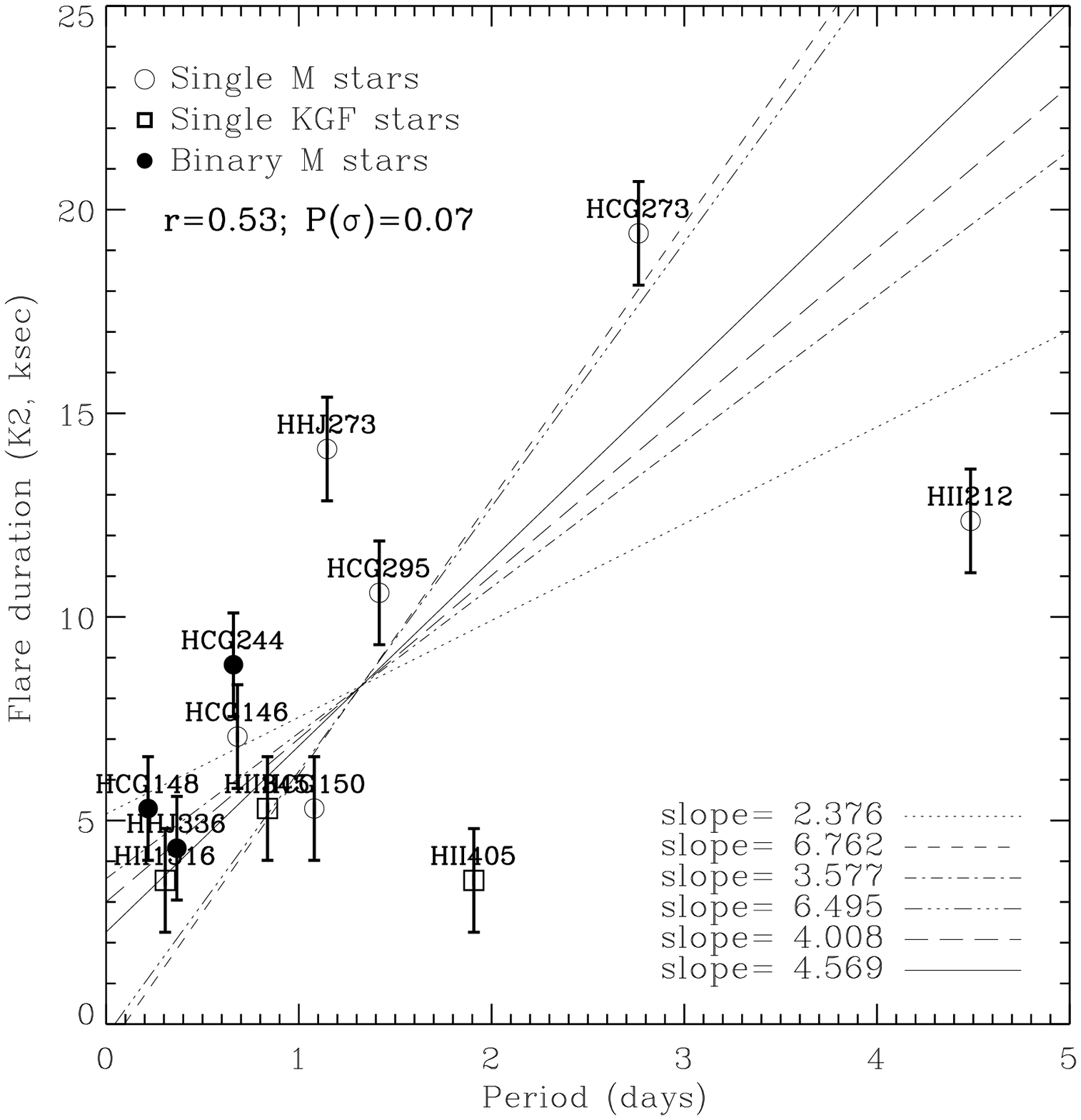}
        \includegraphics[width=9cm]{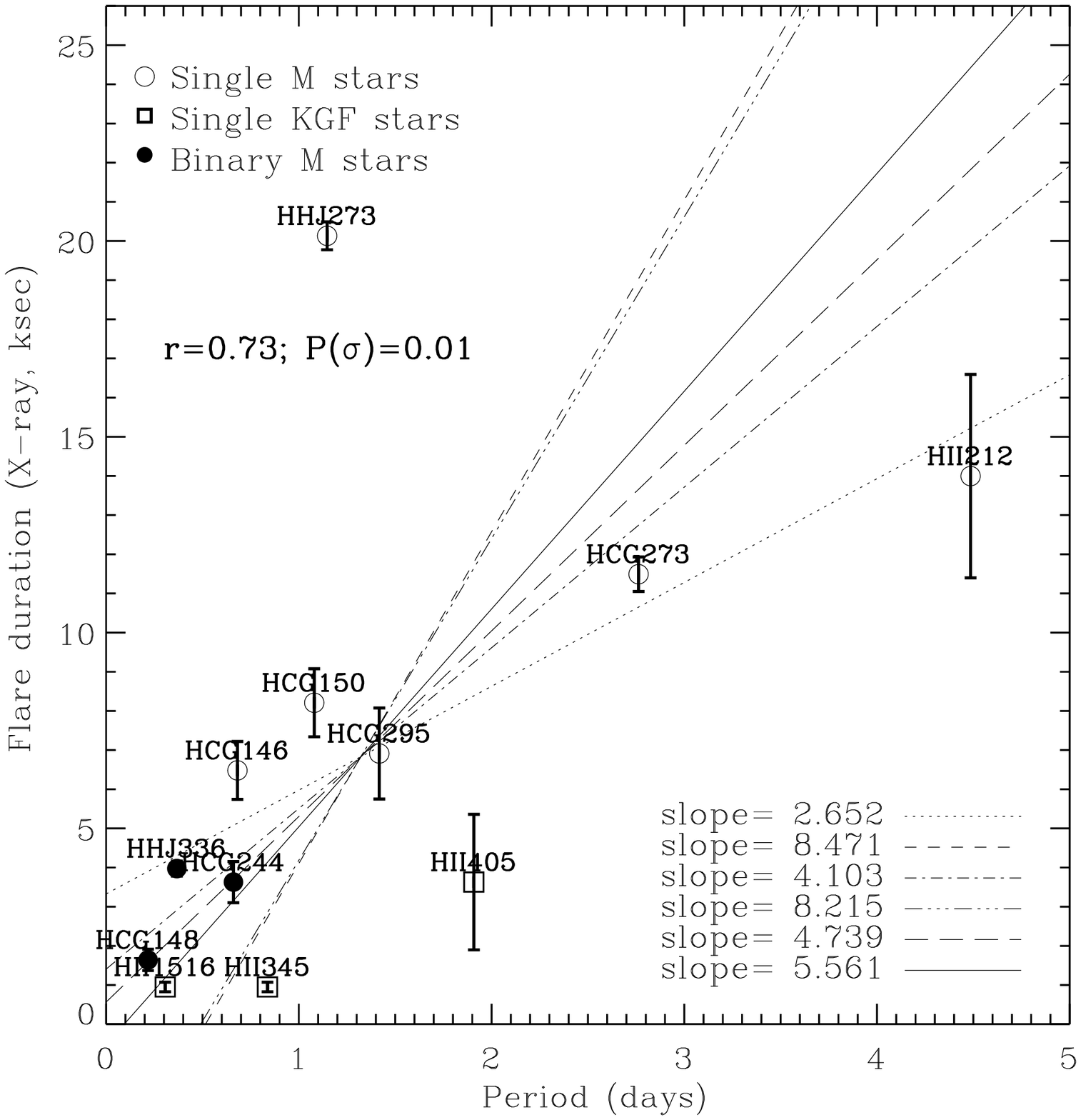}
        \caption{Flare duration in optical (left panel) and X-ray light (right panel) vs. stellar rotation period. For each data point, we show the name of the star. The lines mark six linear fits performed with the \emph{IDL} routine \emph{SIXLIN}. The resulting slopes are indicated in the bottom-right corner. In the top-left corner we also show the Spearman's rank correlation coefficient and the significance of its deviation from zero. The observed correlation between flare duration and stellar rotation period suggests that rapidly rotating stars preferentially host smaller coronal loops.}
        \label{tauvsper_plot}
        \end{figure*}

Figure \ref{tauvsper_plot} shows the flare duration in optical and X-rays as a function of stellar rotation period. In both bands we observe a correlation between the duration of the flares and stellar rotation period, which apparently is not induced by a difficulty in detecting weak flares in the slowly rotating stars. For instance, HCG~273 and HII~212, the two stars with the largest rotation period (2.8 and 4.5 days, respectively), are not the brightest stars in our sample. One might expect it to be more difficult to select weak flares for these stars. However, we note that in the right panel, the bottom-left part of the diagram (rapid rotation and short flares) is populated by binary stars and KGF stars, suggesting that binarity, spectral type, or selection effects due to different quiescent level may affect the observed correlation. If confirmed, this correlation may indicate that rapidly rotating stars predominantly host short flares. Since short flares are typically associated with small flaring structures, and probably short magnetic loops, this might be in general agreement with the scenario where in rapidly rotating stars the intense centrifugal stripping inhibits the formation of long loops and reduces the volume available for stellar coronae \citep[e.g.,][]{Argiroffi2016AA.589A.113A}. However, a detailed analysis of the connection between loop geometry, energy release, and flare rate as a function of the stellar rotation period requires a larger and more complete sample of flares to confirm the existence of a correlation between flare decay time and stellar rotation. \par

It was therefore necessary to verify how our estimate of the X-ray flare duration can be affected by the intensity of the quiescence. To this aim, we repeated the calculation of the flare duration on simulated sequences of photon time detection. We considered five quiescent count rates ranging from 0.05$\,$cnt/s to 0.5$\,$cnt/s, and simulated 100 sequences of photon time detection for each of these assumed quiescence count rates. We also injected into each of these simulated quiescent levels a flare with a count rate of about 0.17$\,$cnt/s (600 photons in 3.63$\,$ks). These values were chosen in order to obtain a photon detection sequence with a quiescent level of 0.05$\,$cnt/s similar to that of HCG~244 (Fig. \ref{phot_arrival_img}). For each of the simulated sequences of photon detection we calculated the start and end time of the flare and thus its duration. The average values corresponding to a given quiescence count rate range from about 4.3$\,$ks (quiescent count rate of 0.05$\,$cnt/s) to 3.0$\,$ks (0.5$\,$cnt/s). This test confirms that our estimate of the duration of the X-ray flare may depend on the intensity of the quiescence, likely because the fainter  the quiescence level the easier we can discern between quiescence and the end of a  flare. However, the $\sim$1$\,$ks difference is not large enough to affect the results obtained from Figs. \ref{duration_plot} and \ref{tauvsper_plot}. In Appendix \ref{AppC_sec} we show five simulated sequences of photon detection time. \par

\section{Time-averaged flare properties}
\label{glob_flare}

   In order to calculate the energy released in the Kepler band, we first computed the equivalent duration (ED) of each flare, which is the integral under the normalized K2 light curve during the flare \citep{Davenport2016ApJ}. The ED is in units of seconds and corresponds to the time it would take the star to emit in quiescence the same amount of energy released by the flare \citep{Gershberg1972ApSS.19.75G,Hunt-Walker2012PASP.124.545H}. Thus, the ED allows us to estimate the total energy released by the flare in the Kepler band by multiplying it by the quiescent stellar luminosity in the Kepler band (E$_{\rm kep,flare}=$ED$\times$L$_{\rm kep,quies}$). We calculated the stellar Kepler luminosity in quiescence (L$_{\rm kep,quies}$) from the magnitude in the Kepler band and individual E$_{\rm B-V}$ value listed in the \emph{K2 Ecliptic Plane Input Catalog}\footnote{http://archive.stsci.edu/k2/epic/search.php} \citep{HuberBHB2016ApJS}, adopting the individual distances listed in Table \ref{sample_table}, a zero point flux in the Kepler band of $3.257\times10^{-9}\,$erg/cm$^2$/s/\AA$\,$ in the AB photometric system\footnote{http://svo2.cab.inta-csic.es/theory/fps/index.php?id=Kepler/Kepler.K \\ \&\&mode=browse\&gname=Kepler\&gname2=Kepler\#filter}, the extinction coefficient in the Kepler band equal to 0.859$^{\rm m}$ from the \citet{Donnell1994ApJ} extinction law, and ignoring any dependency of the extinction law and zero-point flux on stellar spectral type. This approach does not account for the variability of the star and the different spectral shape of the quiescent and flare emission. In principle we could assume a blackbody spectrum for both star and flare emission, the former at the stellar effective temperature and the latter at 10000$\,$K \citep{ShibayamaMNN2013ApJS}. However, this approach is also approximated, since it does not account for the time evolution of the flare spectral energy distribution (SED), the contribution by nonthermal emission to white light flares \citep{KowalskiHWO2013ApJS}, and it assumes the knowledge of the size of the emitting region and stellar radii. \par 
    
In order to calculate the energy released by the X-ray flare and the average flare plasma properties (plasma temperature kT and emission measure EM), we used XSPEC v.~12.8.1 \citep{Arnaud1996} to analyze the X-ray spectrum of each source extracted during the flare. X-ray spectra were fitted with a single temperature (1T) APEC ionization-equilibrium isothermal plasma model \citep{SmithBLR2001ApJL} after subtracting the spectrum extracted during quiescence. In this way, we isolated the contribution to the X-ray spectrum of the flaring plasma. We adopted the element abundances defined by \citet{MaggioFFM2007}. Also, we accounted for interstellar absorption using the TBABS model \citep{WilmsAM2000}, fixing the absorption column N$_{\rm H}$ to the value obtained by multiplying the individual known optical extinctions for 1.8$\times$10$^{21}\,$cm$^2$/mag. We verified that leaving N$_{\rm H}$ as a free parameter did not significantly improve the quality of the fit. Best-fit models were chosen by minimizing the C-statistic and the quality of the fit tested using the XSPEC tool \emph{goodness}. The limit for acceptable fits adopted in this paper, set to a null-hypothesis probability of a good fit (P$_{\%}$) equal to 5\%, was always met using 1T APEC models, given that the number of detected photons did not allow us to resolve the thermal structure of the flaring plasma. The significance of the parameters was verified by analyzing the confidence contours in the C-stat space with the XSPEC tool \emph{steppar}. The total energy released in the broad X-ray band (E$_{\rm xray,flare}$) was calculated using the XSPEC model CFLUX to obtain the flux, which was converted into luminosity adopting the individual stellar distance and multiplying it by the flare duration. The peak luminosity L$_{\rm xray,peak}$ was calculated with the time-resolved X-ray spectral fit explained in Sect. \ref{time_res} and was found to be the largest luminosity obtained in the time blocks defined to sample the flare. In order to estimate plasma properties and X-ray luminosity during quiescence, we fitted the quiescent X-ray spectra using the same set of models adopted for the flaring plasma. For the quiescence, however, background spectra were extracted in suitable background regions selected following the prescription of the SAS guide (see Sect. \ref{sample_sec}). Results are summarized in Table \ref{flareprop_table}, which shows the total energy and peak luminosity of each flare, together with the average plasma properties observed during quiescence and during the flares. We do not show the average flare temperature obtained for the flare that occurred in HII~212 since it is not well determined by the spectral fit that found a temperature smaller than the quiescence value.
\par

    \begin{table*}
    \caption{Absorption, plasma temperature, luminosity, and total energy in optical and X-ray light during quiescence and flares.}
    \label{flareprop_table}
    \centering                       
    \begin{tabular}{|c|c|c|c|c|c|c|c|c|}
    \hline
    \multicolumn{1}{|c|}{Name} &
    \multicolumn{1}{|c|}{N$_{\rm H}$} &
    \multicolumn{1}{|c|}{kT$_{\rm quies}$} &
    \multicolumn{1}{|c|}{kT$_{\rm flare}$} &
    \multicolumn{1}{|c|}{log(L$_{\rm xray,quies}$)} &
    \multicolumn{1}{|c|}{log(L$_{\rm xray,peak}$)} &
    \multicolumn{1}{|c|}{log(E$_{\rm xray,flare}$)} &    
    \multicolumn{1}{|c|}{log(L$_{\rm kep,quies}$)} &
    \multicolumn{1}{|c|}{log(E$_{\rm kep,flare}$)} \\
    \hline
    \multicolumn{1}{|c|}{} &
    \multicolumn{1}{|c|}{$10^{21}\,$cm$^{-2}$} &
    \multicolumn{1}{|c|}{keV} &
    \multicolumn{1}{|c|}{keV} &
    \multicolumn{1}{|c|}{[erg/sec]} &
    \multicolumn{1}{|c|}{[erg/sec]} &
    \multicolumn{1}{|c|}{[erg]} &
    \multicolumn{1}{|c|}{[erg/sec]}  &
    \multicolumn{1}{|c|}{[erg]}\\
   \hline
    HII~212   & 0.6 & $0.77^{0.85}_{0.69}$&                     & $29.05^{29.10}_{29.00}$& $30.30^{30.40}_{30.22}$& $33.52^{33.56}_{33.39}$& 32.02$\pm$0.20 & 34.16$\pm$0.42\\ 
    HHJ~273   & 1.2 & $0.69^{0.77}_{0.62}$& $1.61^{1.70}_{1.52}$& $28.73^{28.78}_{28.69}$& $30.34^{30.37}_{30.30}$& $34.04^{34.06}_{34.02}$& 31.53$\pm$0.23 & 34.12$\pm$0.42\\
    HHJ~336   & 1.1 & $0.35^{0.42}_{0.29}$& $1.56^{1.65}_{1.45}$& $28.86^{28.94}_{28.79}$& $30.43^{30.49}_{30.37}$& $33.65^{33.68}_{33.62}$& 31.88$\pm$0.23 & 33.59$\pm$0.74\\
    HCG~244   & 0.9 & $0.58^{0.61}_{0.55}$& $1.66^{1.98}_{1.43}$& $29.11^{29.12}_{29.09}$& $29.75^{29.86}_{29.67}$& $32.90^{32.95}_{32.85}$& 32.01$\pm$0.24 & 34.04$\pm$0.41\\
    HII~345   & 1.4 & $0.59^{0.60}_{0.59}$& $2.45^{2.69}_{2.23}$& $30.18^{30.19}_{30.18}$& $30.85^{30.90}_{30.80}$& $33.68^{33.71}_{33.66}$& 33.27$\pm$0.20 & 34.66$\pm$0.50\\
    HII~1516  & 0.2 & $0.91^{0.93}_{0.89}$& $2.56^{3.21}_{2.30}$& $29.42^{29.43}_{29.41}$& $30.51^{30.58}_{30.45}$& $33.54^{33.58}_{33.51}$& 32.17$\pm$0.20 & 33.71$\pm$0.56\\
    HCG~146   & 1.6 & $0.74^{0.84}_{0.61}$& $0.78^{0.85}_{0.70}$& $28.54^{28.59}_{28.47}$& $29.48^{29.53}_{29.43}$& $32.99^{33.04}_{32.95}$& 31.53$\pm$0.23 & 33.52$\pm$0.47\\
    HCG~273   & 1.1 & $0.60^{0.64}_{0.56}$& $1.63^{1.69}_{1.57}$& $29.07^{29.09}_{29.06}$& $30.33^{30.36}_{30.29}$& $33.86^{33.87}_{33.84}$& 31.92$\pm$0.23 & 34.34$\pm$0.38\\
    HCG~150   & 0.7 & $0.61^{0.65}_{0.58}$& $2.16^{2.59}_{1.79}$& $28.90^{28.92}_{28.89}$& $29.62^{29.76}_{29.52}$& $33.02^{33.07}_{32.97}$& 31.63$\pm$0.23 & 32.86$\pm$0.53\\
    HCG~295   & 0.7 & $0.69^{0.76}_{0.62}$& $1.96^{3.22}_{1.44}$& $28.96^{29.00}_{28.92}$&                        & $33.36^{33.47}_{33.26}$& 31.44$\pm$0.21 & 33.49$\pm$0.41\\
    HII~405   & 1.0 & $0.47^{0.49}_{0.46}$& $1.29^{1.36}_{1.21}$& $29.66^{29.68}_{29.65}$& $29.87^{29.98}_{29.77}$& $33.50^{33.53}_{33.46}$& 33.75$\pm$0.21 & 33.90$\pm$0.58\\
    HCG~148   & 0.8 & $0.32^{0.36}_{0.29}$& $1.00^{1.15}_{0.89}$& $28.71^{28.77}_{28.66}$&                        & $33.07^{33.13}_{33.01}$& 31.59$\pm$0.22 & 34.07$\pm$0.50\\
     \hline                                                                                                  
    \multicolumn{8}{l}{} \\
    \end{tabular}
    \end{table*}

    \begin{figure*}[]
        \centering      
        \includegraphics[width=8cm]{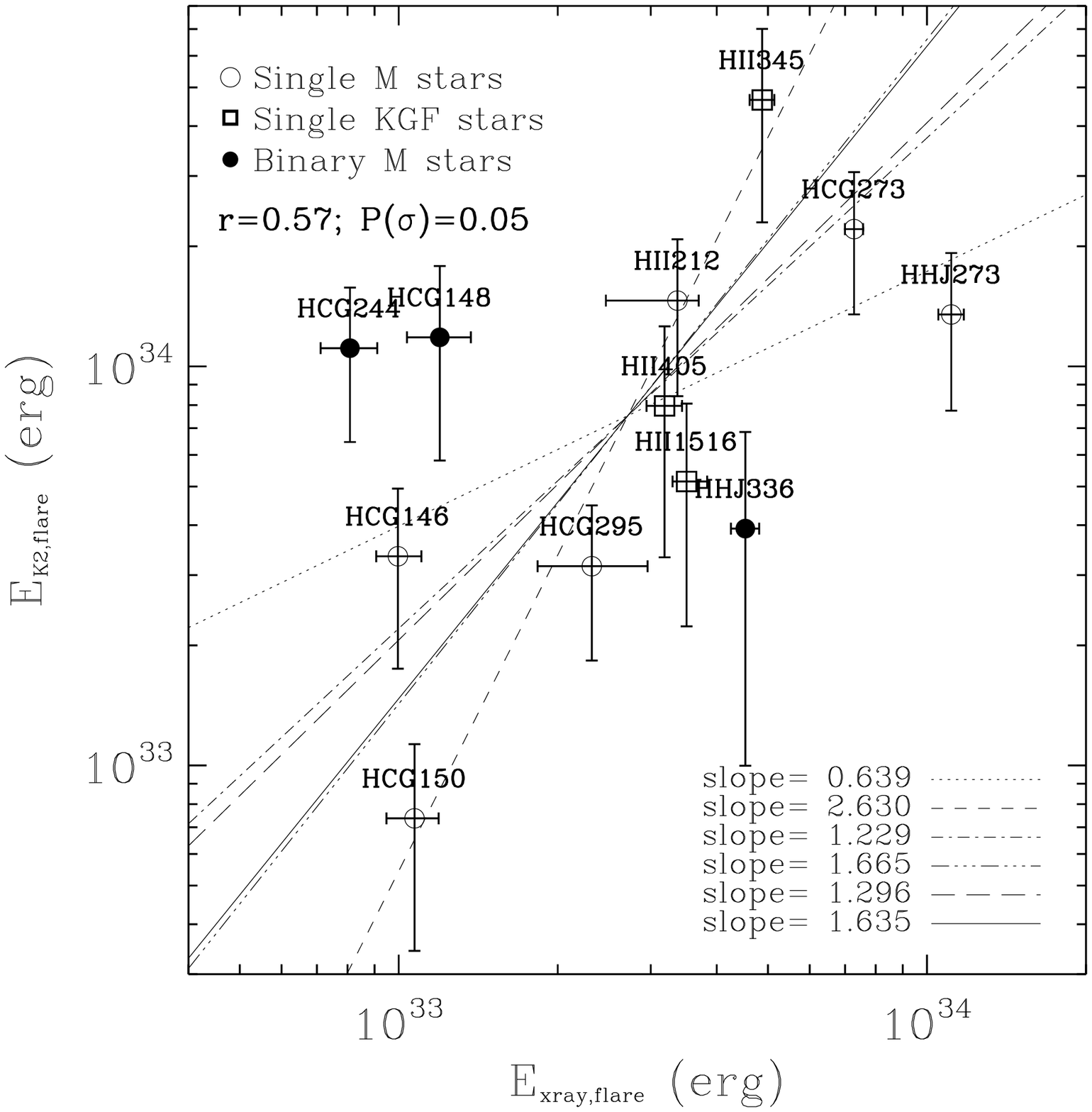}
        \includegraphics[width=8cm]{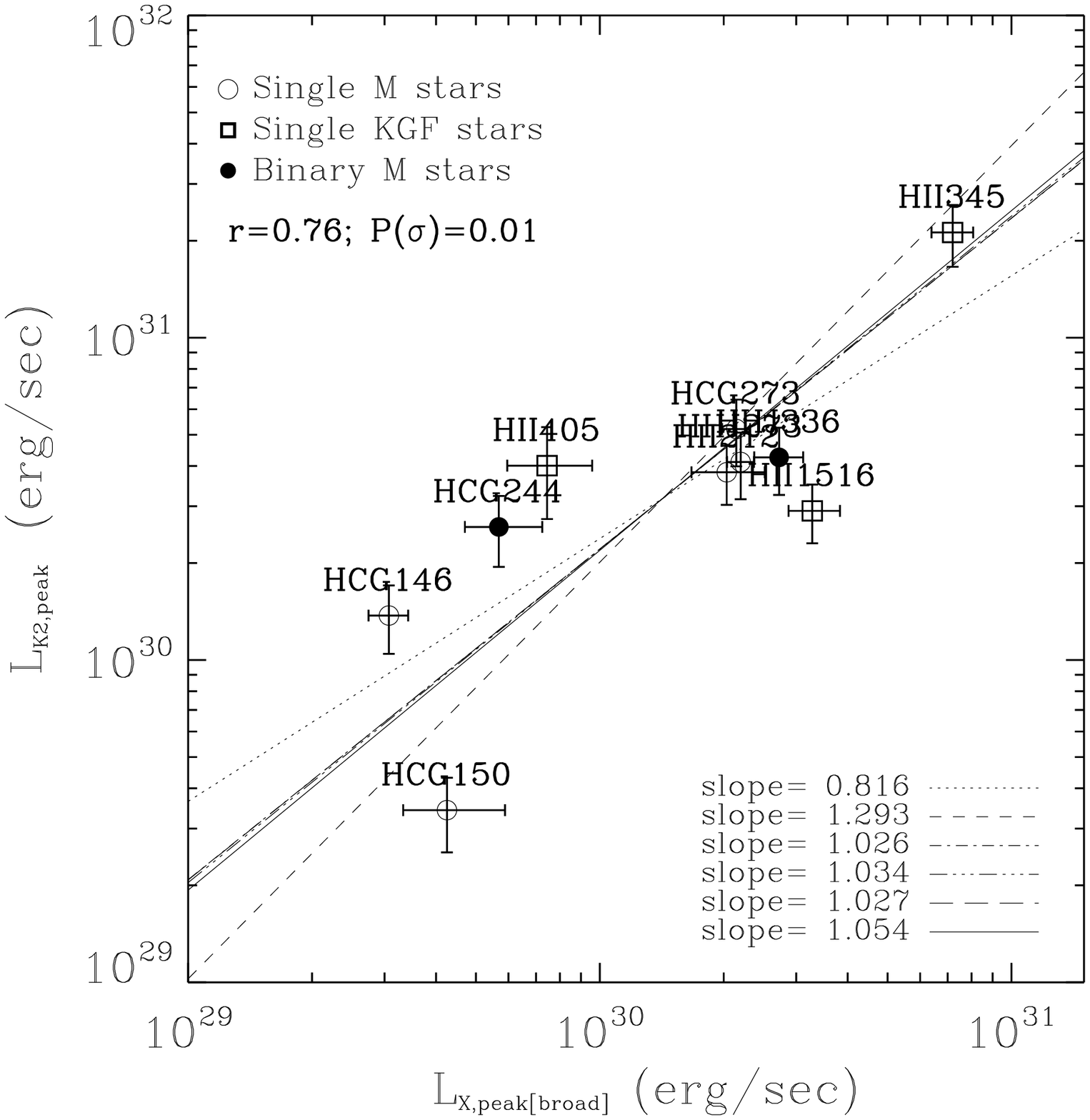}
        \caption{Comparison between the total emitted energy (left panel) and the peak luminosity (right panel) in optical and X-ray light. The name of each star is also indicated. The lines mark six linear fits between the log values performed with the \emph{IDL} routine \emph{SIXLIN}. The resulting slopes are also indicated. We also show the Spearman's rank correlation coefficient and the significance of its deviation from zero. These flares typically released more energy and had a larger peak luminosity in the optical than in X-rays.}
        \label{EkvsLk_plot}
        \end{figure*}

        Figure \ref{EkvsLk_plot}  compares the total energy released in optical and in X-ray light. As typically observed in solar flares \citep[e.g.,][]{Kretzschmar2010NatPh.6.690K,Kretzschmar2011AA.530A.84K}, most of the events analyzed in this paper released more energy in optical light than in X-rays. With the exception of the flares that occurred in HHJ~336 and HCG~150, where the E$_{\rm kep,flare}$/E$_{\rm xray,flare}$ ratio is equal to 0.9 and 0.7, respectively, in the other flares the ratio ranges from 1.2 (HHJ~273) to 13.7 (HCG~244). As indicated by a two-sided Spearman test, the observed correlation between the energy released in the two bands is weak. However, the lack of correlation in the left panel is mainly driven by the stars HCG~244 and HCG~148, the only two spectroscopically confirmed binary stars. Removing these stars, in fact, the correlation test results in r=0.75 and P($\rho$)=0.01. A stronger correlation is instead observed between the peak luminosity observed in optical and X-rays, with, again, larger values observed in the optical band. \par
        
        We searched for correlations between flare properties (such as energy and duration) both in optical and X-rays, with the estimated starspots coverage (A$_{\rm spot}$) of stellar surface. Correlations between starspots coverage and flare properties (energy and flare frequency) have been found in stars observed with \textit{Kepler} \citep{McQuillan2014ApJS.211.24M,Maehara2017PASJ.69.41M} and interpreted as resulting from the relation between A$_{\rm spot}$ and the stellar magnetic activity level \citep[e.g.,][]{McQuillan2014ApJS.211.24M}. To calculate the starspot coverage, we used Eqs. 2 and 3 of \citet{Maehara2017PASJ.69.41M}, which are valid with the hypothesis that stellar spots are organized mainly in large spots and their lifetime is longer than stellar rotation period. Both hypotheses seem reasonable for our sample of active and rapidly rotating stars. These equations require stellar radii and effective temperature, listed in Table \ref{sample_table}, and the normalized amplitude of the optical light curves. Optical amplitude variability is in fact related to the apparent area of starspots, as suggested by \citet{McQuillan2014ApJS.211.24M} and \citet{Maehara2017PASJ.69.41M}. Given the short duration of the XMM observations compared with the length of K2 light curves, we measured the normalized amplitude of the K2 light curve during the XMM observations in order to characterize the starspot coverage occurring during the X-ray observations. The starspot coverage we obtained ranges between 0.01 and 0.1, with only HCG~295 with A$_{\rm spot}$=0.007 and HHJ~273 and HHJ~336 with A$_{\rm spot}$>0.5. However, the flares analyzed here do not show any correlation with the starspots coverage. We interpret this result as a consequence of the rapid rotation of these stars. In fact, correlations between flare properties and starspot coverage are typically observed for stars with rotation periods longer than 3 days, while in more rapidly rotating stars a sort of ``saturation'' of flare properties is found \citep[e.g.,][]{McQuillan2014ApJS.211.24M}, and, as shown in Table \ref{sample_table}, all but one of our stars have rotation periods shorter than 3 days.

\section{Time-resolved X-ray analysis}
\label{time_res}

We studied the time evolution of the emitting plasma during the X-ray flares by sampling the merged MOS+PN light curves with suitable time intervals. Despite being useful for flare detections, the intervals defined in Sect. \ref{sample_sec} did not always sample the flares properly. For this reason, we adopted two additional sets of time blocks: one defined by collecting 100 net photons in the broad band in each block starting from the beginning of the flare, or defining the blocks ad-hoc after visually inspecting the light curve. Among the three sets of intervals defined for each flare, we used the one resulting in the best time sampling, taking into account also the X-ray counts detected in each interval. We thus calculated the time-resolved X-ray properties (plasma temperature and emission measure) in each block, repeating the spectral analysis described in Sect. \ref{glob_flare}. The flares in HCG~295 and HCG~148 were too faint for this analysis. \par

    The main objective of this analysis was to track the time evolution of the X-ray-emitting plasma in the kT-versus-log(EM) plane, where it is possible to distinguish the typical path of the four phases of the evolution of a single loop: the heat pulse with a sudden increase of plasma temperature; the evaporation phase during which the heated plasma fills the magnetic loop until the plasma density reaches its peak; the conductive cooling phase with a decline of temperature while density is still increasing; and the radiative cooling phase after the density reaches its peak, with both temperature and density decreasing \citep{Reale2007AA}. This produces a characteristic path in the temperature-versus-density (or emission measure) plane. This path, and in particular the slope of the cooling phase, can be used as a diagnostic for flare cooling and heating processes. \par

\begin{figure*}[]
        \centering      
        \includegraphics[width=18cm]{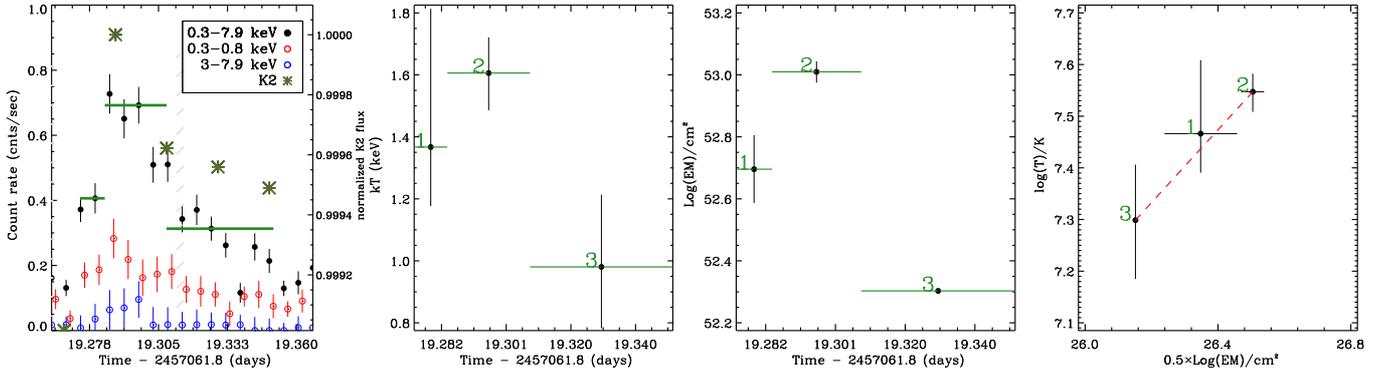}
        \caption{Evolution of the flare observed in HII~405. The left panel shows the merged MOS+PN light curve in broad, soft, and hard energy bands (dots) superimposed on the K2 light curve (green asterisks). The horizontal green lines mark the defined blocks and the vertical dashed areas correspond to the bad time intervals (Sect. \ref{xray_data_sec}). The central panels show the time evolution of plasma temperature and emission measure, while the right panel shows the evolution of the flare in the log(T) vs. 0.5$\times$log(EM) plane. The red dashed line marks the slope of the cooling phase.}
        \label{HII405_flare}
        \end{figure*}

\begin{figure*}[]
        \centering      
        \includegraphics[width=18cm]{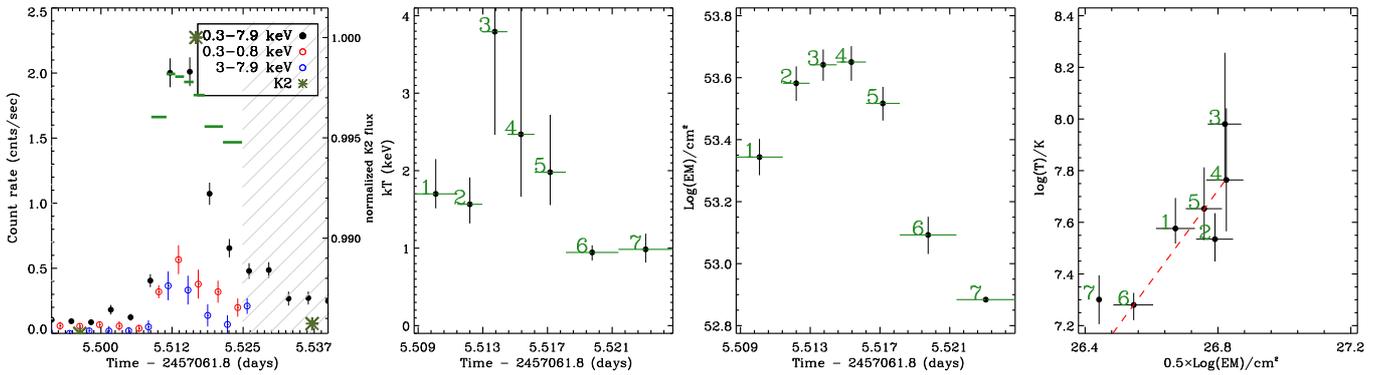}
        \caption{Evolution of the flare observed in HII~1516. Panel layout and content as in Fig.~\ref{HII405_flare}.}
        \label{HII1516_flare}
        \end{figure*}

Figures  \ref{HII405_flare} and \ref{HII1516_flare} show the evolution of the flares observed in HII~1516 and HII~405, respectively. These are the two flares for which  the heating phase could be distinguished with the defined blocks. In each of these figures, the left panel shows the X-ray flare in broad, soft, and hard energy bands, superimposed on the optical flare, together with the defined set of time intervals. Count rates in the soft and hard light curves during the bad time intervals are not shown in these figures, given that their amplitude is typically comparable with background fluctuations. In the flare in HII~1516, the count rate increased by a factor of 40, while a smaller amplitude is observed in HII~405 (a factor of $\sim$15). The two central panels show the time evolution of plasma temperature and EM, which results in the flare evolution pattern in the temperature-versus-EM plane shown in the right panel. Together with the flare observed in HII~345, the flare in HII~1516 is the one for which we measured the highest plasma temperature averaged over the flare duration (Table \ref{flareprop_table}). \par

\begin{figure*}[]
        \centering      
        \includegraphics[width=18cm]{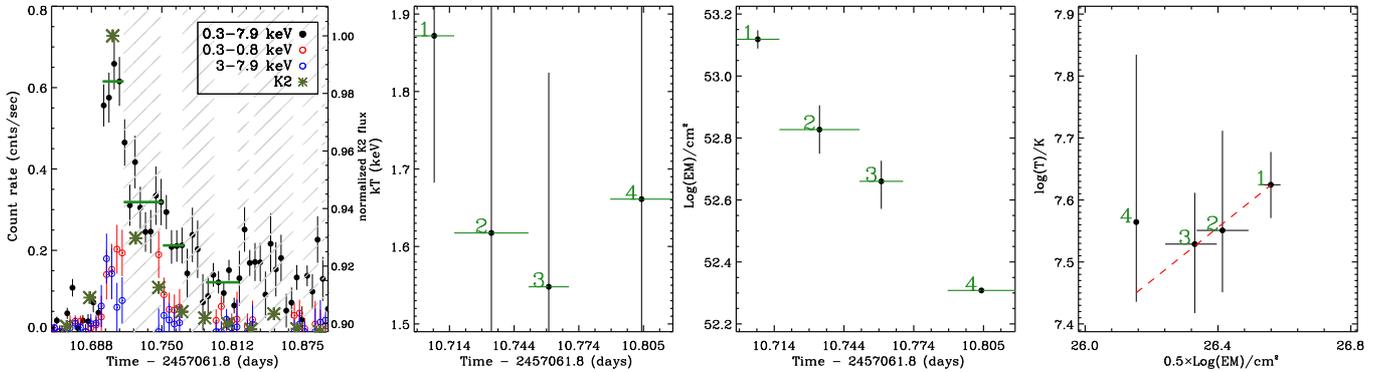}
        \caption{Evolution of the flare observed in HHJ~273. Panel layout and content as in Fig.~\ref{HII405_flare}.}
        \label{HHJ273_flare}
        \end{figure*}

\begin{figure*}[]
        \centering      
        \includegraphics[width=18cm]{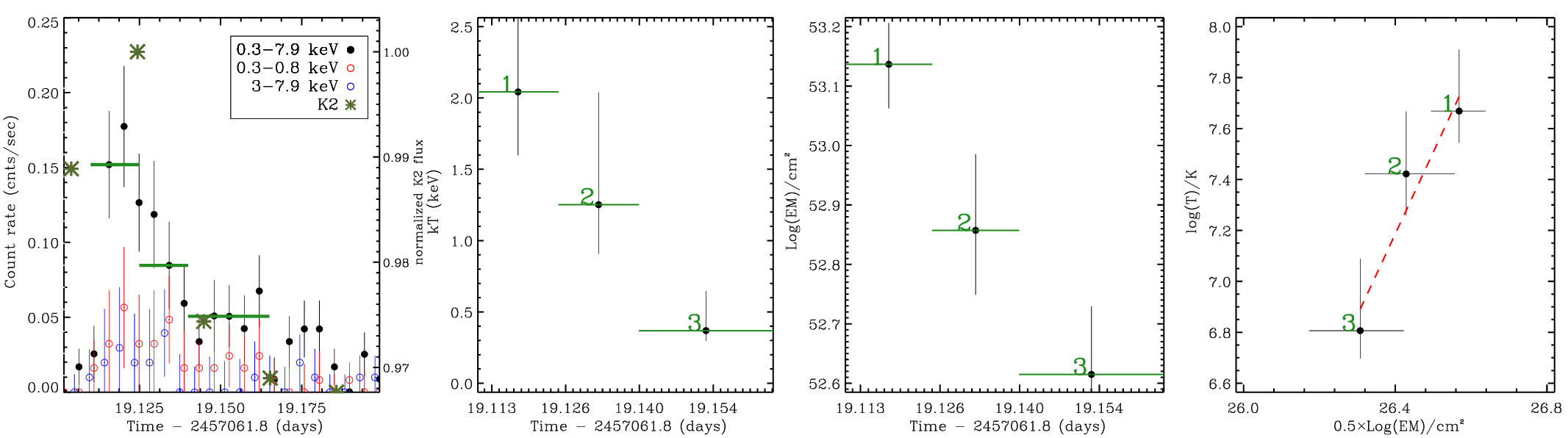}
        \caption{Evolution of the flare observed in HII~212. Panel layout and content as in Fig.~\ref{HII405_flare}.}
        \label{HII212_flare}
        \end{figure*}

\begin{figure*}[]
        \centering      
        \includegraphics[width=18cm]{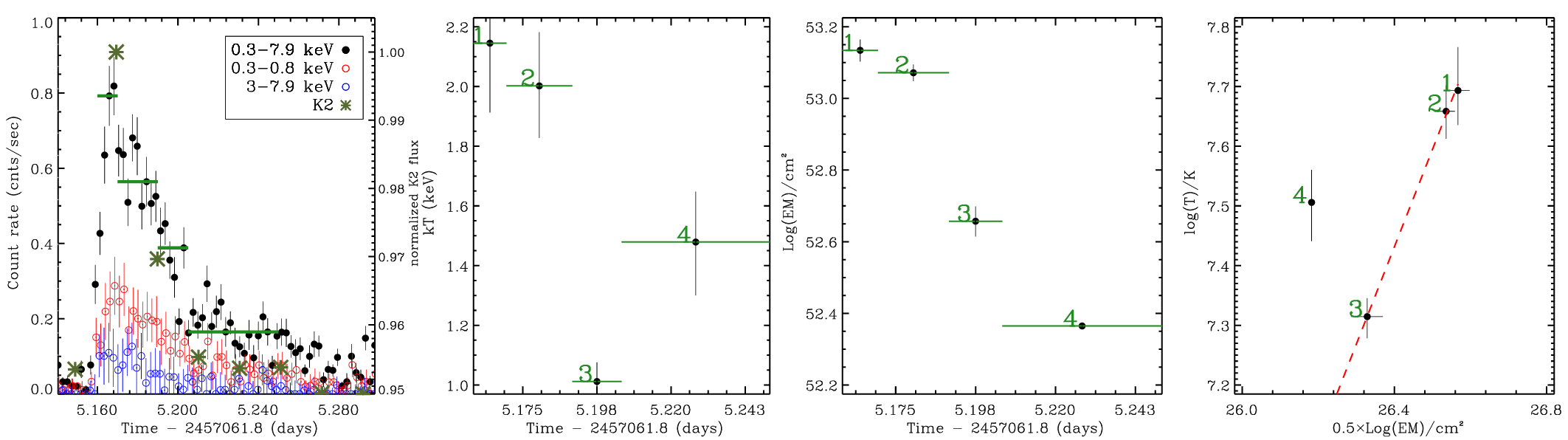}
        \caption{Evolution of the flare observed in HCG~273. Panel layout and content as in Fig.~\ref{HII405_flare}.}
        \label{HCG273_flare}
        \end{figure*}

\begin{figure*}[]
        \centering      
        \includegraphics[width=18cm]{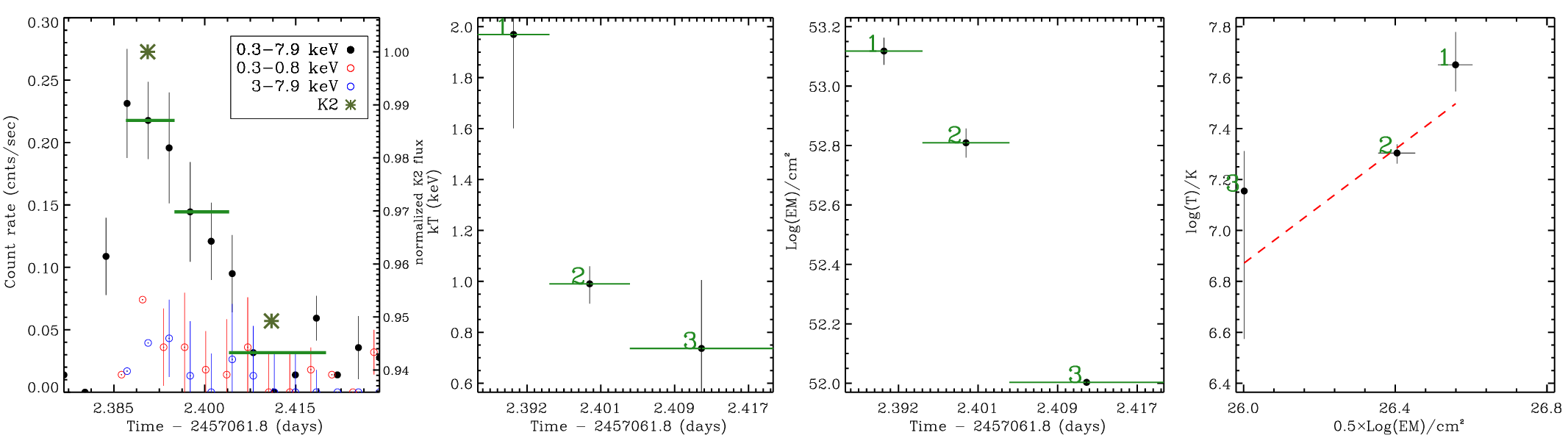}
        \caption{Evolution of the flare observed in HHJ~336. Panel layout and content as in Fig.~\ref{HII405_flare}.}
        \label{HHJ336_flare}
        \end{figure*}

\begin{figure*}[]
        \centering      
        \includegraphics[width=18cm]{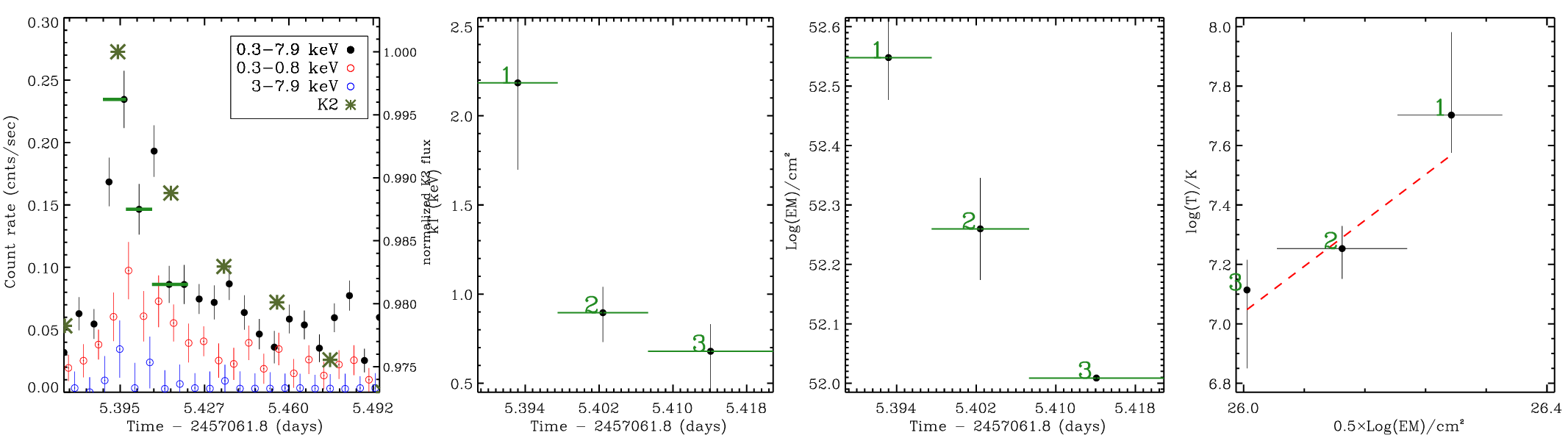}
        \caption{Evolution of the flare observed in HCG~244. Panel layout and content as in Fig.~\ref{HII405_flare}.}
        \label{HCG244_flare}
        \end{figure*}

\begin{figure*}[]
        \centering      
        \includegraphics[width=18cm]{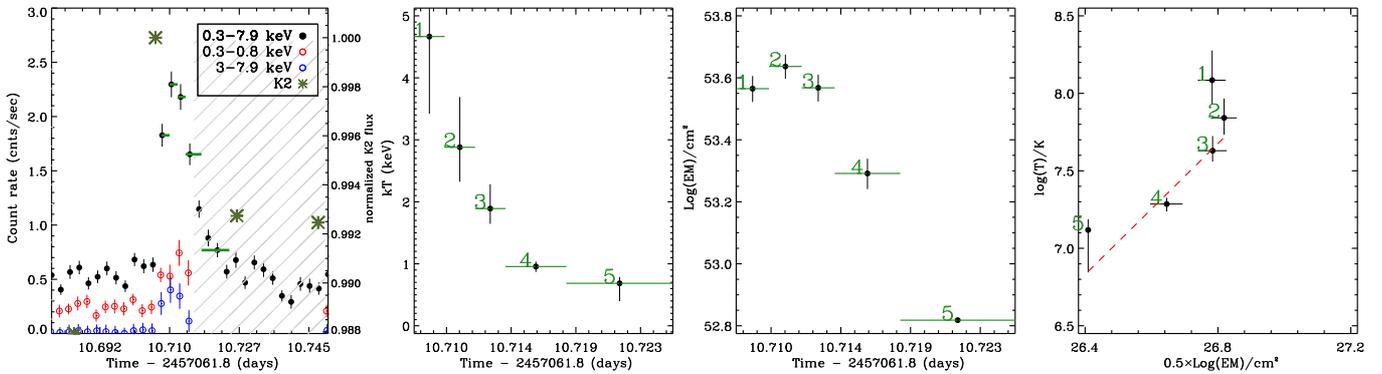}
        \caption{Evolution of the flare observed in HII~345. Panel layout and content as in Fig.~\ref{HII405_flare}.}
        \label{HII345_flare}
        \end{figure*}

In six flares, namely those in HHJ~273 (Fig. \ref{HHJ273_flare}), HII~212 (Fig. \ref{HII212_flare}), HCG~273 (Fig. \ref{HCG273_flare}), HHJ~336 (Fig. \ref{HHJ336_flare}), HCG~244 (Fig. \ref{HCG244_flare}) and the G8 star HII~345 (Fig. \ref{HII345_flare}), the time blocks do not resolve the heating phase, while the cooling phase is sampled with at least three blocks. In all these flares, temperature and density peak during the same interval, with the one exception being HII~345, where the peak plasma temperature is observed during interval \#1 while that of plasma density is during \#2. The largest flare amplitude is observed in HHJ~273, with the count rate increasing by a factor of 115, followed by HCG~273 (32.0), HII~212 (7.4), HHJ~336 (6.9), HCG~244 (6.3) and HII~345 (6.2). The flare occurring in HHJ~273 is also the brightest in X-rays among the whole sample (log(E$_{\rm xray,flare}$[erg])=34.04$^{34.06}_{34.02}$, see Table \ref{flareprop_table}). In these flares, we measured similar values of peak plasma temperature, ranging from 42$^{48}_{37}\,$MK in HHJ~273 to 50$^{96}_{38}\,$MK in HCG~244 (see Table \ref{flareprop_table}). The one exception to this is the flare in the G8 star HII~345, which reached a peak temperature of 122$^{189}_{85}\,$MK (against a quiescent plasma temperature of 6.96$^{7.04}_{6.88}\,$MK). 

It is interesting to compare the variability observed in the broad and hard X-ray bands in HHJ~273 (Fig. \ref{HHJ273_flare}) and HGC~273 (Fig. \ref{HCG273_flare}). In HHJ~273, we observed an evident and narrow peak of hard X-ray emission which precedes the one in the broad and soft band light curve by $\sim$800$\,$s. In HCG~273, the hard X-ray light curve rises for $\sim$600$\,$s before the observed peak in broad and soft bands. In this case however, instead of a single peak, we observe two peaks separated by about 800$\,$s, which could be due to the loop oscillations that we observed in this flare (see Sect. \ref{oscill_sec}). The delayed onset of the soft X-ray flare with respect to the hard band could naturally be explained by the time evolution of flares \citep{Reale2007AA}, with the peak of plasma temperature preceding the density peak. A similar variability in soft and hard X-ray bands was observed in solar flares \citep{Sato2001ApJ.558L.137S,Reale2001ApJ.557.906R} and in a few other young stars \citep{Preibisch2002AJ.123.1613P,Preibisch2003AA.401.543P}. \par

\begin{figure*}[]
        \centering      
        \includegraphics[width=18cm]{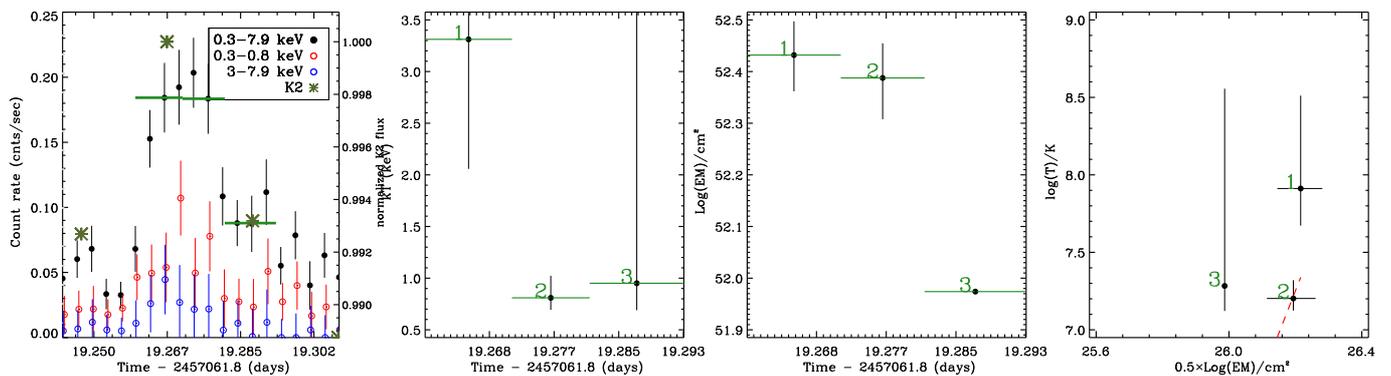}
        \caption{Evolution of the flare observed in HCG~150. Panel layout and content as in Fig.~\ref{HII405_flare}.}
        \label{HCG150_flare}
        \end{figure*}

\begin{figure*}[]
        \centering      
        \includegraphics[width=18cm]{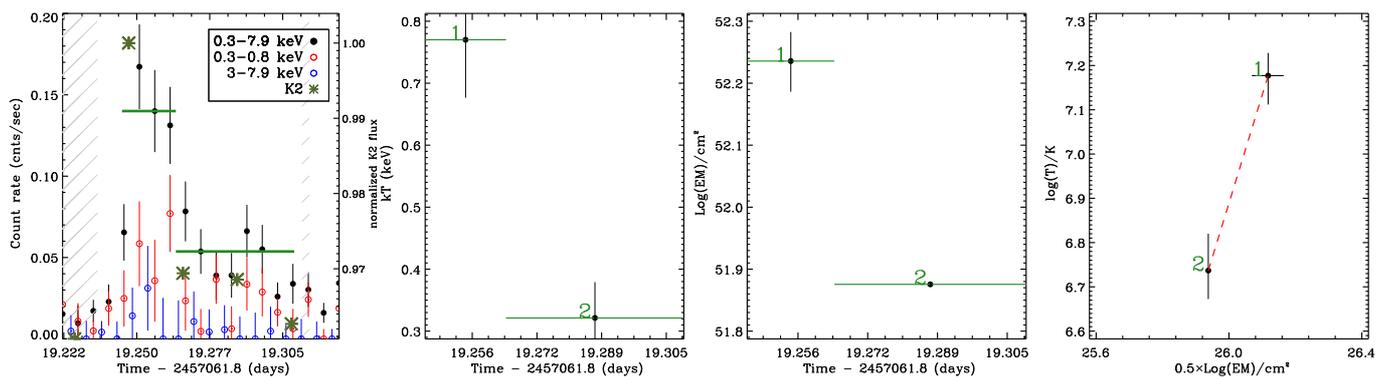}
        \caption{Evolution of the flare observed in HCG~146. Panel layout and content as in Fig.~\ref{HII405_flare}}
        \label{HCG146_flare}
        \end{figure*}

In the remaining flares, we were not able to isolate the heating, while the cooling phase is split into two intervals. During these flares, the count rate increased by a factor 9.7 in HCG~150 (Fig. \ref{HCG150_flare}) and 14 in HCG~146 (Fig. \ref{HCG146_flare}). The flare in HCG~150 also shows a prolonged rise phase lasting about 25 minutes. \par

        We calculated the slope of the cooling pattern ($\rm \zeta$) followed by the flares in the log(kT)-versus-0.5$\times$log(EM) plane by interpolating the points of the cooling phase. As shown by \citet{Jakimiec1992AA.253.269J}, \citet{Sylwester1993AA.267.586S}, and \citet{RealeBPS1997AA}, this value provides hints on the presence of heating during the cooling phase. In fact, $\zeta$ is expected to be $\sim$2 when negligible heating occurs during the cooling phase, otherwise it is $<$2.  As shown in Table \ref{slopecool_table}, slope values suffer large uncertainties. From the analysis of the cooling path of these flares in the log(T)-versus-log(EM) plane, we thus obtained a marginal indication that these flares are associated with single loops without substantial heating during the cooling phase, with the exception of HHJ~273 ($\rm \zeta$=0.4$\pm$1.1, see Fig. \ref{HHJ273_flare}) and HII~405 ($\zeta$=0.7$\pm$0.8, see Fig. \ref{HII405_flare}). However, given the large uncertainties on $\rm \zeta$, we cannot exclude that some of these flares may be associated with more complicated structures, such as flare-loop arcades overlying the active regions, as discussed  by \citet{HeinzelShibata2018ApJ.859.143H}. \par

   \begin{table}
    \caption{Slope of the cooling pattern}
    \label{slopecool_table}
    \centering                       
    \begin{tabular}{|c|c|}
    \hline
    \multicolumn{1}{|c|}{Name} &
    \multicolumn{1}{|c|}{$\zeta$} \\    
    \hline
    \multicolumn{1}{|c|}{} &
    \multicolumn{1}{|c|}{} \\
    \hline
HII 212 &  $3.2\pm2.8$\\
HHJ 273 &  $0.4\pm1.1$\\
HHJ 336 &  $1.1\pm1.2$\\
HCG 244 &  $1.9\pm2.4$\\
HII 345 &  $2.1\pm0.9$\\
HII 1516&  $1.7\pm1.4$\\
HCG 146 &  $2.4\pm1.2$\\
HCG 273 &  $1.6\pm0.5$\\
HCG 150 &  $5.5\pm30$\\
HII 405 &  $0.7\pm0.8$\\
    \hline                      
    \multicolumn{2}{l}{} \\
    \end{tabular}
    \end{table}

\subsection{Loop lengths}
\label{loops_sec}

Most of the slopes listed in Table \ref{slopecool_table} are consistent with pure cooling in a single flaring loop. We can estimate the size of the loops by applying relations derived by \citet{RealeBPS1997AA} and \citet{Reale2007AA} from hydrodynamic loop models. In these models, the plasma is described as a compressible fluid confined in the coronal loops and transporting energy along the magnetic field lines. The only role of the magnetic field is therefore to confine the plasma \citep{Rosner1978ApJ.220.643R}. The flare is triggered by a heat pulse injected inside the loop, and the plasma cools down by radiation and thermal conduction along the field lines.  Compared with other existing models, they have the advantage of being easily connected with observable quantities such as the plasma temperature and density in the loop. Other models, such as that developed by \citet{ShibataYokoyama2002ApJ.577.422S}, present a relation between the loop length and physical quantities that are difficult to estimate from optical and X-ray light curves, such as the coronal pre-flare electron density. \par 

        We estimated the loop lengths using two equations. The former is based on the time of the density peak (t$_{\rm maxden}$), calculated in this paper as the difference between the half time of the block with the largest emission measure and the flare start:

    \begin{equation}
  \rm  L_{\rm loop}=6\times 10^{2.5}\times t_{\rm maxden}\times \Psi^2 \times T^{0.5}_{\rm max}
    \label{loop_eq_2}
    ,\end{equation}
    
    while the latter is based on the duration of the rise of the light curve (i.e., from the flare start to the peak of emission) $\rm t_{rise}$, measured in kiloseconds:
    
    \begin{equation}
   \rm  L_{\rm loop}=0.6 \times \Psi^2 \times T^{0.5}_{\rm max} \times t_{\rm rise}
    \label{loop_eq_3}
    .\end{equation}

  In these equations, T$_{\rm max}$ is the maximum temperature in Kelvin, calculated from the maximum temperature obtained from the time-resolved spectral analysis using T$_{\rm max}$=0.13$\times$T$_{\rm max,fit}^{1.16}$ \citep[specific for EPIC instruments,][]{Reale2007AA}, and {$\rm \psi$ is the ratio between the peak temperature and the temperature at the density peak, which ranges between 1.2 and 2 \citep{Reale2007AA}}. The resulting loop lengths are shown in Table \ref{length_table}. There is a general good agreement between the length estimated with the two equations.    

    \begin{table}
    \caption{Comparison of the loop lengths obtained with Eqs. (\ref{loop_eq_2}) and (\ref{loop_eq_3})}
    \label{length_table}
    \centering                       
    \begin{tabular}{|c|c|c|}
    \hline
    \multicolumn{1}{|c|}{Name} &
    \multicolumn{1}{|c|}{$\rm L_{\rm eq1}$} &
    \multicolumn{1}{|c|}{$\rm L_{\rm eq2}$} \\    
    \hline
    \multicolumn{1}{|c|}{} &
    \multicolumn{1}{|c|}{$\times10^{10}\,$cm} &
    \multicolumn{1}{|c|}{$\times10^{10}\,$cm} \\
    \hline
     HII~212   &    2.1$^{4.4}_{1.0}$  &   1.3$^{4.1}_{0.3}$ \\
     HHJ~273   &    2.5$^{4.2}_{1.3}$  &   3.7$^{7.3}_{1.6}$ \\
     HHJ~336   &    1.1$^{2.0}_{0.5}$  &   0.9$^{2.6}_{0.2}$ \\
     HCG~244   &    1.3$^{2.8}_{0.6}$  &   1.7$^{5.5}_{0.4}$ \\
     HII~345   &    1.3$^{2.7}_{0.6}$  &   1.0$^{3.1}_{0.2}$ \\
     HII~1516  &    2.7$^{5.8}_{1.1}$  &   1.4$^{4.5}_{0.3}$ \\
     HCG~146   &    1.6$^{2.7}_{0.8}$  &   0.9$^{2.3}_{0.2}$ \\
     HCG~273   &    1.4$^{2.5}_{0.7}$  &   2.7$^{5.2}_{1.2}$ \\
     HCG~150   &    2.1$^{6.6}_{0.9}$  &   6.5$^{22.6}_{2.5}$\\
     HII~405   &    5.5$^{9.0}_{2.9}$  &   2.8$^{5.8}_{1.2}$ \\    
    \hline
    \multicolumn{3}{l}{} \\
    \end{tabular}
    \end{table}

\subsection{Oscillations}
\label{oscill_sec}

\begin{figure}[]
        \centering      
        \includegraphics[width=9cm]{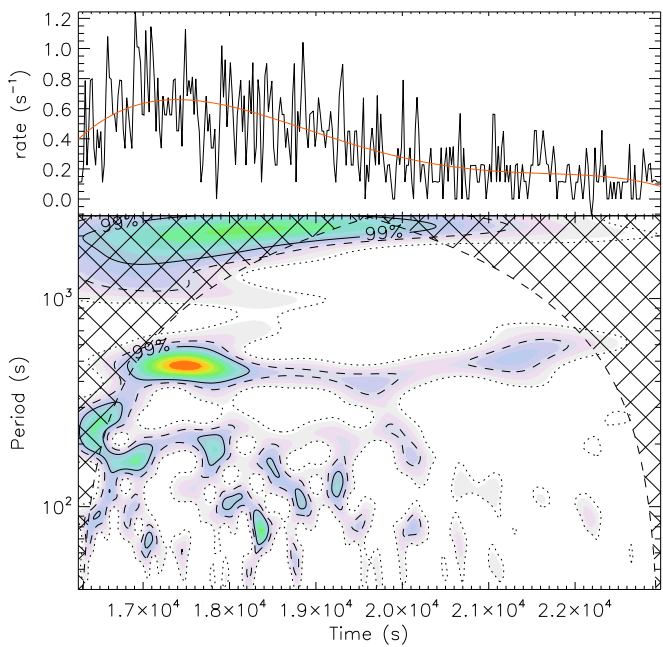}
        \caption{Top panel: Observed light curve during the flare in HCG~273 overplotted with the flare global shape obtained by fitting a third degree polynomial to the binned (10 s) light curve. Bottom panel: Wavelet power spectrum obtained during the flare, with dotted, dashed, and solid contours marking peaks with 68\%, 95\%, and 99\% confidence level, respectively. The hatched area is the ``cone of influence'' (see Sect. \ref{oscill_sec}).}
        \label{wav69_figure}
        \end{figure}

In solar flares, oscillatory patterns in soft X-ray light curves have been observed several times \citep{McLaughlin2018SSRv.214.45M}. They are typically interpreted as MHD oscillations in the magnetic loops \citep[][and references therein]{StepanovZN2012book,VanDoorsselaere2016SoPh.291.3143V}, or global waves traveling across the corona \citep{LiuOfman2014SoPh}. Oscillatory patterns have also been observed in stellar flares, and interpreted as magneto-acoustic waves produced by oscillations of the loop during the evaporation of the chromospheric plasma \citep{ZaitsevStepanov1989SvAL}; fast MHD waves \citep{MathioudakisBJD2006AA}; and fast kink modes in stellar loops \citep{PandeySrivastava2009ApJ}, which however are not expected to produce density oscillations \citep{StepanovZN2012book}. In the alternative model developed by \citet{Reale2016ApJ}, and already applied to flares in young stars \citep{Reale2018ApJ.856.51R}, the oscillations are due to density waves, which are triggered by a heat pulse shorter than the sound crossing time inside the loop at the temperature peak. We followed the method developed by \citet{LopezSantiago2018RSPTA.37670253L} to reveal oscillations in the X-ray flares in our sample. The adopted method is a modified version of the Fourier analysis designed to detect quasi-periodic signals, in which the light curve is transformed from time to frequency domain adopting a Morlet function as a mother function \citep{Torrence98apractical}. Figure \ref{wav69_figure} shows the results of this calculation for the flare in HCG~273, the only one in our sample where significant oscillations were detected\footnote{We are not able to state whether oscillations in the other flares are physically missing or just not detectable.}.

The power spectrum (i.e., the square of the wavelet transform) is two-dimensional (one dimension for time and one for frequency), and oscillation patterns (together with other disturbances in the time series) result in extended peaks with high confidence level. Figure \ref{wav69_figure} shows the wavelet power spectrum obtained from the flare in HCG~273. The cone of influence marked in the diagram is the region of the power spectrum where the edge effects are important \citep{Torrence98apractical}. A significant oscillation with a period of 500$\pm$100$\,$s is detected. The detection of oscillations in the flare observed in HCG~273 supports the hypothesis of a single loop dominating this flare, ignited by a single rapid heat pulse, which is shorter than the sound crossing time across the loop at the maximum plasma temperature. \par

    This result also allows us to obtain an independent estimate of the length of this loop. The observed peak temperature of $49.3^{58.3}_{43.2}\,$MK corresponds to a sound speed v$_{\rm s}$=8.1$^{8.9}_{7.6}\times$10$^7\,$cm/s. As found by \citet{Reale2016ApJ}, during an oscillation period the plasma has traveled twice in the loop. The resulting loop length can be estimated as L$_{\rm loop}$=v$_{\rm s}\times$P/2=2.0$^{2.2}_{1.9}\times10^{10}\,$cm (where P is the oscillation period). This estimate is compatible within errors to those in Table \ref{length_table}. \par

\section{Discussion and conclusions}
\label{thats_all_folks}

We have analyzed 12 bright flares that occurred in Pleiades stars observed simultaneously with XMM-Newton and \textit{Kepler}/K2, with the aim of calculating and comparing the energy released by the flares in the two bands and characterizing the flare evolution and geometry. With a total energy released in the optical band in the 32.9<log(E$_{\rm kep,flare}\,$[erg])<34.7 range (median value 34.0$\,$erg), all but one of the flares in our sample (the one in HCG~150, Fig. \ref{HCG150_flare}) meet the criteria for ``superflares'' defined by \citet{ShibayamaMNN2013ApJS} and \citet{NotsuSMN2013ApJ} (e.g., $10^{33}\,$erg$\leq$E$_{\rm optical}\leq 10^{38}\,$erg). This is not surprising given that our sample is limited to bright flares occurring in young and rapidly rotating stars. In fact, \citet{Maehara2012Natur.485.478M}, in their monitoring of a sample of 83000 stars for 120 days, observed that the frequency of superflares is about 20 times greater in stars with rotation periods shorter than 10 days compared to slower rotators, and \citet{WichmannFWN2014AA} found a larger occurrence of superflares in stars showing the Li~$\rm \lambda6708\,$\AA$\,$absorption line  in their spectra, which is typical of young stars. Besides, \citet{Honda2015PASJ.67.85H} measured Li abundances in 34 stars hosting superflares detected with \textit{Kepler}, showing that half of them are younger than the age of the Hyades cluster (6.25$\times$10$^8\,$yr, \citealp{Perryman1998AA.331.81P}) and most of them are likely younger than the Sun based on their projected rotational velocity. Extending the comparison to a wider range of stellar properties, in their study of the optical flares detected in all available \textit{Kepler} light curves, \citet{Davenport2016ApJ} reported 851168 flares occurring in 4041 stars, with a median flare energy of log(E$_{\rm kep,flare}$[erg])=34.6, which is slightly larger than the mean energy released in optical by the flares in our sample. This difference is likely due to the different distribution of spectral types of the stars included in \citet{Davenport2016ApJ}, that is, mainly G and K dwarfs, and in our paper, where nine in every twelve stars are M dwarfs. \par

        In X-rays, the flares in our sample can be compared with the 130 flares analyzed by \citet{Pye2015AA.581A.28P} occurring in 70 stars included in the XMM-Newton serendipitous source catalog. These stars are mainly within 1$\,$kpc of the Sun, with a few being well-known variable stars and binary systems. In these flares, the peak luminosity in the broad band ranges from $\sim$10$^{29}\,$erg/sec to $\sim$10$^{32}\,$erg/sec, and X-ray energy output ranges from $\sim$10$^{32}\,$erg to $\sim$10$^{35}\,$erg. This catalog thus includes X-ray flares that are brighter than those in our sample, in which the peak luminosity in the broad band is between 3.2$\times$10$^{29}\,$erg/sec and 7.9$\times$10$^{30}\,$erg/sec and the total energy released in the X-ray broad band is between 7.8$\times$10$^{32}\,$erg and 9.8$\times$10$^{33}\,$erg.\par
        
Most (10 over 12) of the flares analyzed in this work released more energy in optical than in X-rays. This is typically observed in solar flares, and naturally explained with the fact that the optical flare traces the heating of plasma in the chromosphere/photosphere, with part of the released energy irradiated by the evaporating plasma confined in the magnetic loop. The two flares releasing more energy in X-rays than in optical (HHJ~336, where E$_{\rm kep,flare}$/E$_{\rm xray,flare} \sim$0.9, and HCG~150, where E$_{\rm kep,flare}$/E$_{\rm xray,flare} \sim$0.7) could be due to more complex events than a single loop or to the uncertainties related to the estimate of the energy of a flare. In the remaining stars, this ratio ranges between 1.2 and 13.8. This is smaller, for instance, than the ratio observed in the bright solar flares by \citet{Woods2006JGRA.11110S14W}, which, converting the GOES band (0.1-0.8$\, \mu$m) into the Chandra/ACIS-I broad band, ranges between 25 and 40. The possibility that a larger fraction of energy is converted into X-ray emission as flares become more energetic is compatible with the relation between E$_{\rm bol,flare}$ and E$_{\rm xray,flare}$ found by \citet{Flaccomio2018AA.620A.55F} in the flares observed in NGC~2264 by \textit{CoRoT} and \textit{Chandra}. \par
        
        Several studies have found a correlation between the energy released by flares and their duration. For instance, \citet{Veronig2002AA.382.1070V}  analyzed an extensive set of 49409 soft-X-ray solar flares observed during the period of 1976-2000 by GOES, thus covering about three solar cycles. This sample contains mainly C-type flares (66\%), with a median duration of 12$\pm$0.2 minutes. In this sample, flare duration, rise time, and decay time are well correlated with both the emitted flux integrated over the whole flare duration and the peak luminosity. Similar results were obtained by \citet{Toriumi2017ApJ.834.56T}. \citet{Namekata2017ApJ.851.91N} extended the existing correlation between the energy released by white flares and their duration to the superflares regime. For the solar flares, they in fact found that t$\rm_{WF}$=E$\rm_{WF}^{A}$, with t$\rm_{WF}$ being the white flare duration, E$\rm_{WF}$ the released energy, and A=0.38 for stellar flares and A=0.39 for superflares. The main difference between the two regimes is in the duration of the flares, which is about one order of magnitude shorter for superflares than for solar flares. The relation between flare duration and released energy was also confirmed by \citet{Hawley2014ApJ.797.121H}, who analyzed more than 1000 optical flares occurring in the M4 star GJ~1243 in about 2 months of \textit{Kepler} observations.
        
   \begin{figure}[]
        \centering      
        \includegraphics[width=8cm]{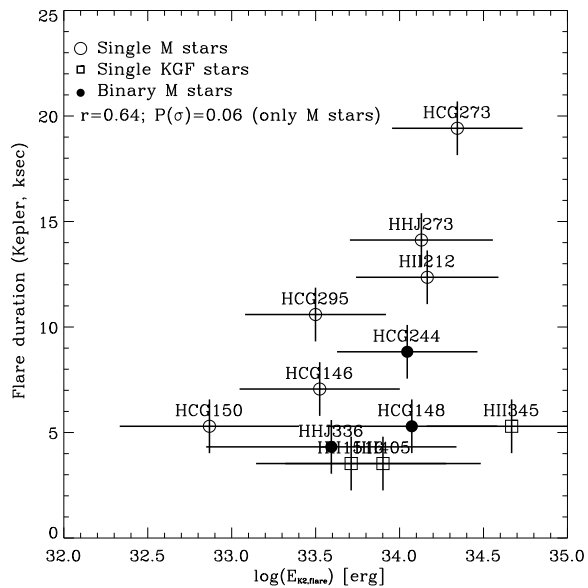}
    \includegraphics[width=8cm]{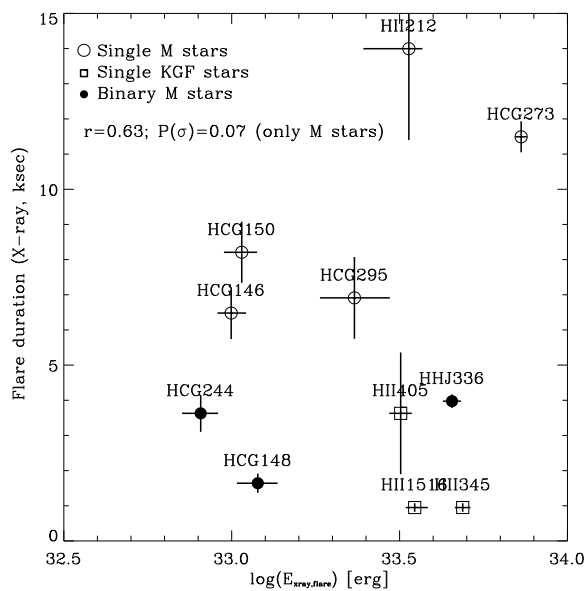}
        
\caption{Flares duration vs. total released energy in optical (upper panel) and X-rays (bottom panel). The different symbols are used to mark M stars, KFG stars, and binary stars. The results of the two-sided Spearman's rank correlation test considering only M stars are shown. In both cases a weak correlation is observed only for M stars.}
        \label{dur_vs_E_fig}
        \end{figure}

        In Fig. \ref{dur_vs_E_fig} we show the observed correlation between duration and energy released in optical and X-rays for the flares in our sample. In both cases, a weak correlation is found considering only the M stars. In the Kepler band we find that t$\rm_{kep}$=E$\rm_{kep,flare}^{0.30}$, which is significantly different from the relation found by \citet{Namekata2017ApJ.851.91N}. In X-rays we find a steeper relation (t$\rm_{xray}$=E$\rm_{xray,flare}^{0.52}$) than the one found for solar flares (t$\rm_{xray}$=E$\rm_{xray,flare}^{0.2-0.33}$; \citealp{Veronig2002AA.382.1070V}). These differences are likely due to the poor statistical sample of superflares analyzed in this paper and the limited range of emitted energy (32.9$\leq$log(E${\rm_{K2,flare}}$)$\leq$34.7 in our sample, 33$\leq$log(E${\rm_{K2,flare}}$)$\leq$36 in \citet{Namekata2017ApJ.851.91N}). We however extend the results obained by \citet{Hawley2014ApJ.797.121H} on the flares occurring in GJ~1243, where the energy of the  observed flares reached 10$^{33}\,$erg in the Kepler band, smaller than most of the flares analyzed in this paper.

        \begin{figure}[]
        \centering      
        \includegraphics[width=8cm]{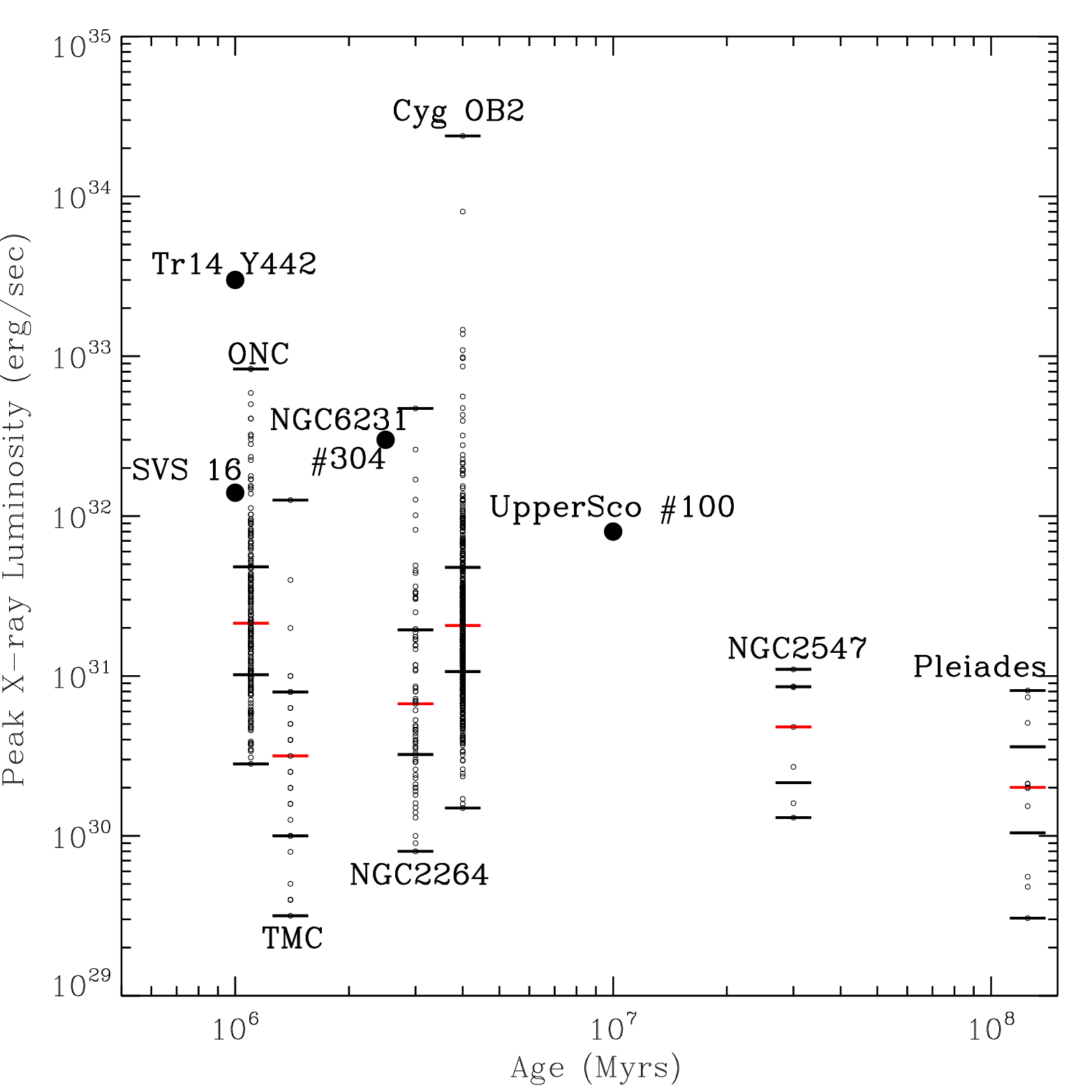}
        \includegraphics[width=8cm]{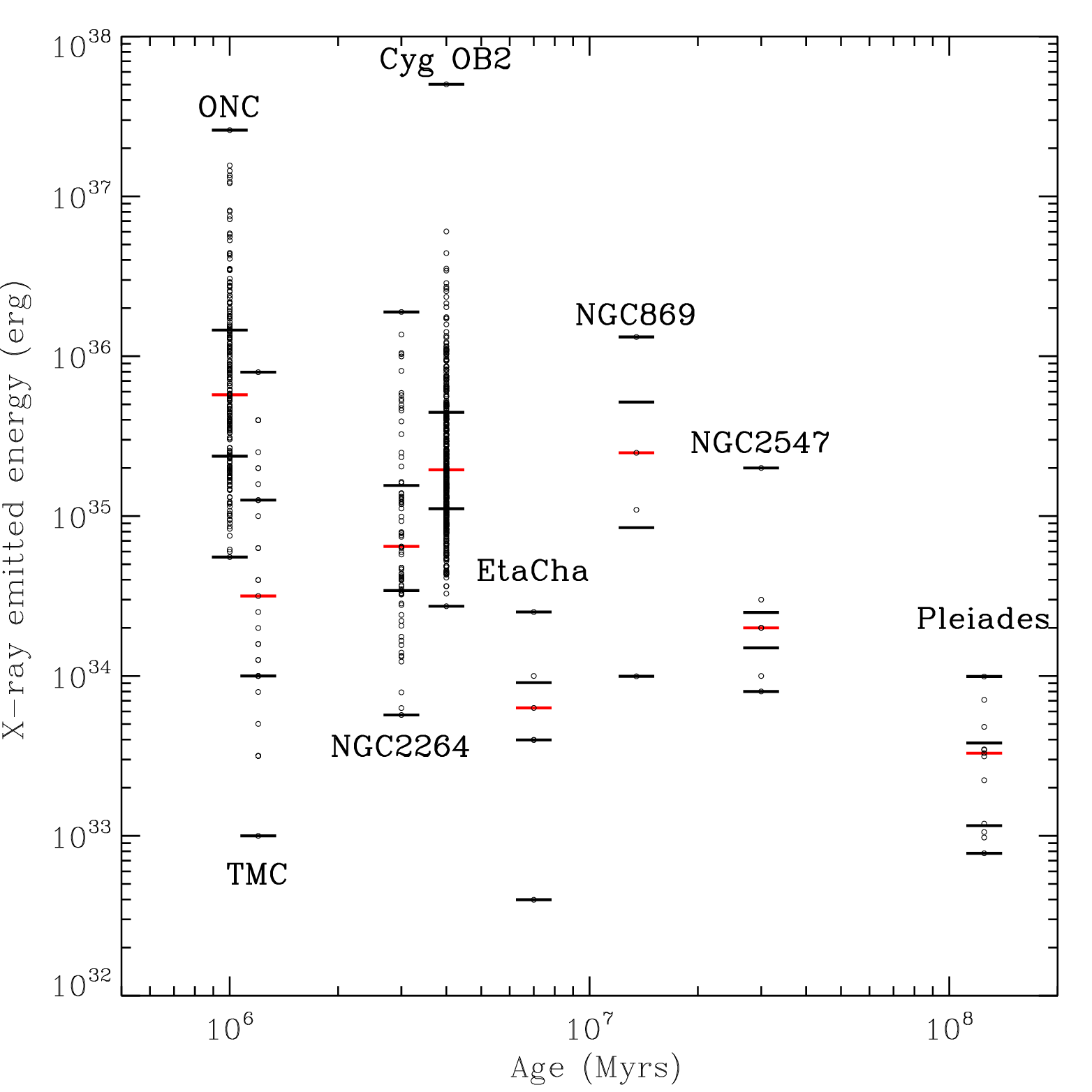}
        \caption{Range of peak luminosity (top panel) and emitted energy (bottom panel) in the X-ray broad band observed in the Pleiades and in samples of stars with different ages. For each data set, the horizontal lines mark the minimum and maximum values, the upper and lower quartiles and, in red, the median value.}
        \label{Lxclusters_figure}
        \end{figure}

        In Fig. \ref{Lxclusters_figure} we attempt a comparison between the energy budget of the X-ray flares observed in stars with different ages. We include data in this figure from the flares detected in NGC~2264 stars as part of the ``Coordinated Synoptic Investigation of NGC~2264'' \citep{CodySBM2014AJ} analyzed by \citet{Flaccomio2018AA.620A.55F}, the Chandra Orion Ultradeep Project (COUP) on the Orion Nebula Cluster \citep[ONC,][]{GetmanFBM2008ApJ}, the Taurus Molecular Cloud \citep[TMC,][]{FranciosiniPSM2007AA,StelzerFBM2007AA}, NGC~2547 \citep{JeffriesEPB2006MNRAS}, Cyg~OB2 \citep{Flaccomio2018arXiv181106769F}, $\eta\,$Chamaleontis \citep[EtaCha,][]{LopezSantiago2010AA.524A.97L}, NGC~869 \citep{Bhatt2014JApA.35.39B}, and the single stars SVS~16 in NGC~1333 \citep{Preibisch2003AA.401.543P}, Trumpler~14~Y442 \citep{Hamaguchi2015ATel.7983.1H}, NGC~6231~\#304 \citep{Sana2007MNRAS.377.945S}, and Upper~Sco~\#100 \citep{Argiroffi2006AA.459.199A}. 
        
        For each data set, we mark each single available measurement with a small circle, the median value with red lines and the quartiles, together with the minimum and maximum values, with black lines. These data sets are not directly comparable with each other without taking into account the different stellar mass spectrum, the different distances of the target stars, the different duration of the X-ray observations and the different number of detected stars, which is beyond the scope of this paper. For instance, the differences between Cygnus~OB2, Orion, and NGC~2264 are likely due to the different size of the stellar samples, and driven mainly by a few stars. In fact, in Cyg~OB2 the number of observed flares is 545 in 501 stars among 7924 detected X-ray sources (\citealp{WrightDGA2014} and \citealp{Flaccomio2018arXiv181106769F}),  compared with $\sim$400 flares occurring in about 1000 X-ray sources detected in NGC~2264 and 216 flares observed in 161 (over a total of 1408) stars in Orion. This results in a larger chance to observe powerful flares in Cygnus~OB2 compared with Orion and NGC~2264. The larger energy released in the flares in the four stars of NGC~869 compared with that observed in stellar samples with similar age can be explained by the fact that this sample is limited to early-type stars (two A and two B-late stars). The stars in Taurus share a similar mass spectrum to the stars studied in this paper. 
        
        \begin{figure}[]
        \centering      
        \includegraphics[width=8cm]{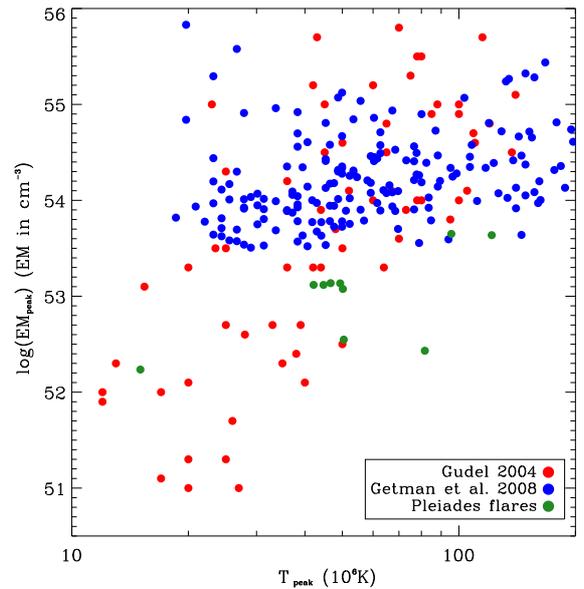}
        \caption{Distribution of the peak EM vs. peak temperature for the flares in our sample compared with those observed in Orion by COUP \citep{GetmanFBM2008ApJ} and those included in the catalog of stellar flares compiled by \citet{Gudel2004AARv}.}
        \label{EMvsT_figure}
        \end{figure}

        Keeping this in mind, Fig. \ref{Lxclusters_figure} is compatible with the scenario in which the energy budget of the most energetic flares observed in clusters declines markedly with stellar age. The values shown in Fig. \ref{Lxclusters_figure} can also be compared with those for solar flares, which typically have L$_{\rm xray,peak}\leq$10$^{28}\,$erg/s \citep{Gudel2004AARv} and total energy released in X-rays ranging from 10$^{29}\,$erg to 10$^{32}\,$erg \citep{ShibataYokoyama2002ApJ.577.422S}, or with the flares observed with XMM in Proxima Centauri (L$_{\rm xray,peak}$=4$\times$10$^{28}\,$erg/s, log(E$_{\rm xray,flare}\,$[ergs])$\sim$32.0--32.5, \citealp{RealeGPA2004AA}). If confirmed, such a decline could also be the consequence of the disappearance of the hottest plasma components in stellar coronae. To test this possibility, in Fig. \ref{EMvsT_figure}, we compared the peak emission measure and plasma temperature observed in the flares of our sample with those observed in Orion stars by COUP \citep{GetmanFBM2008ApJ} and those included in the list of stellar flares compiled by \citet{Gudel2004AARv}. The flares analyzed in this paper, despite being the most powerful flares observed in the Pleiades, are characterized by lower emission measure and plasma temperature than the flares observed in Orion, and than many of the flares analyzed by \citet{Gudel2004AARv}, some of which occurred on main sequence stars. \par

    Finally, we did not observe loops with lengths as large as the loop lengths observed in several COUP stars with disks (e.g., longer than $10^{12}\,$cm). This is compatible with the requirement of a protoplanetary disk to produce and sustain such very long loops \citep{FavataFRM2005ApJs,Reale2018ApJ.856.51R}. \par

\begin{acknowledgements}

We thank the anonymous referee for his/her thoughtful reading and comments, which helped us to improve our paper. For this study we used data from the NASA satellite \textit{Kepler} and the X-ray observatory XMM/Newton, an ESA science mission with instruments and contributions directly funded by ESA Member States and NASA. Funding for the \textit{Kepler} mission is provided by the NASA Science Mission directorate. Funding for the K2 mission is provided by the NASA Science Mission directorate. M.G.G., G.M., S.S., C.A., E.F., F.R., and I.P. acknowledge modest financial support from the agreement ASI-INAF n.2017-14-H.0. J.J.D. was supported by NASA contract NAS8-03060 to the {\it Chandra X-ray Center}. J.D.A.G. was supported by Chandra grants AR4-15000X and GO5-16021X.  

\end{acknowledgements}


\addcontentsline{toc}{section}{\bf Bibliografia}
\bibliographystyle{aa}
\bibliography{biblio}

\begin{onecolumn}
\begin{appendix} 
\section{All K2 targets falling in the XMM fields}
\label{allk2_sec}

    \begin{table}[!h]
    \caption{List of the K2 targets falling in the XMM fields and not included in Table \ref{sample_table}.}
    \label{allk2_table}
    \centering                       
    \begin{tabular}{|c|c|c|c|c|}
    \hline
    \multicolumn{1}{|c|}{Star} &
    \multicolumn{1}{|c|}{Star} &
    \multicolumn{1}{|c|}{Star} &
    \multicolumn{1}{|c|}{Star} &
    \multicolumn{1}{|c|}{Star} \\    
    \hline
    WCZ~185 & HHJ~303   & WCZ~158  & BPL~132  & HHJ~151 \\
    HII~1306& HII~1114  & HII~1100 & BPL~134  & HII~347 \\
    HCG~277 & AKI~2-321 & WCZ~173  & HII~1338 & HII~554 \\
    HII~1640& BPL~121   & BPL~108  & BPL~137  & HII~563 (Taygeta) \\
    WCZ~286 & HII~1266  & HCG~236  & HII~1348 & HCG~194 \\
    WCZ~307 & BPL~136   & BPL~111  & HII~1384 & HII~727 \\
    NPL~24  & WCZ~216   & BPL~118  & HHJ~203  & HII~347 \\
    HII~1969& MH~O13    & HII~1234 & WCZ~259  & HII~554 \\
    WCZ~107 & HII~1485  & JS~9     & HCG~302  & HCG~123 \\
    WCZ~114 & BPL~87    & HCG~253  & HCG~332  &        \\
    BPL~85  & BPL~100   & WCZ~2-3  & WCZ~329  &        \\
    WCZ~137 & HII~1032  & BPL~130  & HCG~343  &        \\
    \hline
    \multicolumn{5}{l}{} \\
    \end{tabular}
    \end{table}

        \begin{figure}[!h]
        \centering      
        \includegraphics[width=8cm]{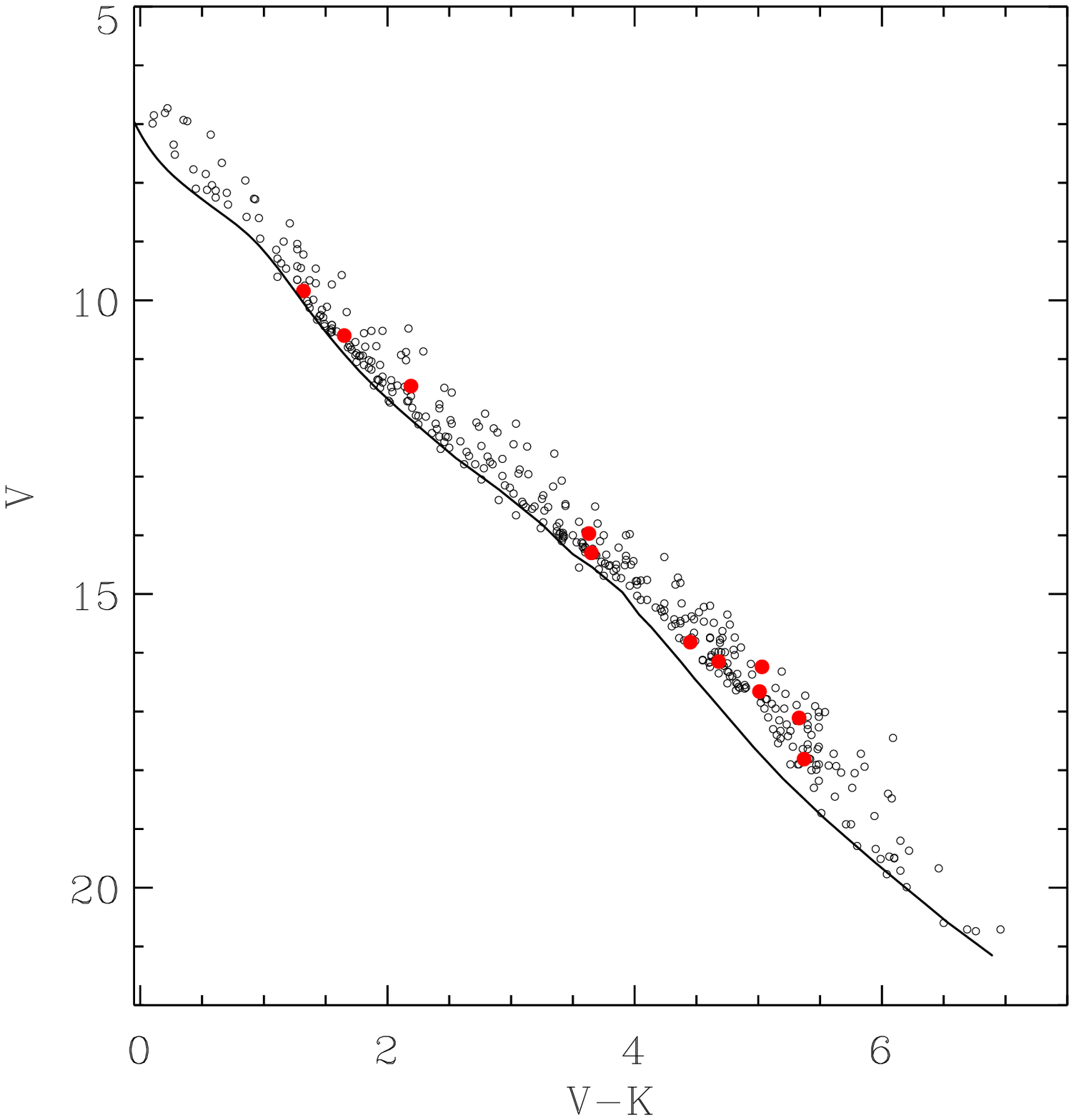}
        \includegraphics[width=8cm]{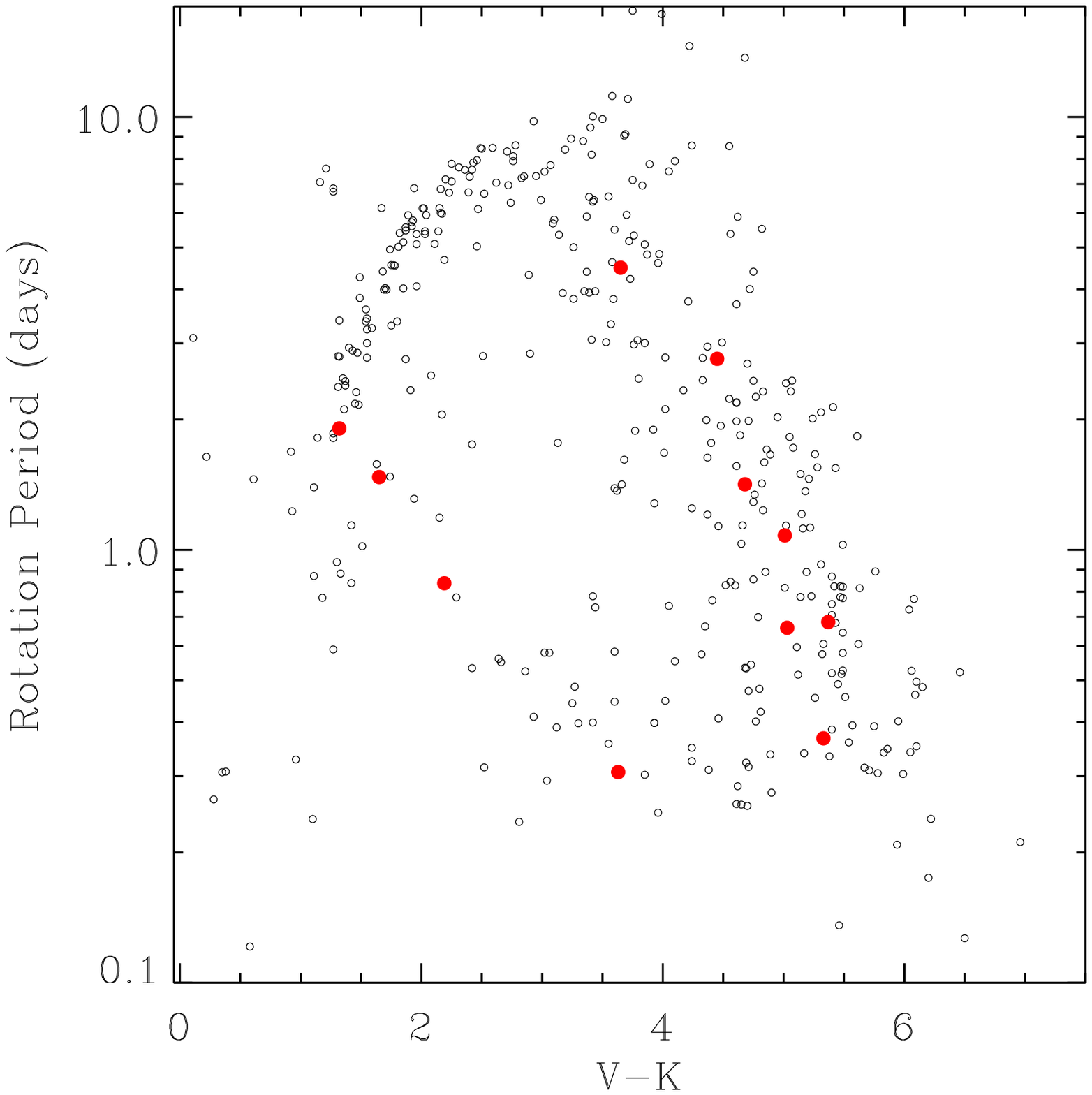}
        \caption{Left panel: $V$ vs. $V-K$ color magnitude diagram of all the Pleiades members observed with K2, with the stars studied in this paper  marked with red circles. The solid line marks the 125$\,$Myr-old isochrone of the cluster. Right panel: Period vs. $V-K$ diagram of all the K2 targets in the Pleiades classified as periodic, with the stars studied in this paper marked with red circles. Only one star lies in the ``slowly rotating sequence'', while five are in the ``rapidly rotating population''}
        \label{appA_fig1}
        \end{figure}
   
Figure \ref{appA_fig1} shows the $V$ versus $V-K$ and the period versus $V-K$ diagrams of all the stars studied in this paper together with all the bona fide members observed with \textit{Kepler} and studied by \citet{RebullSBC2016AJa}. Both periodic and not periodic stars are shown in the left panel, while only periodic stars in the right panel. In the color magnitude diagram the 12 stars hosting the superflares analyzed in this paper do not show peculiar properties compared to the other stars. In the period-versus-$V-K$ plane one star (HII~405) lies in the ``slowly rotating sequence'' (1.1$^m \leq$V-K$\leq$3.7$^m$ and 2$\leq$P[days]$\leq$11), while five (HHJ~336, HCG~244, HCG~146, HCG~150, and HCG~295) lie in the ``rapidly rotating population'' at $V-K\geq$5.0$^m$ and 0.1$\leq$P[days]$\leq$2. Among the remaining stars, HII~1516 and HII~345 rotate faster than most of the stars with similar V-K color (the two points with P<1 day and 2$^m\leq$V-K$\leq$4$^m$). The stars hosting the bright flares analyzed in this paper typically rotate faster than the other members of the Pleiades with similar mass. 
    
\newpage

\section{Entire K2 light curves of the selected stars}
\label{appB_sec}

In Fig. \ref{appB_fig1} we show the entire K2 light curves of the stars selected in this paper. These stars show a significant activity with a few evident and several probable flares occurring during K2 observations. A more detailed analysis of all optical flares occurred in the Pleiades will be the subject of a forthcoming paper, with an additional selection and analysis of faint flares. 

\begin{figure}[!h]
\centering      
\includegraphics[width=9cm]{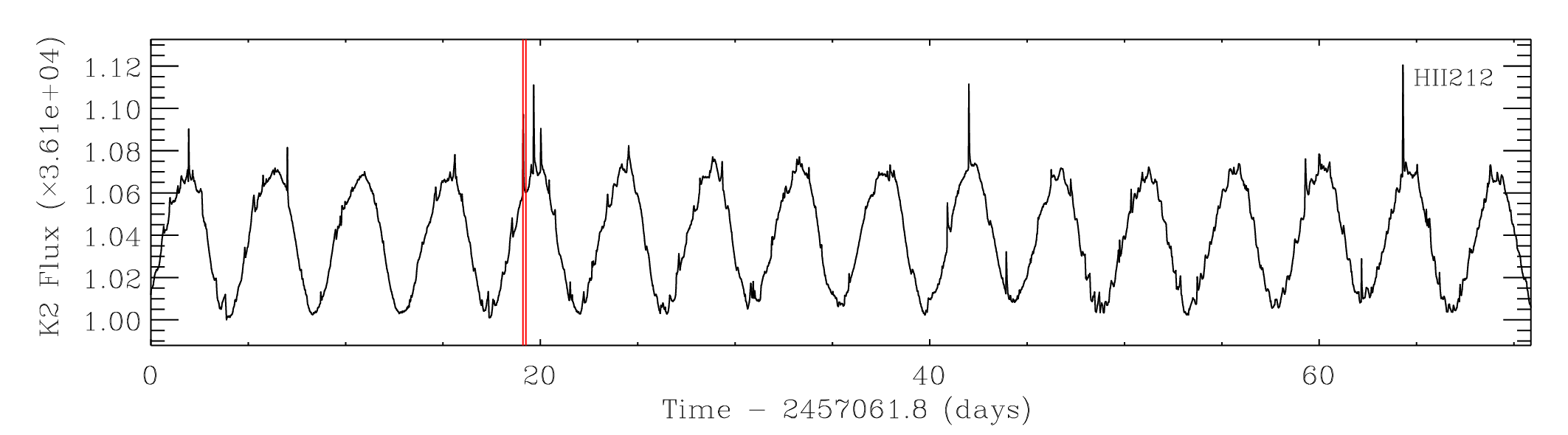}
\includegraphics[width=9cm]{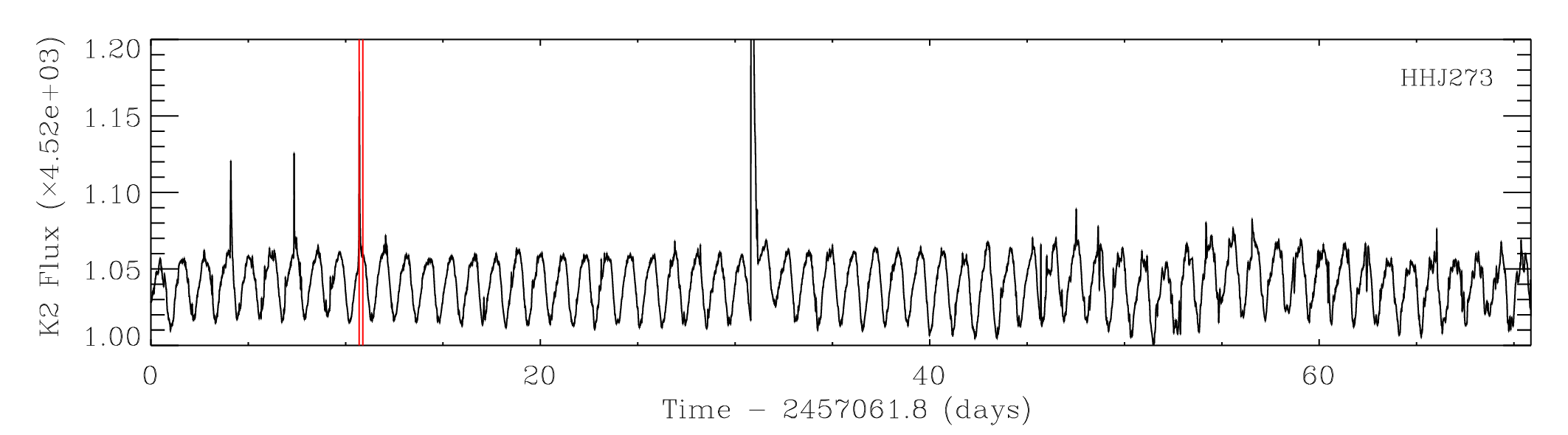}
\includegraphics[width=9cm]{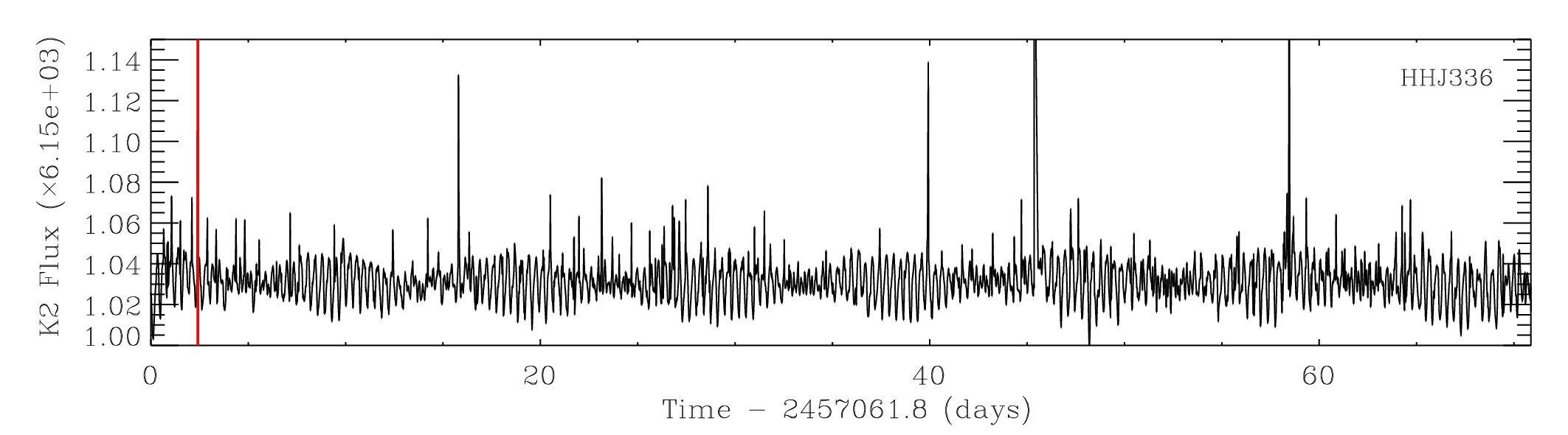}
\includegraphics[width=9cm]{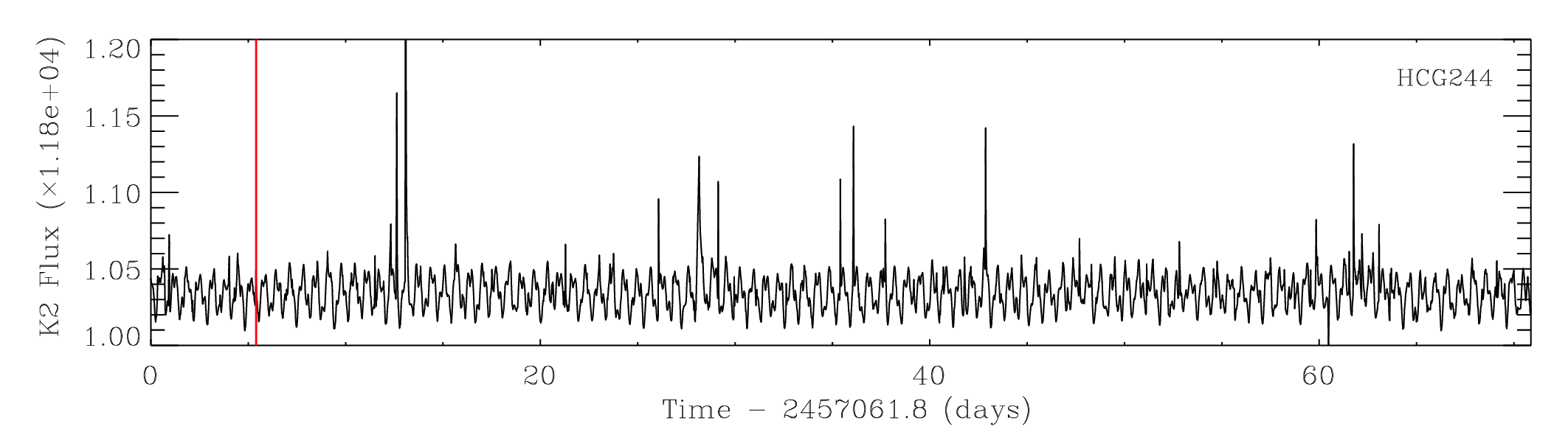}
\includegraphics[width=9cm]{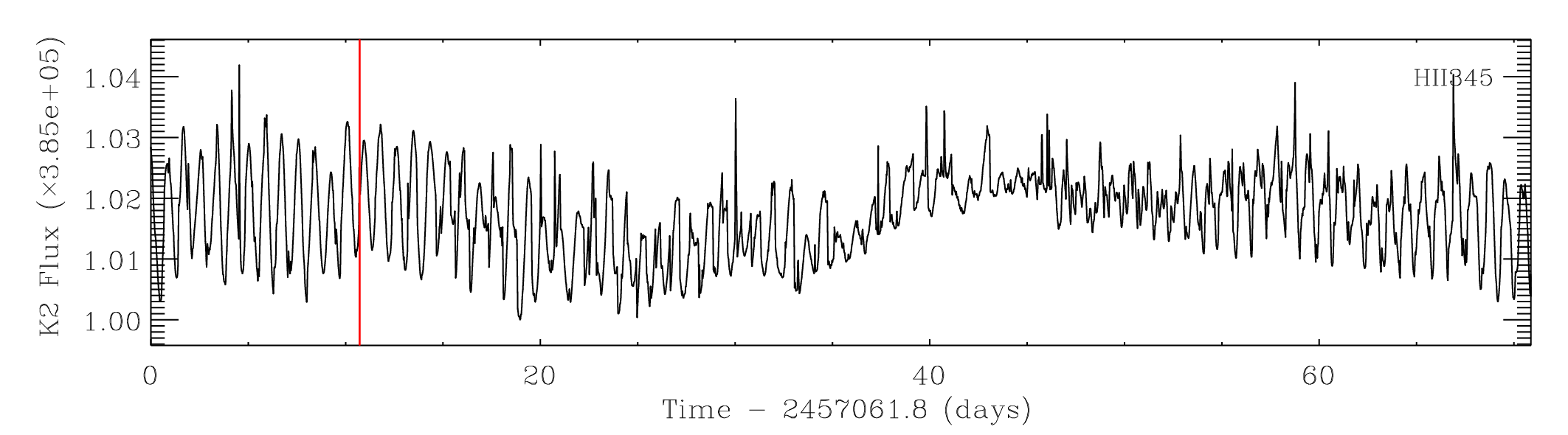}
\includegraphics[width=9cm]{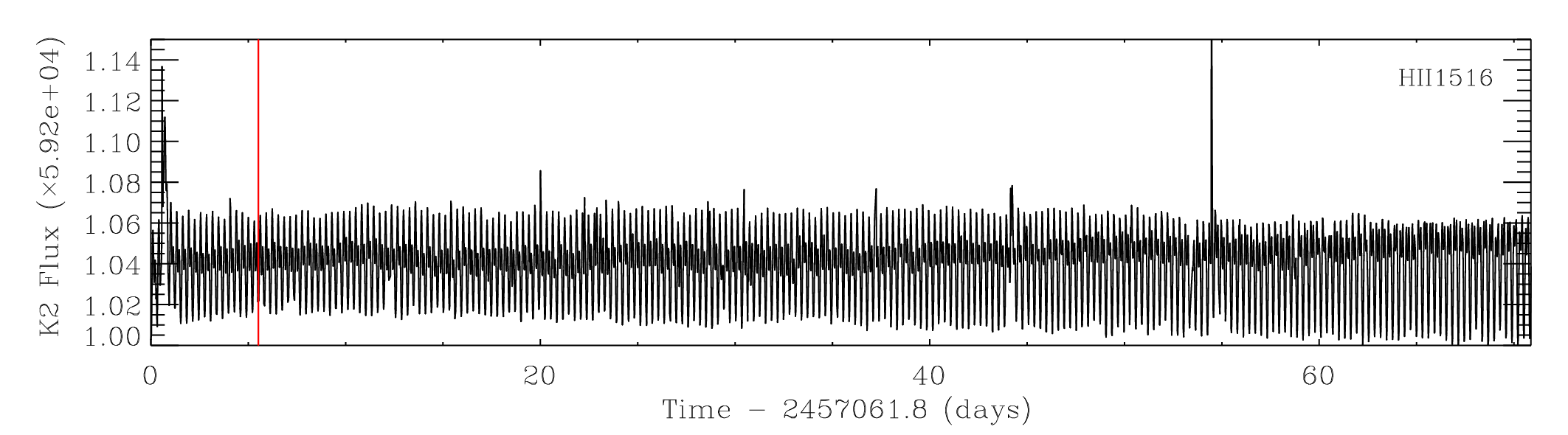}
\includegraphics[width=9cm]{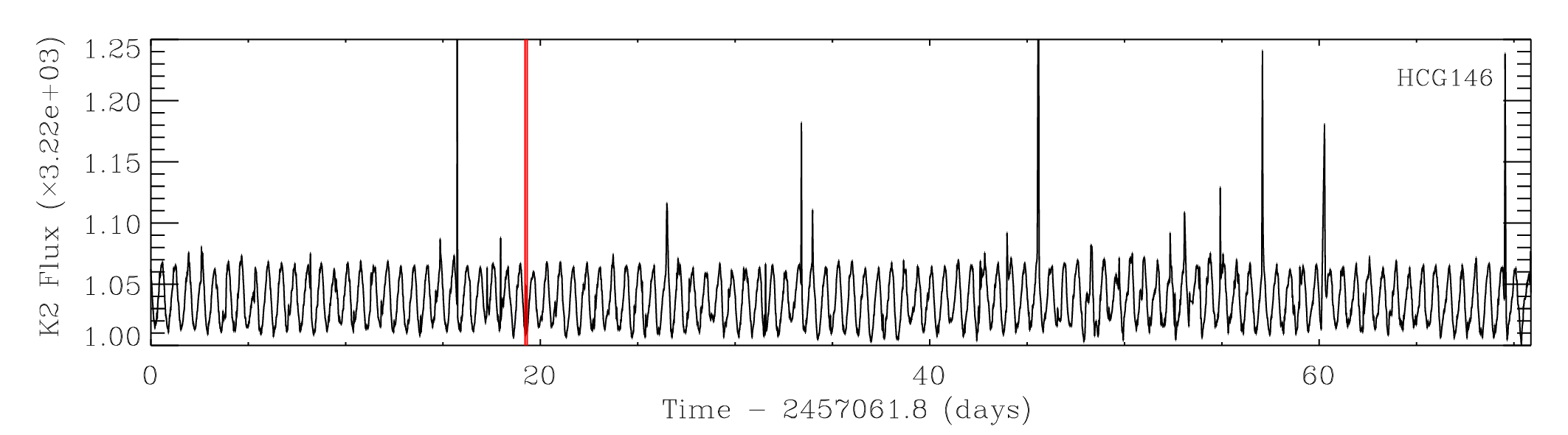}
\includegraphics[width=9cm]{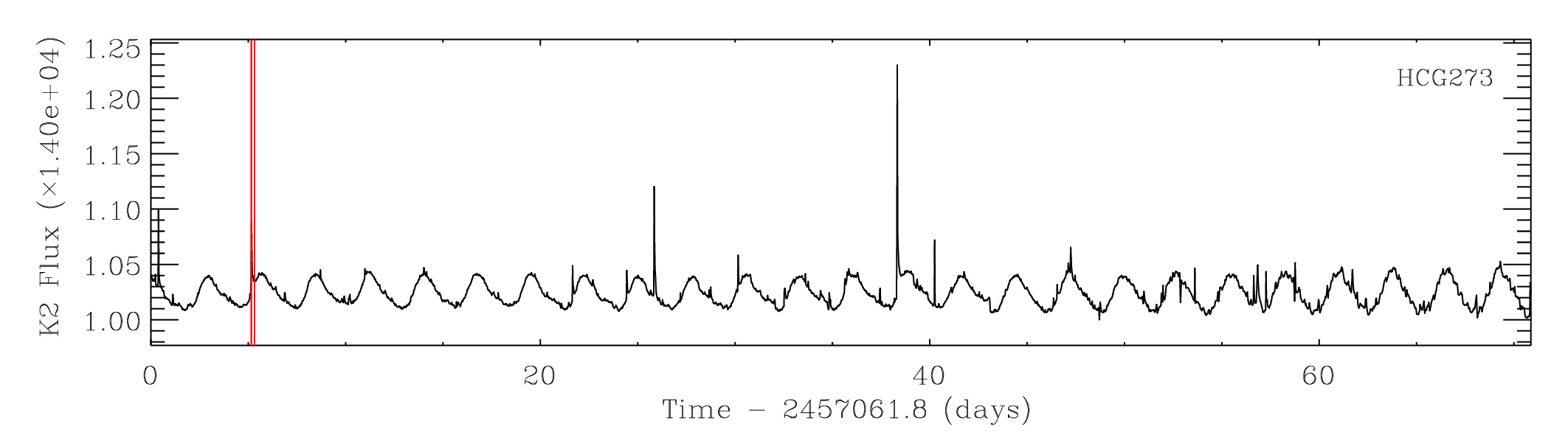}
\includegraphics[width=9cm]{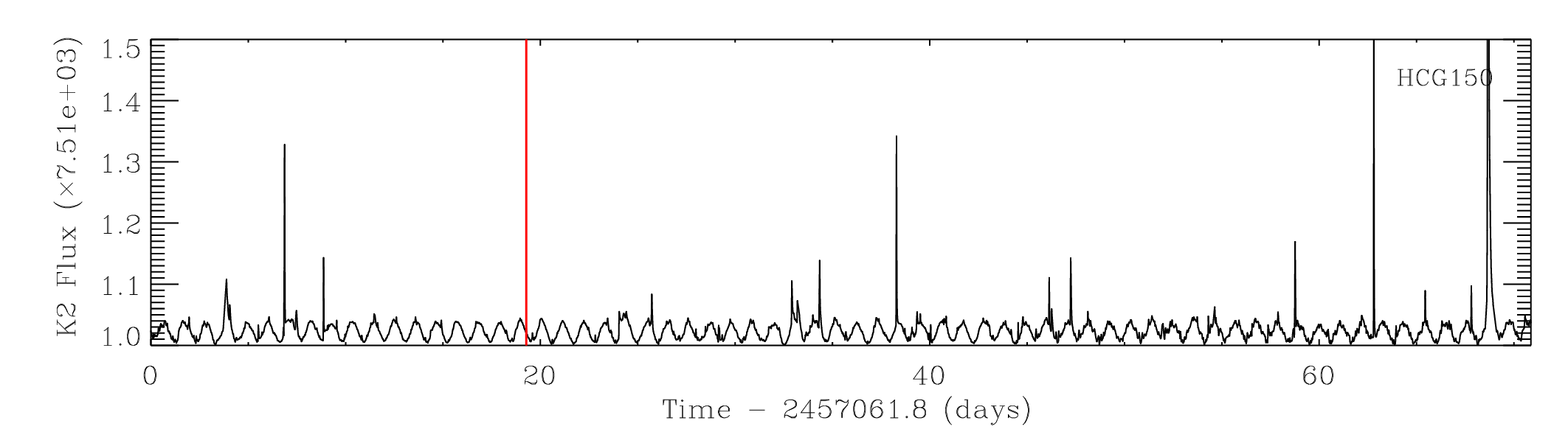}
\includegraphics[width=9cm]{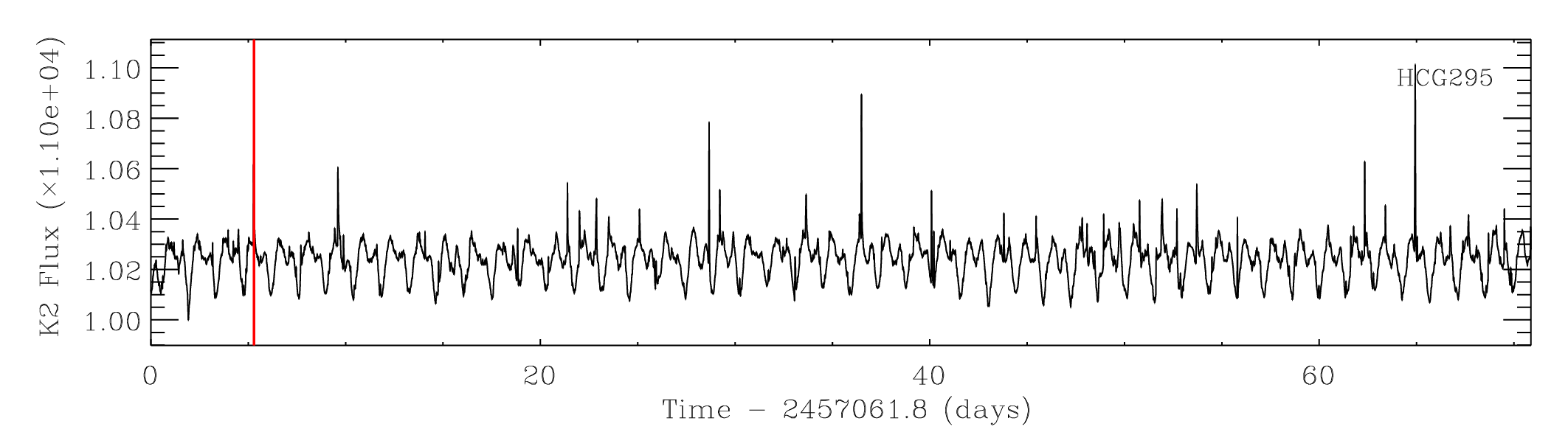}
\includegraphics[width=9cm]{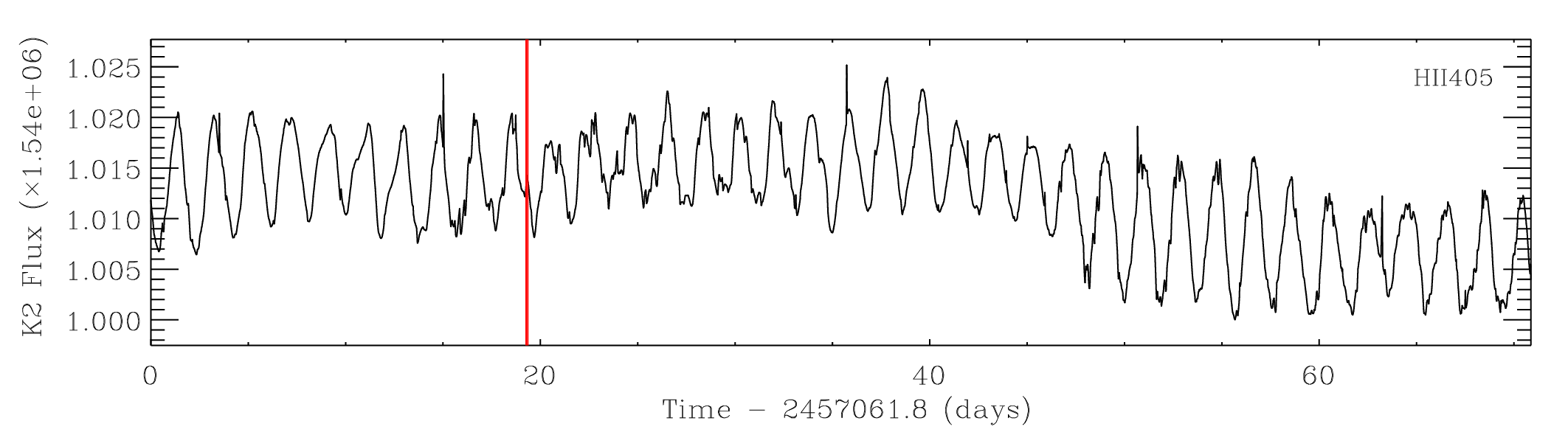}
\includegraphics[width=9cm]{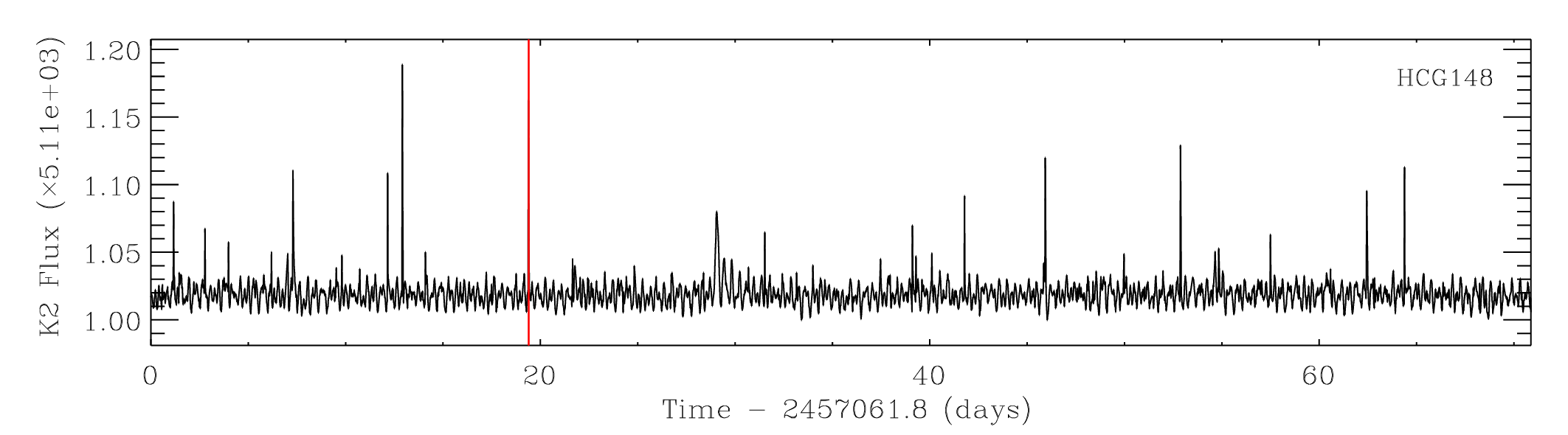}
\caption{Entire K2 light curves of the stars selected in this paper}
\label{appB_fig1}
\end{figure}

Four very bright flares occurred in the stars HHJ~273, HHJ~336, HCG~146, and HCG~150. In Figs. \ref{appB_fig2} to \ref{appB_fig5} we show these flares and the continuum subtracted light curves. We calculated the total energy released in the Kepler band during these flares as described in Sect. \ref{glob_flare}. Results are shown in Table \ref{fourflares_table}. All these flares released more than 7.9$\times$10$^{34}\,$erg of energy in the optical, and were therefore more energetic than the flares occurring during XMM observations. Furthermore, if plotted in the plane of stellar period versus optical flare duration, they share the same behavior as the 12 flares analyzed in this paper (Fig. \ref{tauvsper_plot}), with the shortest flare (4.3$\,$ks) occurring in the most rapidly rotating star (HHJ~336), with a similar duration to the other flares observed in stars with similar rotation periods. A forthcoming complete analysis of the flares observed by K2 in the Pleiades will confirm this relation. 

\begin{figure}[!h]
\centering      
\includegraphics[width=9cm]{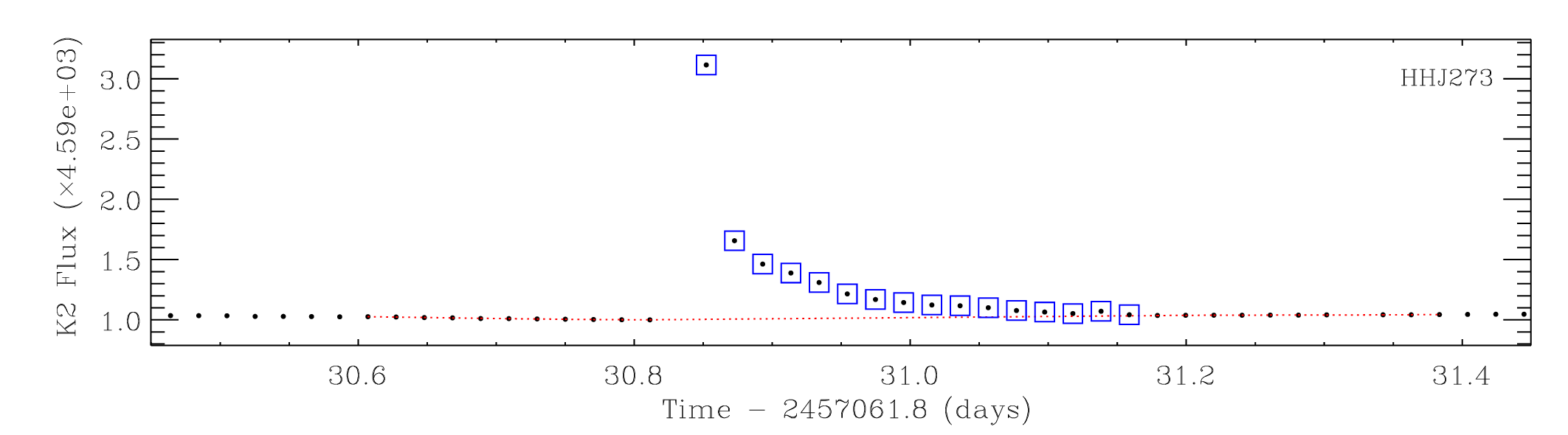}
\includegraphics[width=9cm]{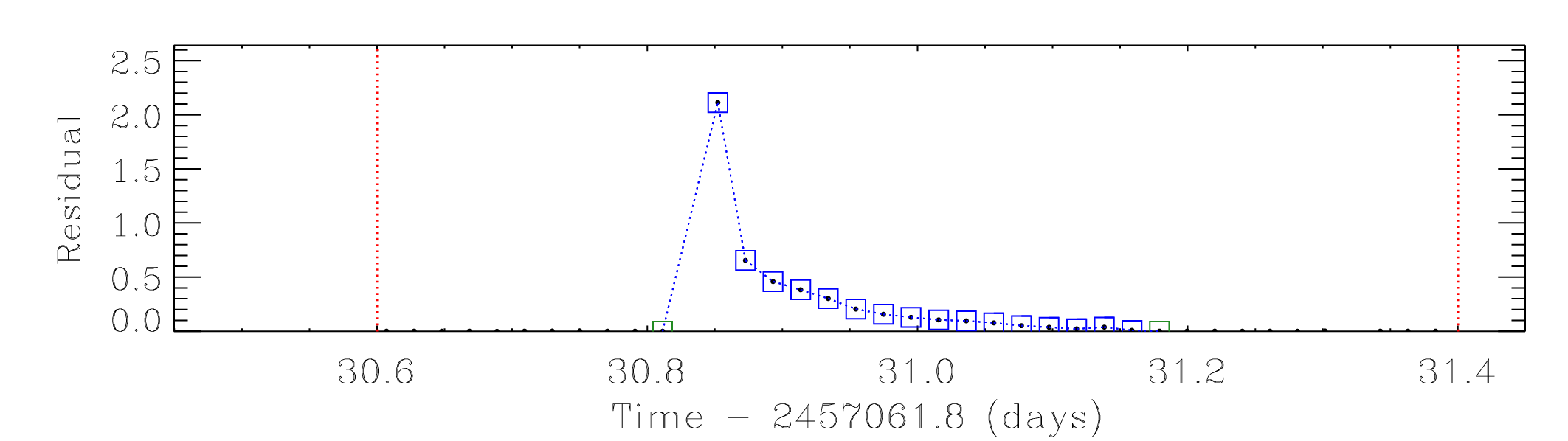}
\caption{Bright flare occurring in HHJ~273 (left panel) and continuum-subtracted light curve during the flare (right panel). Panel layout and content as in the top panels of Fig.~\ref{HHJ336_lc}.}
\label{appB_fig2}
\end{figure}

\begin{figure}[!h]
\centering      
\includegraphics[width=9cm]{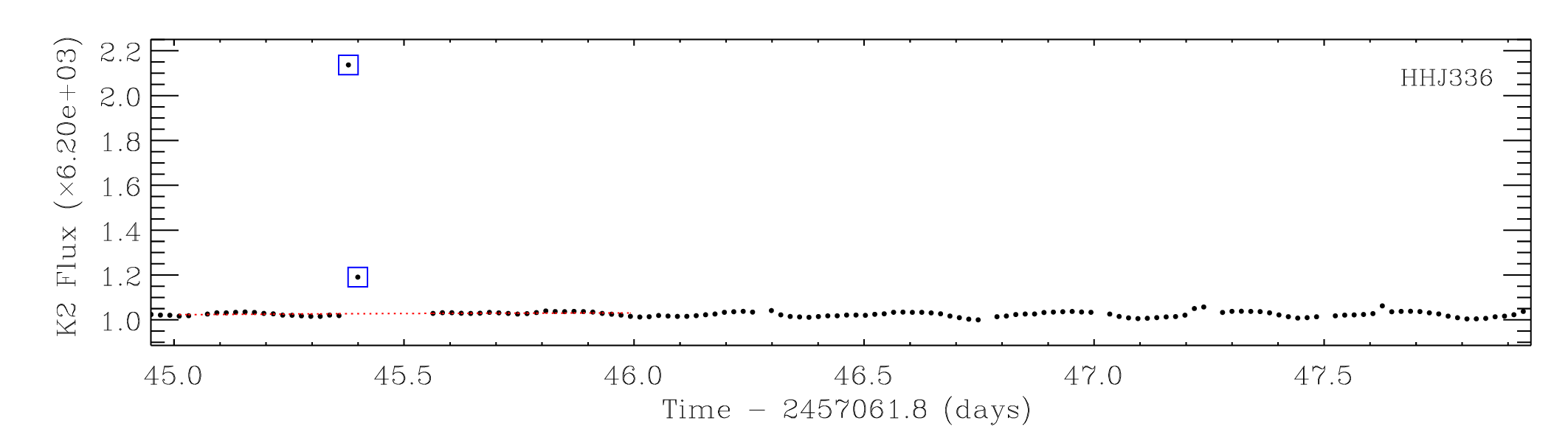}
\includegraphics[width=9cm]{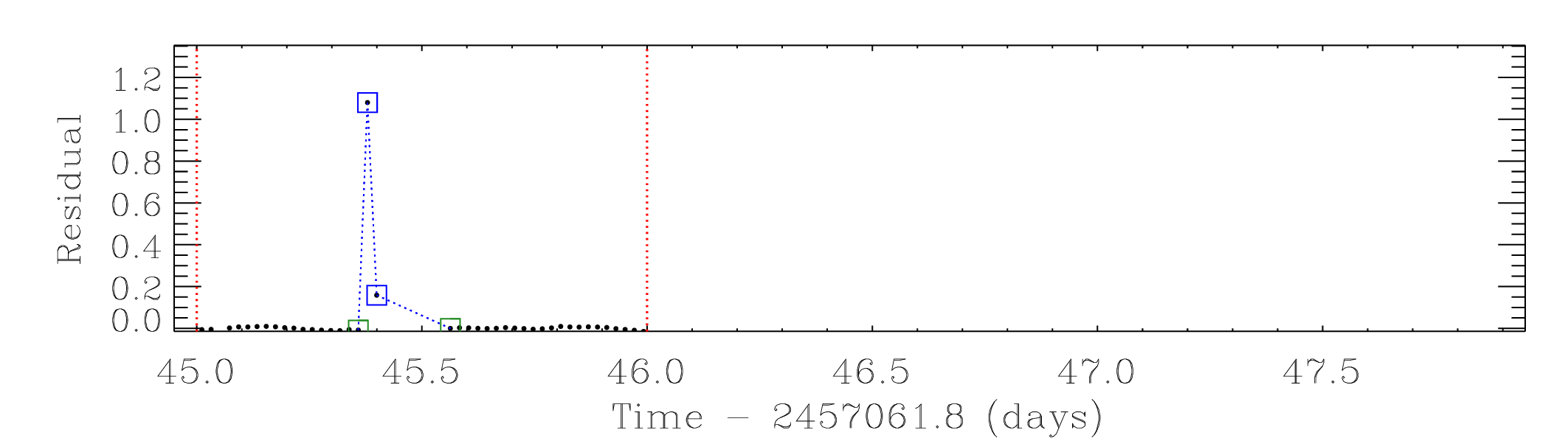}
\caption{Bright flare occurring in HHJ~336 (left panel) and continuum-subtracted light curve during the flare (right panel). Panel layout and content as in the top panels of Fig.~\ref{HHJ336_lc}.}
\label{appB_fig3}
\end{figure}

\begin{figure}[!h]
\centering      
\includegraphics[width=9cm]{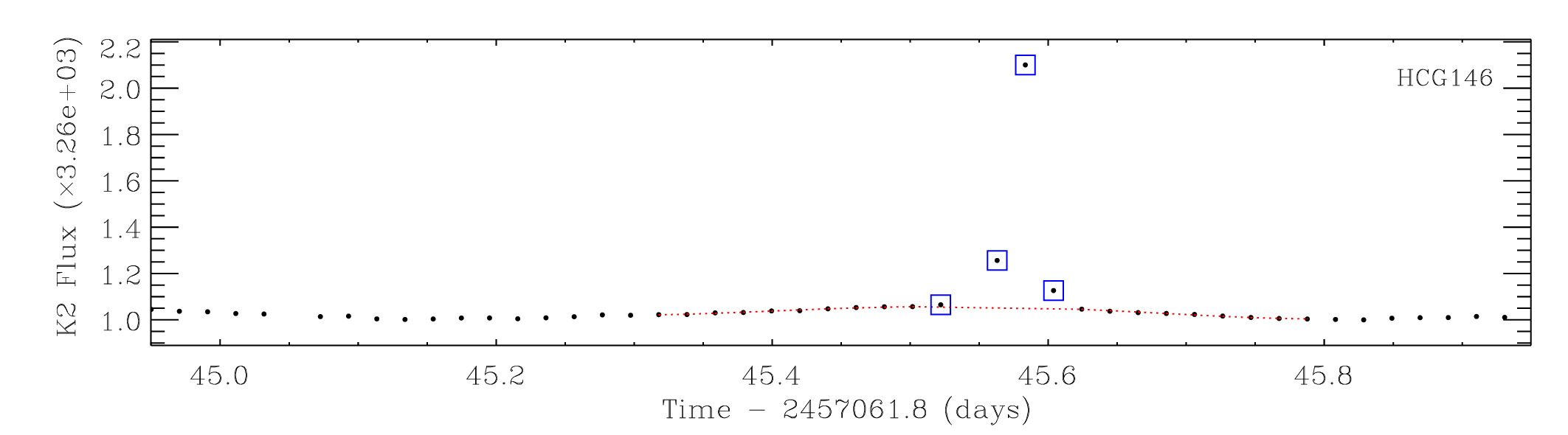}
\includegraphics[width=9cm]{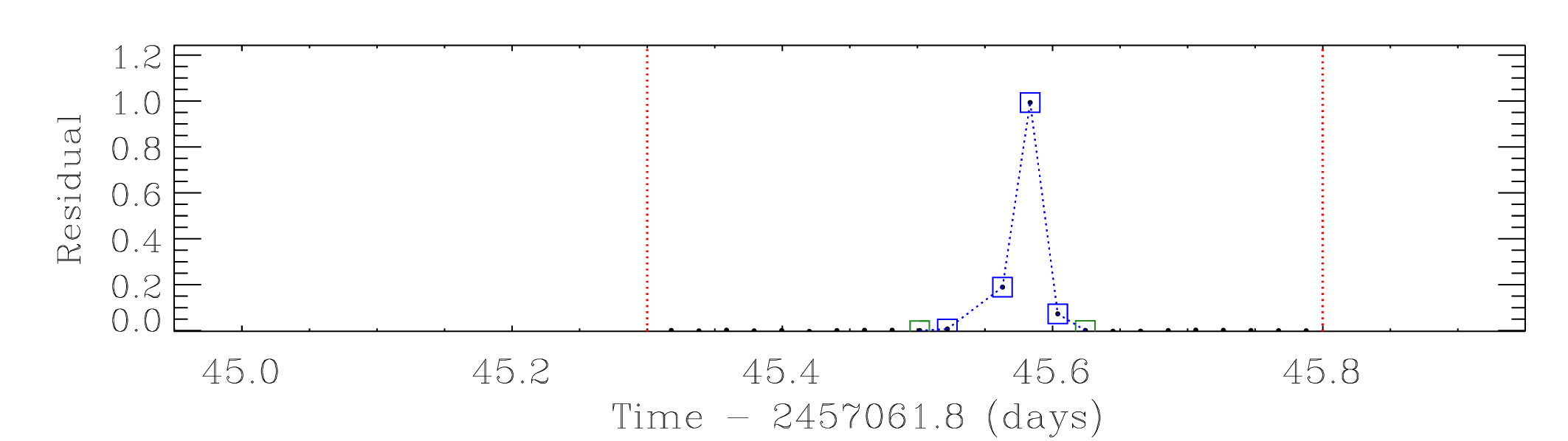}
\caption{Bright flare occurring in HCG~146 (left panel) and continuum-subtracted light curve during the flare (right panel). Panel layout and content as in the top panels of Fig.~\ref{HHJ336_lc}.}
\label{appB_fig4}
\end{figure}

\begin{figure}[!h]
\centering      
\includegraphics[width=9cm]{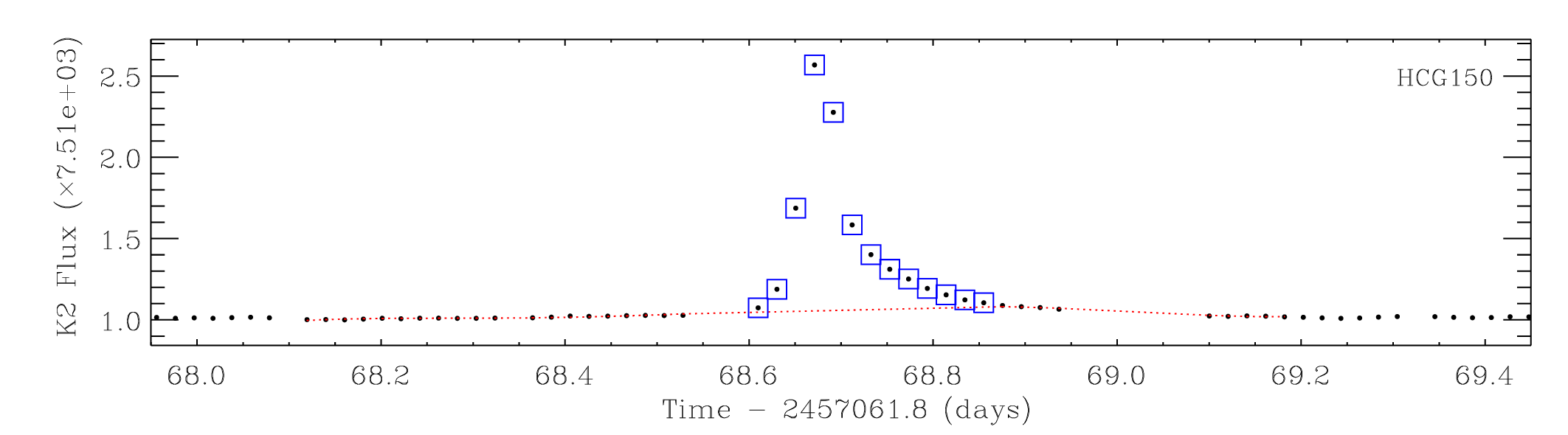}
\includegraphics[width=9cm]{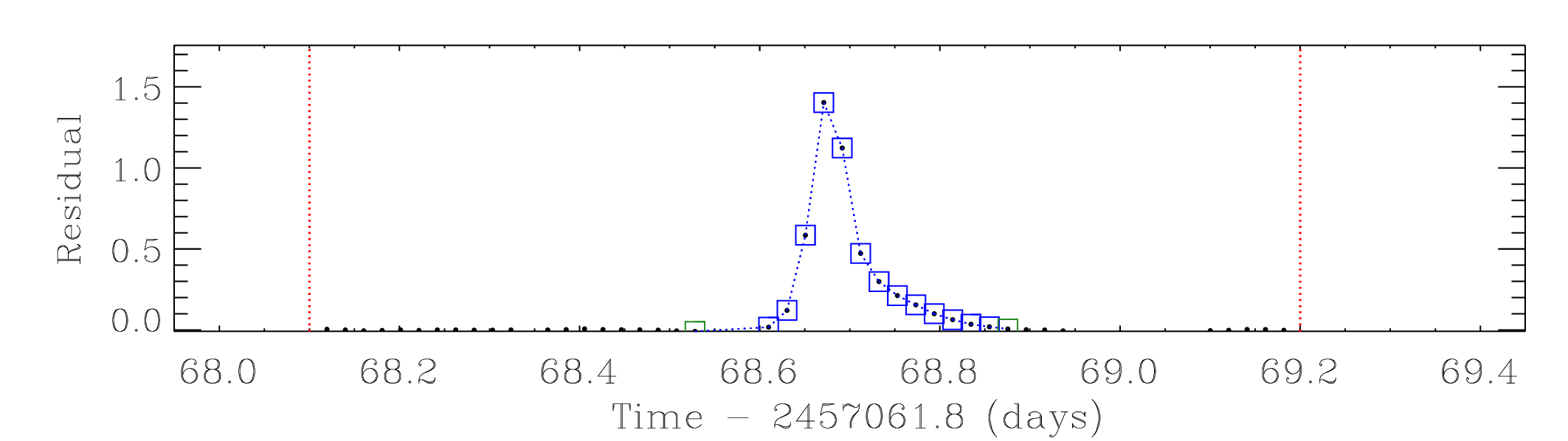}
\caption{Bright flare occurring in HCG~150 (left panel) and continuum-subtracted light curve during the flare (right panel). Panel layout and content as in the top panels of Fig.~\ref{HHJ336_lc}.}
\label{appB_fig5}
\end{figure}

    \begin{table}[!h]
    \caption{Properties of the four bright flares observed in HHJ~273, HHJ~336, HCG~146, and HCG~150.}
    \label{fourflares_table}
    \centering                       
    \begin{tabular}{|c|c|c|c|}
    \hline
    \multicolumn{1}{|c|}{Star} &
    \multicolumn{1}{|c|}{log(E$_{\rm kep,flare}$)} &
    \multicolumn{1}{|c|}{t$_{\rm kep}$} &
    \multicolumn{1}{|c|}{Period} \\    
    \hline
    \multicolumn{1}{|c|}{} &
    \multicolumn{1}{|c|}{[erg]} &
    \multicolumn{1}{|c|}{ksec} &
    \multicolumn{1}{|c|}{days} \\    
    \hline
    HHJ~273 & $35.6\pm0.4$ & $31.8\pm1.3$ & 1.15 \\
    HHJ~336 & $35.4\pm0.5$ & $4.3\pm1.3$  & 0.37 \\
    HCG~146 & $34.9\pm0.5$ & $10.6\pm1.3$ & 0.68 \\
    HCG~150 & $35.5\pm0.4$ & $30.0\pm1.3$ & 1.08 \\
    \hline
    \multicolumn{4}{l}{} \\
    \end{tabular}
    \end{table}

\newpage
\section{Simulated sequences of photons detection time}
\label{AppC_sec}

Here we show five simulated sequences of photon-detection time as explained in Sect. \ref{duration_sect}. 

   \begin{figure}[!h]
        \centering      
    \includegraphics[width=6cm]{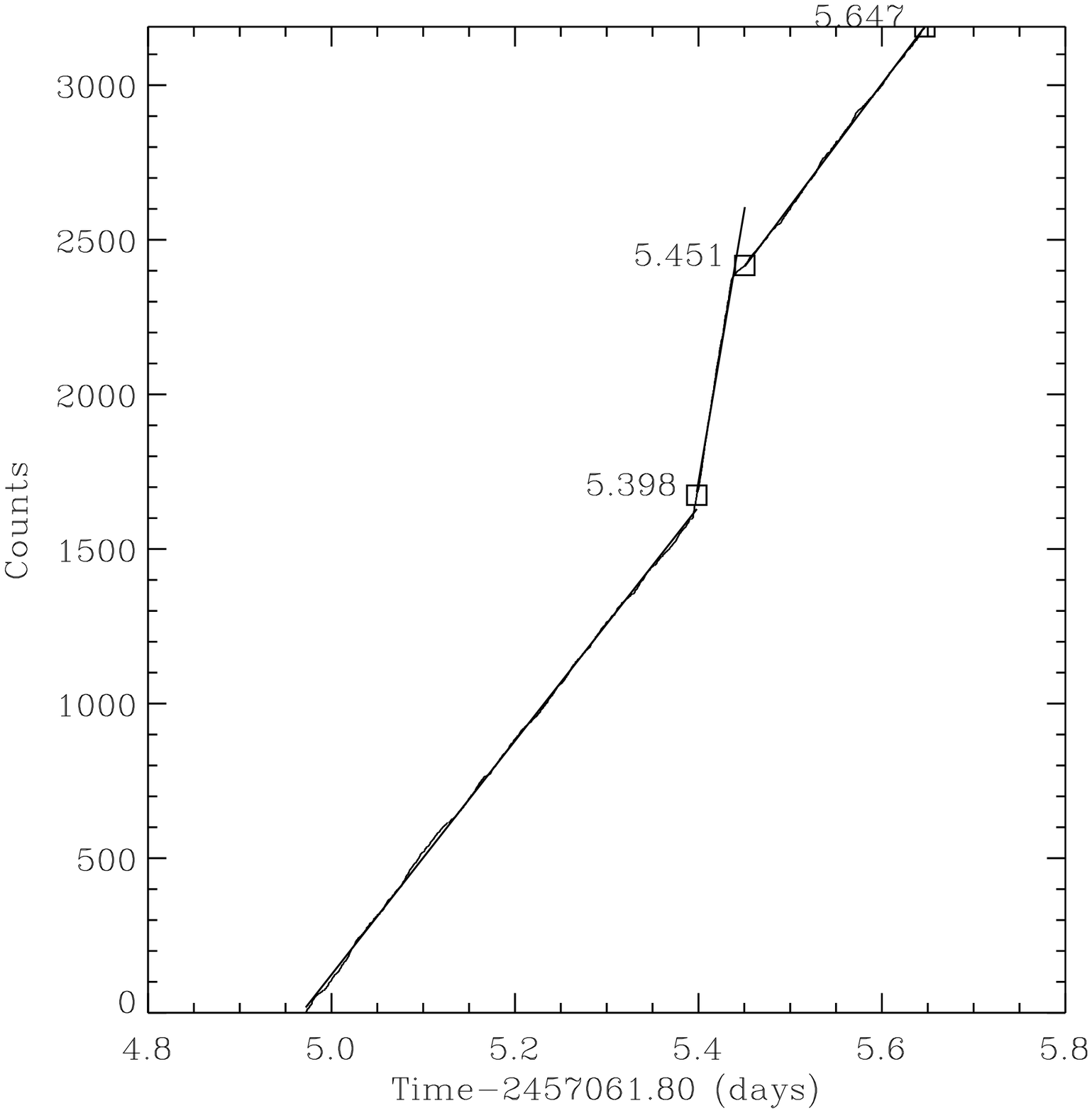}
    \includegraphics[width=6cm]{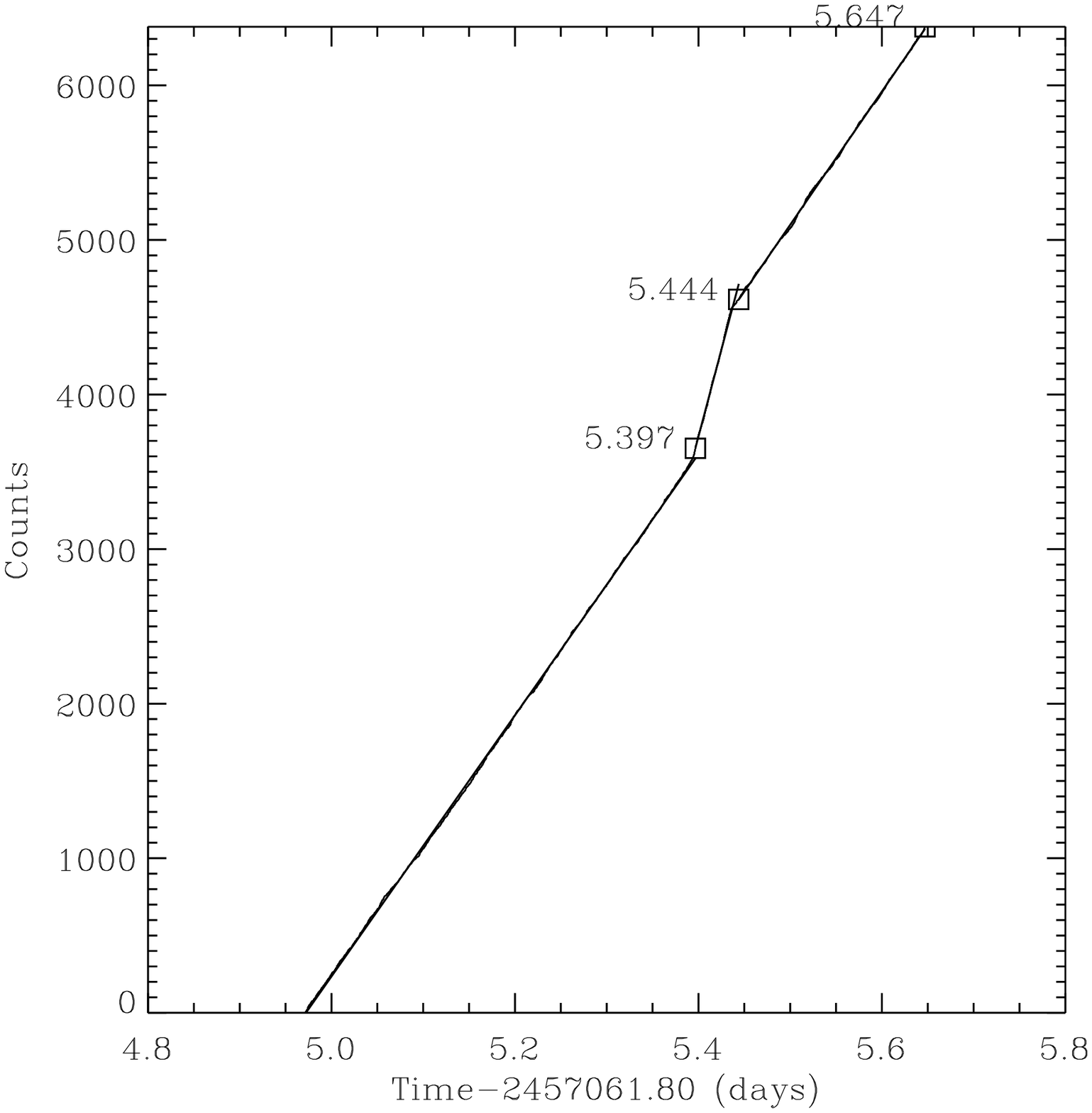}
    \includegraphics[width=6cm]{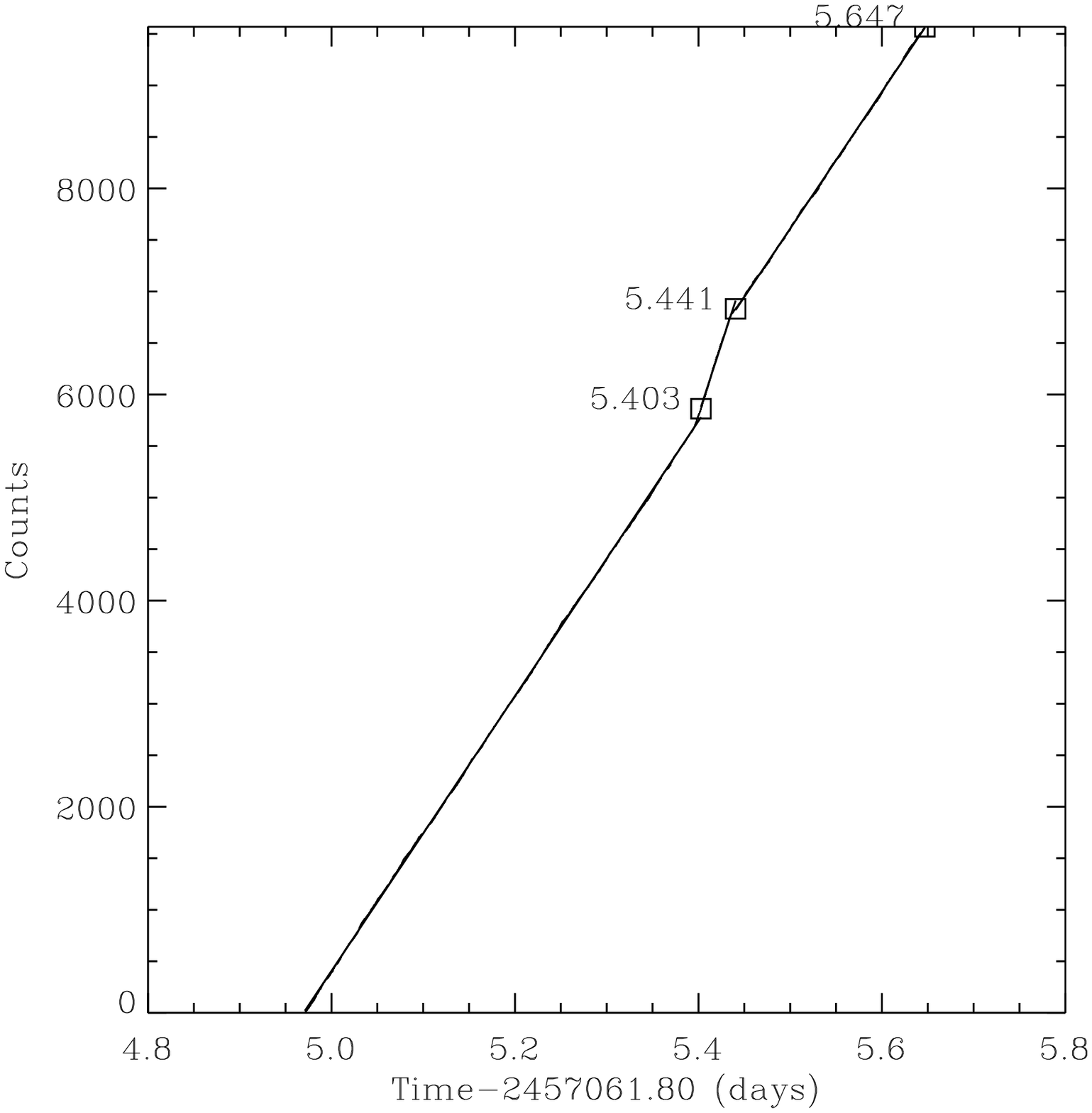}
    \includegraphics[width=6cm]{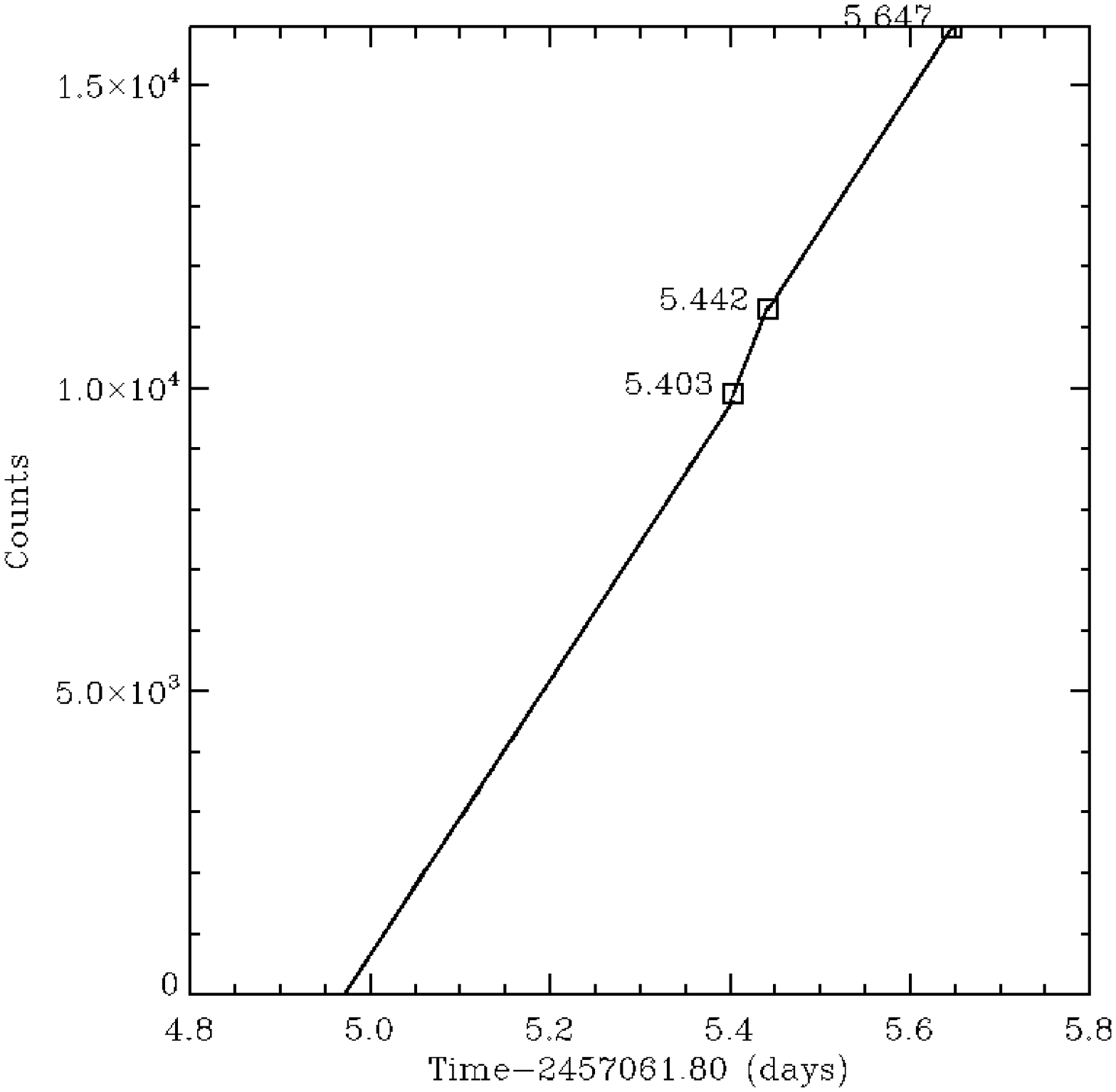}
    \includegraphics[width=6cm]{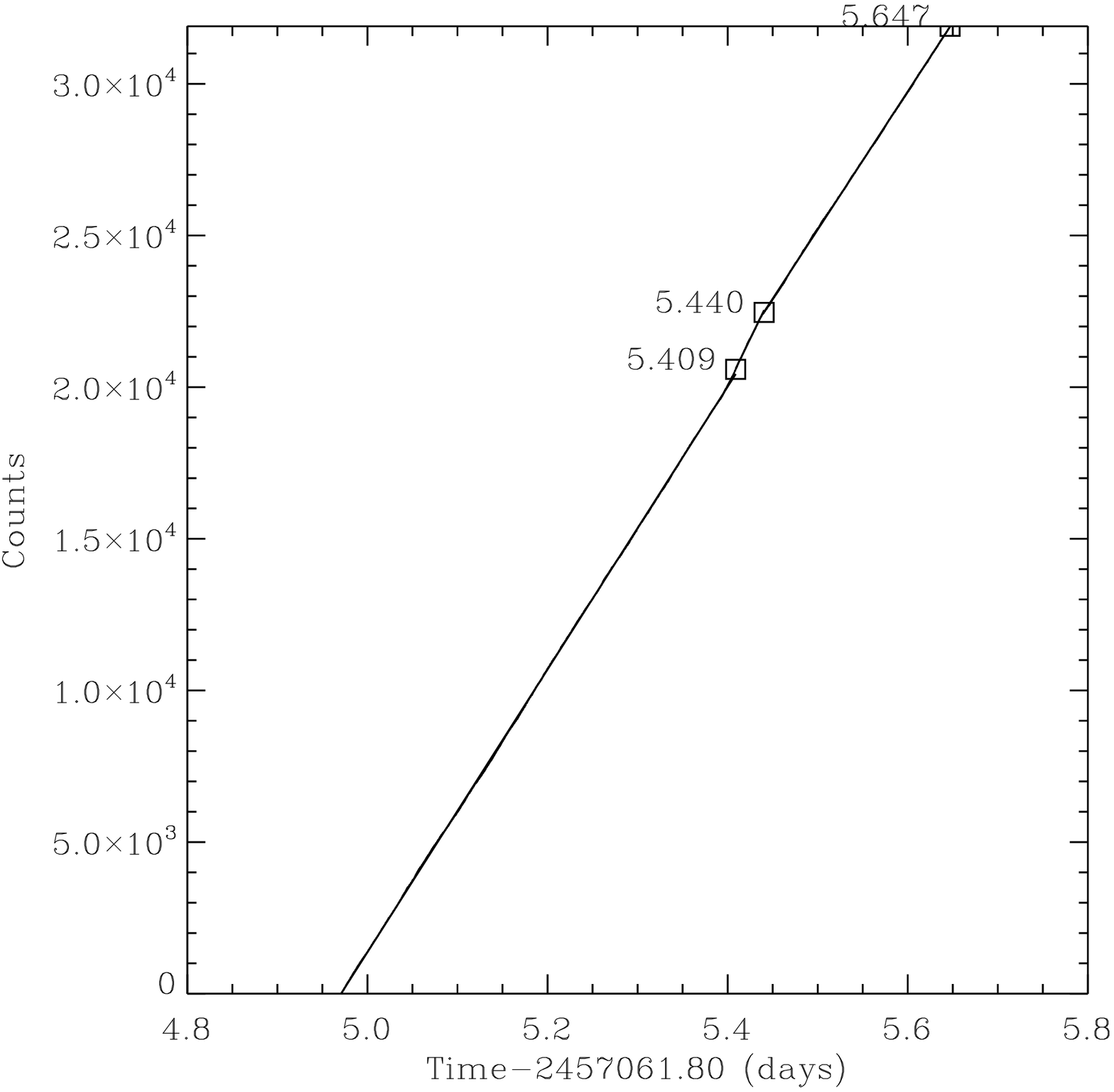}
    \caption{Simulated sequence of the arrival time of X-ray photons, adopting (from the top-left panel to the bottom-right panel) a quiescent count rate of 0.05$\,$cnt/s, 0.1$\,$cnt/sec, 0.15$\,$cnt/s, 0.25$\,$cnt/s, and 0.5$\,$cnt/s.}
        \label{AppC_fig1}
        \end{figure}

\end{appendix}
\end{onecolumn}
\end{document}